\documentclass[journal,onecolumn,final]{IEEEtran}%
\usepackage{amssymb}
\usepackage{amsmath}
\usepackage{amsfonts}
\usepackage{graphicx}
\usepackage{epsfig}
\usepackage{epstopdf}
\usepackage{theorem}
\usepackage{color}
\providecommand{\U}[1]{\protect\rule{.1in}{.1in}}
\newtheorem{theorem}{Theorem}

\newtheorem{lemma}{Lemma}

\newtheorem{proposition}{Proposition}

\newtheorem{remark}{Remark}

\begin{document}

\title{On the Resolution Probability of Conditional and Unconditional Maximum
Likelihood DoA Estimation}
\author{Xavier Mestre$^{\ast}$\thanks{This work was partially supported by the Catalan
and Spanish grants 2017-SGR-01479 and RTI2018-099722-B-I00. This paper was
presented in part at the European Signal Processing Conference EUSIPCO'13.},
Pascal Vallet$^{\ast\ast}$\\$^{\ast}$Centre Tecnol\`{o}gic de Telecomunicacions de Catalunya, Av. Carl
Friedrich Gauss, 7, 08860 Castelldefels, Barcelona (Spain), e-mail:
\texttt{xavier.mestre@cttc.cat}\\$^{\ast\ast}$Institut Polytechnique de Bordeaux, Laboratoire IMS, 351, Cours
de la Lib\'{e}ration, 33405 Talence (France), e-mail:
\texttt{pascal.vallet@bordeaux-inp.fr}}
\maketitle

\begin{abstract}
After decades of research in Direction of Arrival (DoA) estimation, today
Maximum Likelihood (ML) algorithms still provide the best performance in terms
of resolution capabilities. At the cost of a multidimensional search, ML
algorithms achieve a significant reduction of the outlier production mechanism
in the threshold region, where the number of snapshots per antenna and/or the
signal to noise ratio (SNR) are low. The objective of this paper is to
characterize the resolution capabilities of ML algorithms in the threshold
region. Both conditional and unconditional versions of the ML algorithms are
investigated in the asymptotic regime where both the number of antennas and
the number of snapshots are large but comparable in magnitude. By using random
matrix theory techniques, the finite dimensional distributions of both cost
functions are shown to be Gaussian distributed in this asymptotic regime, and
a closed form expression of the corresponding asymptotic covariance matrices
is provided. These results allow to characterize the asymptotic behavior of
the resolution probability, which is defined as the probability that the cost
function evaluated at the true DoAs is smaller than the values that it takes
at the positions of the other asymptotic local minima.

\end{abstract}

\begin{IEEEkeywords}
Conditional Maximum Likelihood, Unconditional Maximum Likelihood, DoA Estimation, Random Matrix Theory, Central Limit Theorem.
\end{IEEEkeywords}

\section{Introduction\label{sec:intro}}

The determination of the direction of arrival (DoA) of one or multiple
far-field sources continues to be a highly relevant problem in multiple
fields, such as radar, sonar, seismology or radioastronomy. Among all the DoA
determination methods available today, classical Maximum Likelihood (ML)
procedures remain to be the ones that offer the best performance in terms of
both precision and spatial resolution. The extraordinary performance of ML
methods comes at the expense of an increased computational complexity, since
they require a non-linear multi-dimensional search instead of a
one-dimensional one (as it is the case in subspace-based algorithms). However,
ML methods still represent the only valid alternative in challenging scenarios
with closely located and/or highly correlated sources. They are also the most
attractive solution in offline processing applications.

Several alternatives have been proposed in the literature in order to
alleviate the high computational complexity associated with the non-linear
multidimensional search of ML methods. One can differentiate between
algorithms that try to simplify the local search procedure and algorithms that
aim to simplify the global search. Among the first group, we can include the
alternating projection method \cite{ZISK88A}, the IQML algorithm\ for Uniform
Linear Arrays (ULA) \cite{Bresler86}, some methods based on large sample
volume approximations of the ML objective function (such as MODE
\cite{Stoica90b} or related alternatives \cite{Swindlehurst94, Stoica96}), and
some recently proposed approximate ML methods for the specific case where only
two closely spaced sources are present in the scenario \cite{Vincent14,
Vincent14b}. All these methods achieve a significant reduction of the
computational complexity at the expense of a certain performance loss, caused
by the fact that the ML objective function is in fact approximated. The second
family of ML-based DoA detection algorithms are aimed at simplifying the
global multidimensional search, avoiding the computational burden associated
with an evaluation of the ML objective function in a uniform multidimensional
fine grid. One may include here genetic algorithms for global search
\cite{Sharman89, Li07}, or two-stage methods that first select a set of
candidate DoAs with simpler one-dimensional search methods and then refine
these initial estimations by a multidimensional ML search procedure
\cite{Stoica99, abramovich2010}. The aim of all these global search methods is
to avoid convergence to a local extremum of the ML objective function while
avoiding the need to evaluate this function at a high number of points in the
multi-dimensional grid.

One must point out that there exist two alternative ML DoA estimation
procedures with different objective functions, depending on whether the
signals are modeled as stochastic or as unknown deterministic parameters. The
Unconditional ML (UML) method is based on the assumption that the source
signals are temporally white Gaussian random variables \cite{BOHM860,
stoica95}, whereas the Conditional ML (CML) method simply treats them as
unknown deterministic parameters that need to be estimated. It was early
recognized \cite{STOI90A} that the UML method asymptotically outperforms CML
when the number of samples tends to infinity for a fixed array dimension. In
fact, it was shown in \cite{stoica89a} that, contrary to UML, the CML method
is statistically inefficient, in the sense that the algorithm does not achieve
the Cram\'{e}r-Rao Bound (CRB) corresponding to the deterministic signal
assumption. This effect is caused by the fact that the total number of
parameters that need to be estimated by the CML\ method increases with the
sample size, whereas this is not the case in the UML algorithm. On the other
hand, it was demonstrated in \cite{renaux07} that CML and UML are in fact
equivalent at large signal to noise ratio (SNR), meaning that the difference
between their estimates converges to zero in probability, even for finite
values of the sample volume.

All of the above performance\ results for CML\ and UML are only valid in the
small error regime, that is when the estimated DoAs are very close to the true
ones, either because of a relatively large sample volume with respect to the
array dimension \cite{STOI90A} or a relatively low noise power level with
respect to the source signals \cite{Renaux06,renaux07}. Unfortunately, none of
the above insights carry over to the more relevant case where the number of
samples is comparable to the array dimension and the SNR\ is moderately low.
Note that this is precisely the regime where ML offers the main advantages
with respect to other one-dimensional DoA estimation techniques, such as more
conventional subspace based approaches. The main objective of this paper will
be the performance characterization of the CML and UML techniques in this
\textquotedblleft threshold region\textquotedblright, whereby both the number
of samples per antenna and the SNR take moderate values. This region is
typically characterized by a systematic appearance of outliers in the DoA
estimates, which are mainly caused by the incapability of resolving closely
spaced sources.

Unfortunately, the performance of CML\ and UML DoA estimation methods in this
threshold region has received less attention in the literature. One of the
most relevant contributions in this direction was the work by Athley in
\cite{athley05}, which characterized the probability of resolving closely
spaced sources for finite values of the sample size and SNR. Some additional
insights into the problem were given in \cite{abramovich11}, where the outlier
production mechanism was related to the asymptotic eigendecomposition of the
observation covariance matrix. Recently, the approach in \cite{athley05} has
been used to study the resolution probability of both CML and UML in the
multi-frequency case \cite{Filippini19}. In this paper, we also follow the
approach in \cite{athley05} and characterize the resolution probability of
these two ML methods by studying the stochastic behavior of the CML and UML
objective functions under the assumption that the observations are Gaussian
random variables. Our approach will be asymptotic in both the number of
antennas and the number of samples, although we will always assume that both
the SNR and the number of samples per antenna are finite and bounded
quantities. In practice, this asymptotic regime models the situation where the
number of antennas is of the same order of magnitude as the number of
available snapshots.

The approach followed by Athley in \cite{athley05} conforms to the classical
Method of Interval Errors (MIE), which predicts the threshold region
performance of estimation methods based on minimizing a certain cost function.
This method has also been used in \cite{Richmond05} to characterize the
threshold performance of the Capon method and in \cite{Richmond06} to study
the threshold of the ML estimation method corresponding to one signal in
spatially colored noise. The main idea is that the mean squared error (MSE) of
the estimated parameters can be decomposed into a sum of two weighted terms, a
local error term (the small-error variance), and an outlier term for global
errors (outliers), that is
\begin{equation}
MSE=P_{res}MSE_{\text{small}}+(1-P_{res})MSE_{\text{large}}
\label{eq:Pred_MSE}%
\end{equation}
where $MSE_{\text{small}}$ is the small error MSE (usually the Cram\'{e}r-Rao
Bound), $MSE_{\text{large}}$ is the large error MSE (typically approximated as
the average MSE\ under a uniformly distributed parameter choice) and $P_{res}$
is the resolution probability. In order to characterize the resolution
probability, the MIE\ identifies all the local minima of the asymptotic
(deterministic) cost function. Clearly, only one of these local minima will be
associated to the true value of the parameters\ (the global minimum, assuming
consistency), whereas the rest will contain the typical values taken by
outliers, which can all be associated to other local minima. The resolution
probability is defined as the probability that the value of the cost function
at the position of the deterministic global minima (true DoAs) is smaller than
the values at the rest of the local ones. Observe that the global behavior of
the random cost function is summarized as the behavior in a finite set of
deterministic points.

The rest of the paper is structured as follows. Section \ref{sec:signal_model}
introduces the conditional and unconditional ML DoA estimation methods and
presents the two main results of the paper, namely (i) almost sure convergence
of the ML cost functions and (ii) convergence in law of the associated finite
dimensional distributions. The proof of these two main results is given in
Section \ref{sec_prooftheorem} and Section \ref{sec:proof_theorem_CLT}
respectively. Section \ref{sec:NumericalEval} provides a numerical evaluation
of the asymptotic probability of resolution that is established according to
the results in Section \ref{sec:signal_model} and finally Section
\ref{sec:Conclusions} concludes the paper.

\section{\label{sec:signal_model}DoA ML Methods and Main\ Asymptotic Results}

Let us denote by $\mathbf{y}(n)\in\mathbb{C}^{M\times1}$ a column vector that
contains the complex samples received by an array of $M$ elements at the $n$th
time instant. This signal contains the contribution from $K<M$ far field
sources with DoAs $\bar{\theta}=\left[  \bar{\theta}(1),\ldots,\bar{\theta
}(K)\right]  ^{T}$, so that we can express $\mathbf{y}(n)$ as
\[
\mathbf{y}(n)=\mathbf{A}(\bar{\theta})\mathbf{s}(n)+\mathbf{n}(n).
\]
In this expression, $\mathbf{s}(n)$ $\in\mathbb{C}^{K\times1}\,$\ is a column
vector with the source signal samples, $\mathbf{A}(\bar{\theta})$
$\in\mathbb{C}^{M\times N}$ is a matrix that contains as columns the steering
vectors corresponding to the different directions of arrival and
$\mathbf{n}(n)\in\mathbb{C}^{M\times1}$ is a column vector with background
noise entries, assumed independent and identically distributed (i.i.d.)
according to a\ circularly symmetric Gaussian law with with zero mean and
unknown variance $\sigma^{2}$. Assume that $N$ different snapshots or
realizations of $\mathbf{y}(n)$ are available, and denote\textbf{\ }%
$\mathbf{Y}=\left[  \mathbf{y}(1),\ldots,\mathbf{y}(N)\right]  $. As mentioned
above, two different ML methods for the estimation of $\bar{\theta}$ can be
derived depending on the nature of the signal vectors $\mathbf{s}(n)$. In both
cases, the minimization is carried out in a multidimensional compact domain
$\Theta_{K}\subset\mathbb{R}^{K}$ for which $\mathbf{A}(\theta)$ has full
column rank, for example%
\begin{equation}
\Theta_{K}=\{\theta_{1}<\ldots<\theta_{K},\theta_{k}\geq\theta_{k-1}%
+\varepsilon,\theta_{1}\geq-\pi+\varepsilon,\theta_{K}\leq\pi-\varepsilon\}
\label{eq:feasiblity_set}%
\end{equation}
for some small $\varepsilon>0$.

\subsection{Conditional ML}

In the conditional model, the signals are assumed to be deterministic unknown
parameters, denoted by $\mathbf{S=}\left[  \mathbf{s}(1),\ldots,\mathbf{s}%
(N)\right]  $. The corresponding ML estimator is the minimizer of the
normalized negative log-likelihood
\[
L_{C}\left(  \sigma^{2},\mathbf{S,}\theta\right)  =M\log\sigma^{2}+\frac
{1}{N\sigma^{2}}\left\Vert \mathbf{Y-A}(\theta)\mathbf{S}\right\Vert _{F}^{2}%
\]
where $\left\Vert \text{\textbf{\textperiodcentered}}\right\Vert _{F}$ is the
Frobenius norm and where we constrain $\sigma^{2}>0$. The minimum is achieved
at $\mathbf{\hat{S}(}\theta)=(\mathbf{A}^{H}(\theta)\mathbf{A}(\theta
))^{-1}\mathbf{A}^{H}(\theta)\mathbf{Y}$ and, regardless of $\sigma^{2}$, the
CML estimator of $\theta$ is obtained by minimizing the cost function
$L_{C}(\sigma^{2},\mathbf{\hat{S}}(\theta)\mathbf{,}\theta)$ or, equivalently,
the function
\begin{equation}
\hat{\eta}_{C}\left(  \theta\right)  =\frac{1}{M}\operatorname*{tr}\left[
\mathbf{P}_{A}^{\perp}(\theta)\mathbf{\hat{R}}\right]
\label{eq_CML_costfunction}%
\end{equation}
where $\mathbf{\hat{R}}$ is the sample covariance matrix, i.e. $\mathbf{\hat
{R}}=\frac{1}{N}\mathbf{YY}^{H}$, and $\mathbf{P}_{A}^{\perp}(\theta)$ is the
orthogonal projection matrix onto the null column space of $\mathbf{A}%
(\theta)$, namely $\mathbf{P}_{A}^{\perp}(\theta)=\mathbf{I}_{M}%
-\mathbf{P}_{A}(\theta)$, where
\[
\mathbf{P}_{A}(\theta)=\mathbf{A}(\theta)\left(  \mathbf{A}^{H}(\theta
)\mathbf{A}(\theta)\right)  ^{-1}\mathbf{A}^{H}(\theta).
\]

\subsection{Unconditional ML}

The unconditional model assumes the signals $\mathbf{s}(n)$ are independent
circularly symmetric Gaussian vectors with zero mean and unknown covariance
matrix $\mathbf{P}_{s}=\mathbb{E}\left[  \mathbf{s}(n)\mathbf{s}%
^{H}(n)\right]  $. In this case, the ML estimator can be formulated as the
minimizer of normalized negative log-likelihood%
\[
L_{U}\left(  \mathbf{P}_{s},\sigma^{2},\theta\right)  =\log\det\mathbf{R}%
\left(  \mathbf{P}_{s},\sigma^{2},\theta\right)  +\mathrm{tr}\left[
\mathbf{R}^{-1}\left(  \mathbf{P}_{s},\sigma^{2},\theta\right)  \mathbf{\hat
{R}}\right]
\]
where here again $\mathbf{\hat{R}}$ is the sample covariance matrix defined
above and where $\mathbf{R}\left(  \mathbf{P}_{s},\sigma^{2},\theta\right)
=\mathbf{\mathbf{A}}(\theta\mathbf{)P}_{s}\mathbf{A}^{H}(\theta)+\sigma
^{2}\mathbf{I}_{M}$. The above minimization is carried with the constraints
$\mathbf{P}_{s}\geq0,$ $\sigma^{2}>0$ and $\theta\in\Theta_{K}$. The solution
to the above optimization problem can be formulated as follows. Let us denote
by $\hat{\alpha}_{1}(\theta\mathbf{)}\leq\ldots\leq\hat{\alpha}_{K}%
(\theta\mathbf{)}$ the eigenvalues of the $K\times K$ matrix $\mathbf{U}%
_{A}^{H}(\theta)\mathbf{\hat{R}\mathbf{U}}_{A}(\theta)$, where $\mathbf{U}%
_{A}(\theta)=\mathbf{A}(\theta)(\mathbf{A}^{H}(\theta)\mathbf{A}%
(\theta))^{-1/2}$ and let $\mathbf{\hat{q}}_{1}(\theta),\ldots,\mathbf{\hat
{q}}_{K}(\theta)$ denote the associated eigenvectors. For $k=1,\ldots,K $,
denote by $\hat{\Lambda}_{k}(\theta)$ a $k\times k$ diagonal matrix that
contains the $k$ highest eigenvalues and let $\mathbf{\hat{Q}}_{k}(\theta)$ be
defined as a $K\times k$ matrix that contains the associated eigenvectors,
that is $\hat{\Lambda}_{k}(\theta)=\mathrm{diag}(\hat{\alpha}_{K-k+1}%
(\theta),\ldots,\hat{\alpha}_{K}(\theta))$ and $\mathbf{\hat{Q}}_{k}%
(\theta)=[\mathbf{\hat{q}}_{K-k+1}(\theta),\ldots,\mathbf{\hat{q}}_{K}%
(\theta)]$. Furthermore, we define $\hat{\sigma}_{k}^{2}(\theta)$ as
\[
\hat{\sigma}_{k}^{2}(\theta)=\frac{1}{M-k}\mathrm{tr}\left[  \mathbf{\hat{R}%
}\left(  \mathbf{I}-\mathbf{U}_{A}(\theta)\mathbf{\hat{Q}}_{k}(\theta
)\mathbf{\hat{Q}}_{k}^{H}(\theta)\mathbf{\mathbf{U}}_{A}^{H}(\theta)\right)
\right]  .
\]
Now, let $m$ denote the maximum integer such that\footnote{It can be seen that
if this inequality holds for a particular index $m$, it must hold for all
integers smaller than $m$.} $\hat{\Lambda}_{m}(\theta)>\hat{\sigma}_{m}%
^{2}(\theta)\mathbf{I}_{m}$. If there does not exist such an integer (meaning
that $\hat{\alpha}_{K}(\theta\mathbf{)}<\hat{\sigma}_{1}^{2}(\theta)$) then
the optimum is achieved at $\mathbf{P}_{s}=0$ and $\sigma^{2}=\frac{1}%
{M}\mathrm{tr}\mathbf{\hat{R}}$. Otherwise, the function $L_{U}(\mathbf{P}%
_{s},\sigma^{2},\theta)$ reaches its minimum at $\sigma^{2}=\hat{\sigma}%
_{m}^{2}(\theta)$ and $\mathbf{P}_{s}=\mathbf{\hat{P}}_{s}^{(m)}$, where%
\[
\mathbf{\hat{P}}_{s}^{(m)}=\left(  \mathbf{A}^{H}(\theta)\mathbf{A}%
(\theta)\right)  ^{-1/2}\mathbf{\hat{Q}}_{m}(\theta)\left(  \hat{\Lambda}%
_{m}(\theta)-\hat{\sigma}_{m}^{2}(\theta)\mathbf{I}_{m}\right)  \mathbf{\hat
{Q}}_{m}^{H}(\theta)\left(  \mathbf{A}^{H}(\theta)\mathbf{A}(\theta)\right)
^{-1/2}.
\]
Furthermore, the corresponding negative log-likelihood takes the minimum value%
\[
L_{U}\left(  \mathbf{\hat{P}}_{s}^{(m)},\hat{\sigma}_{m}^{2}(\theta
),\theta\right)  =\left(  M-m\right)  \log\hat{\sigma}_{m}^{2}(\theta
)+\log\det\hat{\Lambda}_{m}(\theta)+M.
\]
The UML estimator can therefore be obtained by exploring the above cost
function over the domain $\Theta_{K}$. The problem here is that every time we
need to evaluate the above expression at a particular $\theta$ we need to find
the full eigendecomposition of the matrix $\mathbf{U}_{A}^{H}(\theta
)\mathbf{\hat{R}\mathbf{U}}_{A}(\theta)$, so finding the true UML estimator is
quite complex from the computational complexity. For this reason, typical
implementations of the UML DoA estimator assume that $N>K$ and $\hat{\alpha
}_{1}(\theta\mathbf{)}>\hat{\sigma}_{K}^{2}(\theta)$ \cite{BOHM860, stoica95},
implying that $m=K$ and turning the problem into the minimization of
\begin{equation}
\hat{\eta}_{U}\left(  \theta\right)  =\frac{M-K}{M}\log\hat{\sigma}_{K}%
^{2}(\theta)+\frac{1}{M}\log\det\left(  \mathbf{U}_{A}^{H}(\theta
)\mathbf{\hat{R}\mathbf{U}}_{A}(\theta)\right)  \label{eq:UMLcost0}%
\end{equation}
where
\begin{equation}
\hat{\sigma}_{K}^{2}\left(  \theta\right)  =\frac{1}{M-K}\operatorname*{tr}%
\left[  \mathbf{P}_{A}^{\perp}(\theta)\mathbf{\hat{R}}\right]  .
\label{eq:def_sigmasquare_tilde}%
\end{equation}
By using the identity $\det(\mathbf{I}+\mathbf{AB})=\det(\mathbf{I}%
+\mathbf{BA})$ (valid for matrices of compatible dimensions), this cost
function can also be expressed in the more conventional form
\begin{equation}
\hat{\eta}_{U}\left(  \theta\right)  =\frac{1}{M}\log\det\left[  \hat{\sigma
}_{K}^{2}(\theta)\mathbf{P}_{A}^{\perp}(\theta)+\mathbf{P}_{A}(\theta
)\mathbf{\hat{R}P}_{A}(\theta)\right]  \label{eq_UML_costfunction}%
\end{equation}
which can be evaluated without the need of computing an eigendecomposition at
each $\theta$. Furthermore, this cost function is well defined with
probability one even if $\hat{\alpha}_{1}(\theta\mathbf{)}\leq\hat{\sigma}%
_{K}^{2}(\theta)$ (because $\hat{\alpha}_{1}(\theta\mathbf{)}>0$ almost
surely). For this reason, the minimizer of the above cost function is usually
taken to be, by definition, the UML\ estimator.

When $N<K$ (undersampled regime), we will always have $\hat{\alpha}_{1}%
(\theta\mathbf{)}=\ldots=\hat{\alpha}_{K-N}(\theta\mathbf{)}=0$ and therefore
the cost function in (\ref{eq:UMLcost0})-(\ref{eq_UML_costfunction}) is not
well defined. Noting however that $\hat{\alpha}_{K-N+1}(\theta\mathbf{)}>0$
with probability one, we can instead assume that $\hat{\alpha}_{K-N+1}%
(\theta\mathbf{)}\geq\hat{\sigma}_{N}^{2}(\theta)$, implying that $m=N$ in the
original UML\ problem, which can readily be turned into the minimization of
\[
\hat{\eta}_{U}\left(  \theta\right)  =\frac{M-N}{M}\log\hat{\sigma}_{N}%
^{2}(\theta)+\frac{1}{M}\log\det\left(  \frac{\mathbf{Y}^{H}\mathbf{P}%
_{A}(\theta)\mathbf{Y}}{N}\right)
\]
with $\hat{\sigma}_{N}^{2}(\theta)$ being expressible as in
(\ref{eq:def_sigmasquare_tilde}) simply replacing $K$ by $N$. This cost
function can also be evaluated without the need for eigenvalue decompositions
and is always well defined even if the hypothesis $\hat{\alpha}_{K-N+1}%
(\theta\mathbf{)}\geq\hat{\sigma}_{N}^{2}(\theta)$ does not hold. For this
reason, this is the natural extension of the simplified UML cost function to
the undersampled regime ($N<K$). From now on, we will consider this natural
extension of the UML method to the undersampled regime.

In conclusion, for both undersampled and oversampled regimes, we will assume
that the the minimum positive eigenvalue of $\mathbf{U}_{A}^{H}(\theta
)\mathbf{\hat{R}\mathbf{U}}_{A}(\theta)$ is larger than the associated noise
power estimate, taken to be $\hat{\sigma}_{\tilde{K}}^{2}\left(
\theta\right)  $ as defined in (\ref{eq:def_sigmasquare_tilde}) with $K$
replaced with $\tilde{K}=\min\left(  K,N\right)  $. In both cases, this
assumption allows to evaluate the corresponding cost function without the need
for a parameter-dependent eigenvalue decomposition.

As it can be readily seen, both CML and UML\ methods are based on the
minimization of highly nonlinear multidimensional objective functions
$\hat{\eta}_{C}\left(  \theta\right)  $, $\hat{\eta}_{U}\left(  \theta\right)
$ that typically present multiple local minima. In order to obtain the ML
estimators, one should first identify all the local minima in the parameter
space $\Theta_{K}$ and then select the lowest one as the corresponding
estimated DoAs. The randomness of these ML cost functions will lead to
fluctuations of the local minima, which will generate a loss in both precision
and resolution. Fluctuations in the position of the global minimum around the
true DoAs $\bar{\theta}$ will result in a loss of precision, which will lead
to fluctuations of the estimated DoAs around the true ones $\bar{\theta}$. The
objective of this paper is the characterization of this effect, following the
approach established in \cite{athley05}.

\subsection{First order asymptotic behavior}

In order to overcome the difficulties in the statistical characterization of
the ML cost functions, we will take an asymptotic approach and characterize
the behavior of the two ML\ cost functions when both sample size ($N$) and
array dimension ($M$) are large but comparable in magnitude. Furthermore, we
will assume that the source signals follow the unconditional model, as
formalized in the following.

$\mathbf{(As1)}$ The number of elements of the array depends on the sample
size, $M=M(N)$ and $M(N)\rightarrow\infty$ when $N\rightarrow\infty$.
Furthermore, the quotient $M/N$ converges to a positive constant as
$N\rightarrow\infty$, namely $M/N\rightarrow c$, $0<c<\infty$. From now on the
expressions \textquotedblleft large $N$\textquotedblright\ and
\textquotedblleft large $M$\textquotedblright\ will be indistinctly used to
refer to this asymptotic regime.

$\mathbf{(As2)}$ The observations $\mathbf{y}(n)$, $n=1,\ldots,N$, are
complex, circularly symmetric, independent and identically distributed
Gaussian random variables with zero mean and covariance matrix $\mathbf{R}$.
Furthermore, the eigenvalues of $\mathbf{R}$ are allowed to fluctuate with
increasing $M$ but are always located inside a compact interval of
$\mathbb{R}_{\star}^{+}$ (the real positive axis) independent of $M.$

$\mathbf{(As3)}$ The number of sources $K$ is strictly lower than the number
of sensors, that is $K<M$. Furthermore, $K$ may increase with $N$, so that we
can either have\footnote{The case $K=N$ requires more specific asymptotic
tools and is left for future reserarch. On the other hand, observe that the
case of $K$ constant (independent of $N$) is included in (\ref{eq:oversampled}%
).}
\begin{equation}
0\leq\liminf\frac{K}{N}\leq\limsup\frac{K}{N}<1 \label{eq:oversampled}%
\end{equation}
(oversampled regime) or
\begin{equation}
1<\liminf\frac{K}{N}\leq\limsup\frac{K}{N}<\infty\label{eq:undersampled}%
\end{equation}
(undersampled regime).

We consider here a family of $L$ sequences ($L$ fixed and independent of $M$)
of $K$-dimensional points in $\Theta_{K}$, which will be denoted by
$\{\theta_{M}^{(\ell)}\}_{\ell=1,\ldots,L}$. Our objective is to characterize
the asymptotic behavior of the CML and the UML cost functions evaluated at
these $L$ distinct point sequences. Observe that according to $\mathbf{(As3)}
$ the dimensionality of these points will scale up with the array dimension.
Consider the $M\times M$ deterministic matrix
\[
\mathbf{R}_{A}^{(\ell)}=\mathbf{P}_{A}\left(  \theta_{M}^{(\ell)}\right)
\mathbf{RP}_{A}\left(  \theta_{M}^{(\ell)}\right)
\]
and note that this matrix is always singular due to the fact that $K<M$. We
will denote by $\bar{M}_{\ell}+1$ the total number of distinct eigenvalues of
this matrix, which will be written as $0=\gamma_{0}^{(\ell)}<\gamma_{1}%
^{(\ell)}<\ldots<\gamma_{\bar{M}_{\ell}}^{(\ell)}$. The multiplicity of the
$m$th positive eigenvalue is denoted as $K_{m}^{(\ell)}$, so that, assuming
that $\mathbf{A}(\theta_{M}^{(\ell)})$is full column rank, we will have
$K_{0}^{(\ell)}=M-K$, and $\sum_{m=1}^{\bar{M}_{\ell}}K_{m}^{(\ell)}=K$. Note
that, for the sake of notational simplicity, we omit the dependence on the
observation dimension $M$ in all these quantities. The following assumption is
not really needed for the derivations in this paper, but greatly simplifies
the exposition and derivation of the results. It basically ensures that all
the matrices $\mathbf{R}_{A}^{(\ell)}$ have exactly $K$ positive eigenvalues.
Thanks to $\mathbf{(As2)}$, it is sufficient to ensure that the $\mathbf{A}%
(\theta_{M}^{(\ell)})$ is full column rank at all $\theta_{M}^{(\ell)}$
uniformly in $M$.

$\mathbf{(As4)}$\ For each $\ell=1,\ldots,L$, the sequence of points
$\{\theta_{M}^{(\ell)}\}$ is such that the minimum eigenvalue of
$\mathbf{A}^{H}(\theta_{M}^{(\ell)})\mathbf{A}(\theta_{M}^{(\ell)})$ is
bounded away from zero uniformly in $M$.

We are now ready to introduce the first result of the paper, which
characterizes the first order behavior of the UML\ and the CML cost functions.

\begin{theorem}
\label{theorem:first_order}Under $\mathbf{(As1)}$--$\mathbf{(As4)}$, and for
each $\ell=1,\ldots,L$, we have $|\hat{\eta}_{C}(\theta_{M}^{(\ell)}%
)-\bar{\eta}_{C}(\theta_{M}^{(\ell)})|\rightarrow0$ and $|\hat{\eta}%
_{U}(\theta_{M}^{(\ell)})-\bar{\eta}_{U}(\theta_{M}^{(\ell)})|\rightarrow0$
almost surely as $N\rightarrow\infty$, where $\bar{\eta}_{C}\left(
\theta\right)  $ and $\bar{\eta}_{U}\left(  \theta\right)  $ are two
deterministic equivalent objective functions, defined as follows. For the
CML\ method, we define $\bar{\eta}_{C}\left(  \theta\right)  =M^{-1}%
\mathrm{tr}[\mathbf{P}_{A}^{\perp}(\theta)\mathbf{R]}$. For the UML method, we
must distinguish between undersampled and oversampled scenarios. More
specifically, if (\ref{eq:oversampled}) holds, we have%
\begin{equation}
\bar{\eta}_{U}\left(  \theta\right)  =\frac{1}{M}\log\det\left[  \bar{\sigma
}^{2}\left(  \theta\right)  \mathbf{P}_{A}^{\perp}(\theta)+\mathbf{P}%
_{A}(\theta)\mathbf{RP}_{A}(\theta)\right]  +\frac{N-K}{M}\log\left(  \frac
{N}{N-K}\right)  -\frac{K}{M} \label{eq_eta_uml_bar}%
\end{equation}
where
\[
\bar{\sigma}_{K}^{2}\left(  \theta\right)  =\frac{1}{M-K}\operatorname*{tr}%
\left[  \mathbf{P}_{A}^{\perp}(\theta)\mathbf{R}\right]  .
\]
Conversely, when (\ref{eq:undersampled}) holds, we have%
\begin{equation}
\bar{\eta}_{U}\left(  \theta\right)  =\frac{1}{M}\log\det\left[  \left\vert
\phi_{0}(\theta)\right\vert \mathbf{I}_{M}+\mathbf{P}_{A}(\theta
)\mathbf{RP}_{A}(\theta)\right]  +\frac{M-N}{M}\log\bar{\sigma}_{N}^{2}\left(
\theta\right)  -\frac{M-N}{M}\log\left\vert \phi_{0}(\theta)\right\vert
-\frac{N}{M} \label{eq_eta_uml_bar2}%
\end{equation}
where $\phi_{0}(\theta)$ is the only negative solution to the following
equation in $\phi$
\begin{equation}
\frac{1}{N}\operatorname*{tr}\left[  \mathbf{P}_{A}(\theta)\mathbf{RP}%
_{A}(\theta)\left(  \mathbf{P}_{A}(\theta)\mathbf{RP}_{A}(\theta
)-\phi\mathbf{I}_{M}\right)  ^{-1}\right]  =1. \label{eq:negative_sol}%
\end{equation}

\end{theorem}

\begin{IEEEproof}%
See Section \ref{sec_prooftheorem}.%
\end{IEEEproof}

In the light of the above theorem, one may clearly establish a different
asymptotic behavior for the CML and the UML cost functions. The CML cost
function is asymptotically equivalent to its large-sample approximation
$M^{-1}\operatorname*{tr}\left[  \mathbf{P}_{A}^{\perp}(\theta)\mathbf{R}%
\right]  $, whereas the behavior of the UML cost function is somewhat
different. In the oversampled situation, the UML cost function is
asymptotically close to the large sample equivalent (the first term in
(\ref{eq_eta_uml_bar})) plus a constant factor that does not play any role in
the parameter optimization process. In the undersampled case, however, the
UML\ cost function presents quite a different behavior, in the sense that it
becomes equivalent to a cost function that does not appear to bear any
similarity with its large sample approximation. We can see, however, that in
all these cases, the maximum of the objective function is achieved at the true
DoAs, as formally established in the following lemma.

\begin{lemma}
\label{lemma:global_max}Assume that the array does not present ambiguities, so
that whenever $\mathbf{A}(\theta_{1})=\mathbf{A}(\theta_{2})\mathbf{T}$ with
$\mathbf{T}$ an invertible matrix, we have $\theta_{1}=\theta_{2}$. Then,
regardless of whether $N>K$ or $N<K$, both $\bar{\eta}_{C}\left(
\theta\right)  $ and $\bar{\eta}_{U}\left(  \theta\right)  $ achieve their
minimum at the true DoAs, namely%
\[
\bar{\theta}=\arg\min_{\theta\in\Theta_{K}}\bar{\eta}_{C}\left(
\theta\right)  =\arg\min_{\theta\in\Theta_{K}}\bar{\eta}_{U}\left(
\theta\right)
\]
where $\Theta_{K}$ is as defined above.
\end{lemma}

\begin{IEEEproof}%
See Appendix \ref{sec:proof_Lemma1}.%
\end{IEEEproof}

The fact that both ML cost functions are asymptotically close some
deterministic functions with a global minimum at the true parameters
$\bar{\theta}$ supports the conjecture that both methods provide consistent
estimates even under a finite number of samples per observation dimension.
However, this is not formally proven here, since it would require extending
the above pointwise convergence result to uniform convergence on the parameter
space $\Theta_{K}$. In this paper, we are more interested in the behavior of
the cost functions themselves, and more specifically in their fluctuations
around their deterministic equivalents, which will eventually lead to the
presence of outliers. As mentioned above, the presence of outliers is
typically a consequence of the cost function achieving its minimum at a local
extremum that is far away from the one associated with the true parameters. In
the next subsection, we will further characterize this effect by investigating
the nature of the fluctuations of the CML and the UML cost functions at these
local minima.

\subsection{Second order behavior}

In this subsection, we will prove that the two ML cost functions
asymptotically fluctuate around their deterministic equivalents as Gaussian
random variables. We define the two $L\times1$ column vectors
\begin{align*}
\underline{\hat{\eta}}_{C}  &  =\left[  \hat{\eta}_{C}\left(  \theta_{M}%
^{(1)}\right)  ,\ldots,\hat{\eta}_{C}\left(  \theta_{M}^{(L)}\right)  \right]
^{T}\\
\underline{\bar{\eta}}_{C}  &  =\left[  \bar{\eta}_{C}\left(  \theta_{M}%
^{(1)}\right)  ,\ldots,\bar{\eta}_{C}\left(  \theta_{M}^{(L)}\right)  \right]
^{T}%
\end{align*}
and take equivalent definitions for the UML cost function. We will assume that
the $K$-dimensional points $\theta_{M}^{(\ell)}$ are such that the asymptotic
covariance is uniformly invertible. In order to formulate this point more
precisely, let us define (for $\ell=1,\ldots,L$) the matrix
\begin{equation}
\mathcal{Q}_{\ell}=\mathbf{R}^{1/2}\mathbf{A}_{\ell}\left[  \mathbf{A}_{\ell
}^{H}\left(  \mathbf{R}-\phi_{0}^{(\ell)}\mathbf{I}_{M}\right)  \mathbf{A}%
_{\ell}\right]  ^{-1}\mathbf{A}_{\ell}^{H}\mathbf{R}^{1/2} \label{eq:def_Qcal}%
\end{equation}
where $\phi_{0}^{(\ell)}$ is the only non-positive solution to the equation in
(\ref{eq:negative_sol}) when $\theta=\theta_{M}^{(\ell)}$ and $\mathbf{A}%
_{\ell}\mathbf{=A}(\theta_{M}^{(\ell)}).$

$\mathbf{(As5)}$ Given the $L$ sequences of $K$-dimensional points
$\{\theta_{M}^{(\ell)}\}$ we consider two $L\times L$ matrices $\mathbf{W}_{P}
$ and $\mathbf{W}_{Q}$ with $(i,j)$th entry respectively defined as%
\begin{align}
\left\{  \mathbf{W}_{P}\right\}  _{i,j}  &  =\frac{1}{N}\mathrm{tr}\left[
\mathbf{P}_{A}^{\perp}\left(  \theta_{M}^{(i)}\right)  \mathbf{P}_{A}^{\perp
}\left(  \theta_{M}^{(j)}\right)  \right] \label{eq:defWp}\\
\left\{  \mathbf{W}_{Q}\right\}  _{i,j}  &  =\frac{1}{N}\mathrm{tr}\left[
\mathcal{Q}_{i}\mathcal{Q}_{j}\right]  \label{eq:defWq}%
\end{align}
where $\mathcal{Q}_{i}$ is as in (\ref{eq:def_Qcal}). We assume that the
minimum eigenvalue of both $\mathbf{W}_{P}$ and $\mathbf{W}_{Q}$ is bounded
away from zero uniformly in $M$.

Under the above set of assumptions, it is possible to characterize the
asymptotic pointwise convergence of the two ML cost functions, which we
summarize in the following result. The following result establishes the fact
that $\underline{\hat{\eta}}_{C}$ and $\underline{\hat{\eta}}_{U}$
asymptotically fluctuate around their deterministic equivalents as Gaussian
random vectors. We formulate the result so that it holds for both the
undersampled ($N<K$) and the oversampled ($N>K$) regimes.

\begin{theorem}
\label{theorem:second_order}Consider the quantity $\sigma_{\ell}%
^{2}=(M-\widetilde{K})^{-1}\mathrm{tr}[\mathbf{P}_{A}^{\perp}(\theta
_{M}^{(\ell)})\mathbf{R]}$ where $\widetilde{K}=\min(K,N)$ and define the
$M\times M$ matrices%
\[
\mathcal{P}_{\ell}=\mathbf{R}^{1/2}\mathbf{P}_{A}\left(  \theta_{M}^{(\ell
)}\right)  \mathbf{R}^{1/2}\quad\mathcal{P}_{\ell}^{\perp}=\mathbf{R}%
^{1/2}\mathbf{P}_{A}^{\perp}\left(  \theta_{M}^{(\ell)}\right)  \mathbf{R}%
^{1/2}.
\]
Under $\mathbf{(As1)-(As5)}$, and as $N\rightarrow\infty$, the random vectors
$M\Gamma_{C}^{-1/2}(\underline{\hat{\eta}}_{C}-\underline{\bar{\eta}}_{C})$
and $M\Gamma_{U}^{-1/2}(\underline{\hat{\eta}}_{U}-\underline{\bar{\eta}}%
_{U})$ converge in law to a multivariate standardized Gaussian distribution,
where
\[
\left\{  \Gamma_{C}\right\}  _{\ell,m}=\frac{1}{N}\operatorname*{tr}\left[
\mathcal{P}_{\ell}^{\perp}\mathcal{P}_{m}^{\perp}\right]
\]
and
\begin{align*}
\left\{  \Gamma_{U}\right\}  _{\ell,m}  &  =\frac{1}{\sigma_{\ell}^{2}%
\sigma_{m}^{2}}\frac{1}{N}\operatorname*{tr}\left[  \mathcal{P}_{\ell}^{\perp
}\mathcal{P}_{m}^{\perp}\right]  +\frac{1}{\sigma_{m}^{2}}\frac{1}%
{N}\operatorname*{tr}\left[  \mathcal{P}_{m}^{\perp}\mathcal{Q}_{\ell}\right]
\\
&  +\frac{1}{\sigma_{\ell}^{2}}\frac{1}{N}\operatorname*{tr}\left[
\mathcal{P}_{\ell}^{\perp}\mathcal{Q}_{m}\right]  -\log\left\vert 1-\frac
{1}{N}\operatorname*{tr}\left[  \mathcal{Q}_{\ell}\mathcal{Q}_{m}\right]
\right\vert
\end{align*}
with $\mathcal{Q}_{\ell}\,$\ denoting the matrix defined in (\ref{eq:def_Qcal}).
\end{theorem}

\begin{IEEEproof}%
See Section \ref{sec:proof_theorem_CLT}.
\end{IEEEproof}

The above results provide a means of establishing the asymptotic probability
of resolution of the UML and the CML methods. Before turning to its proof, let
us draw some conclusions that can be derived from it. First of all, it is
interesting to observe that when $M,N\rightarrow\infty$, the two cost
functions are asymptotically close to the two deterministic counterparts
$\bar{\eta}_{C}\left(  \theta\right)  $ and $\bar{\eta}_{U}\left(
\theta\right)  $. Even if the two asymptotic equivalents present a single
global minimum at the true value of the DoAs, these functions are in practice
highly multimodal, i.e. they present several local minima. As shown in Lemma
\ref{lemma:global_max}, one of these local minima will coincide with the true
DoAs, namely $\bar{\theta}$. Let $L\,\ $denote the number of additional local
minima inside the feasibility region, other than $\bar{\theta}$, and let
$\theta_{M}^{(1)},\ldots,\theta_{M}^{(L)}$ denote the $K$-dimensional points
where these minima are achieved. In general, both $L$ and the points
$\theta_{M}^{(\ell)}$ may be different in $\bar{\eta}_{C}\left(
\theta\right)  $ and $\bar{\eta}_{U}\left(  \theta\right)  $. The probability
of resolution is defined as the probability that the original cost functions
$\hat{\eta}_{C}\left(  \theta\right)  $ and $\hat{\eta}_{U}\left(
\theta\right)  $ at any of these additional local minima takes a lower value
than the corresponding function at $\bar{\theta}$, namely
\begin{equation}
P_{res}^{CML}=\mathbb{P}\left[  \bigcap\limits_{\ell=1}^{L}\left\{  \hat{\eta
}_{C}\left(  \theta_{M}^{\text{(}\ell)}\right)  >\hat{\eta}_{C}\left(
\bar{\theta}\right)  \right\}  \right]  \label{eq_Pres}%
\end{equation}
and equivalently for $\hat{\eta}_{U}$. As explained above, this definition of
the resolution probability provides a very accurate description of both the
breakdown effect and the expected mean squared error (MSE) of the DoA
estimation process. Unfortunately, in our ML setting, (\ref{eq_Pres}) is
difficult to analyze for finite values of $M,N$ due to the complicated
structure of the cost functions (\ref{eq_CML_costfunction}%
)-(\ref{eq_UML_costfunction}). For this reason, previous studies
\cite{athley05, Richmond05, Richmond06} focused instead on the union bound of
the complementary of (\ref{eq_Pres}), i.e. the outlier probability, obtained
by assuming independent events. Theorem \ref{theorem:second_order} provides a
very simple way of approximating (\ref{eq_Pres}), by simply using the
asymptotic distributions (as $M,N\rightarrow\infty$) instead of the actual
ones. It will be shown below via simulations that the result provides a very
accurate description of the actual probability, even for very low $M,N$.

It should be mentioned here that the above results are merely concerned with
finite dimensional distributions and do not formally imply convergence of the
resolution probability of the CML and UML methods. This is because the number
of local minima $L$ of $\bar{\eta}\left(  \theta\right)  $ may in practice
increase with $M$, and this substantially complicates the asymptotic behavior
of (\ref{eq_Pres}). We conjecture that this will be the case for reasonably
well behaved $\mathbf{A}(\theta)$, but a more rigorous study of this problem
is left for future research.

\section{\label{sec_prooftheorem}Proof of Theorem \ref{theorem:first_order}}

The proof is based on the study of the eigenvalues of matrices of the type
$\mathbf{\hat{R}}_{A}^{(\ell)}=\mathbf{P}_{A}(\theta_{M}^{(\ell)}%
)\mathbf{\hat{R}P}_{A}(\theta_{M}^{(\ell)})$ and strongly relies on random
matrix theory techniques. In order to introduce these methods, consider the
complex random function $\hat{m}_{\ell}(z)$, $\ell=1,\ldots,L$, $z\in
\mathbb{C}^{+}=\left\{  z\in\mathbb{C}:\operatorname{Im}z>0\right\}  $,
defined as%
\begin{equation}
\hat{m}_{\ell}(z)=\frac{1}{M}\operatorname*{tr}\left[  \left(  \mathbf{\hat
{R}}_{A}^{(\ell)}-z\mathbf{I}_{M}\right)  ^{-1}\right]  .
\label{eq:Stieltjes_def}%
\end{equation}
This function is the Stieltjes transform of the empirical distribution
function of the eigenvalues of $\mathbf{\hat{R}}_{A}^{(\ell)}$, and it is
extremely important in order to characterize the asymptotic behavior of the
eigenvalues of this matrix. We will also identify $\mathbf{\hat{R}}_{A}%
^{(0)}=\mathbf{\hat{R}}$ and therefore denote $\hat{m}_{0}(z)$ as the
Stieltjes transform of the empirical eigenvalue distribution of $\mathbf{\hat
{R}}$, that is
\[
\hat{m}_{0}(z)=\frac{1}{M}\operatorname*{tr}\left[  \left(  \mathbf{\hat{R}%
}-z\mathbf{I}_{M}\right)  ^{-1}\right]  .
\]

The Stieltjes transforms defined above allow to characterize a number of
quantities that bear some dependence with the eigenvalues through the Cauchy
integration formula. For example, one can readily express the CML objective
function as%
\begin{equation}
\hat{\eta}_{C}\left(  \theta_{M}^{(\ell)}\right)  =\frac{1}{2\pi
\operatorname*{j}}{\displaystyle\oint\nolimits_{\mathcal{C}_{0}^{-}}}z\hat
{m}_{0}(z)dz-\frac{1}{2\pi\operatorname*{j}}{\displaystyle\oint
\nolimits_{\mathcal{C}_{\ell}^{-}}}z\hat{m}_{\ell}(z)dz \label{eq:eta_CML_int}%
\end{equation}
where $\mathcal{C}_{\ell}^{-}$ ($\ell\geq0$) is a negatively (clockwise)
oriented contour enclosing all the positive eigenvalues of $\mathbf{\hat{R}%
}_{A}^{(\ell)}$ and not zero. Hence, we can easily study the asymptotic
behavior of $\hat{\eta}_{C}(\theta_{M}^{(\ell)})$ by examining the asymptotic
behavior of the function $\hat{m}_{\ell}(z)$. A similar observation is
possible for the UML cost function. Indeed, we can express $\hat{\eta}%
_{U}\left(  \theta\right)  $ in (\ref{eq_UML_costfunction}) as%
\[
\hat{\eta}_{U}\left(  \theta_{M}^{(\ell)}\right)  =\frac{M-\widetilde{K}}%
{M}\log\hat{\sigma}_{\widetilde{K}}^{2}\left(  \theta_{M}^{(\ell)}\right)
+\frac{1}{M}\log\operatorname*{pdet}\left[  \mathbf{\hat{R}}_{A}^{(\ell
)}\right]
\]
where $\widetilde{K}=\min(K,N)$ and where $\operatorname*{pdet}\left(
\text{\textperiodcentered}\right)  $ here denotes the pseudo-determinant
(product of positive eigenvalues), so that the above expression makes sense
even if $\mathbf{\hat{R}}_{A}^{(\ell)}$ is singular. By using the definition
of the Stieltjes transform in (\ref{eq:Stieltjes_def}), we may write%
\begin{equation}
\hat{\eta}_{U}\left(  \theta_{M}^{(\ell)}\right)  =\frac{M-\widetilde{K}}%
{M}\log\left[  \frac{M}{M-\widetilde{K}}\hat{\eta}_{C}\left(  \theta
_{M}^{(\ell)}\right)  \right]  +\frac{1}{2\pi\operatorname*{j}}%
{\displaystyle\oint\nolimits_{\mathcal{C}_{\ell}^{-}}}\log z~\hat{m}_{\ell
}(z)dz \label{eq:eta_UML_int}%
\end{equation}
where now $\log z$ is the principal branch of the complex logarithm, analytic
on $\mathbb{C}\backslash\mathbb{R}^{\mathbb{-}}$, and where $\mathcal{C}%
_{\ell}^{-}$ is as defined above. In particular, since $\mathcal{C}_{\ell}%
^{-}$ encloses the positive eigenvalues of $\mathbf{\hat{R}}_{A}^{(\ell)}$ and
not zero, the above representation is valid for both $N>K$ and $N<K$. Hence,
we conclude here again that we can infer the asymptotic properties of
$\hat{\eta}_{U}(\theta_{M}^{(\ell)})$ from the asymptotic behavior of $\hat
{m}_{\ell}(z)$.

It is well known \cite{silverstein98, bai99b} that, under
$\mathbf{(As1)-(As4)}$ the positive eigenvalues of $\mathbf{\hat{R}}%
_{A}^{(\ell)}$ are almost surely located inside a fixed compact interval
$\mathcal{S}_{\ell}\subset\mathbb{R}_{\star}^{+}$ for all $N$ large enough.
This implies that for sufficiently large $N$ we can fix the contour
$\mathcal{C}_{\ell}^{-}$ in (\ref{eq:eta_CML_int})-(\ref{eq:eta_UML_int}) so
that it does not depend either on $N$ or the realization of the eigenvalues of
$\mathbf{\hat{R}}_{A}^{(\ell)}$, while still enclosing $\mathcal{S}_{\ell}$
and not $\left\{  0\right\}  $. This means that, for all $N$ large enough, the
only source of randomness in (\ref{eq:eta_CML_int})-(\ref{eq:eta_UML_int}) is
through the integrand function $\hat{m}_{\ell}(z)$, which has been well
studied in the random matrix theory literature. In particular, the following
result establishes that this function is asymptotically close to a
deterministic counterpart, which will be referred to as the asymptotic
deterministic equivalent. We will use a uniform notation for $\mathbf{R}%
_{A}^{(\ell)}$ and $\mathbf{R}$ by identifying $\mathbf{R=R}_{A}^{(0)}$ and
defining $\bar{M}_{0}+1$ as the total number of different eigenvalues of
$\mathbf{R}$, which will be denoted by $0<\gamma_{0}^{(0)}<\ldots<\gamma
_{\bar{M}_{0}}^{(0)}$. The $m$th eigenvalue of $\mathbf{R}$ will have
multiplicity $K_{m}^{(0)}$, so that in particular $M=\sum_{m=0}^{\bar{M}_{0}%
}K_{m}^{(0)}$.

\begin{theorem}
\label{theorem:first_order_general}\cite{silverstein95, girko98} Let
$z\in\mathbb{C}^{\mathbb{+}}$ and assume that $\mathbf{(As1)-(As4)}$ hold.
Then, for $\ell=0,\ldots,L$, $|\hat{m}_{\ell}(z)-\bar{m}_{\ell}(z)|\rightarrow
0$ almost surely as $N\rightarrow\infty$, where $\bar{m}_{\ell}(z)$ is a
deterministic complex function defined as
\begin{equation}
\bar{m}_{\ell}(z)=\frac{\omega_{\ell}\left(  z\right)  }{z}\frac{1}{M}%
\sum_{m=0}^{\bar{M}_{\ell}}K_{m}^{(\ell)}\frac{1}{\gamma_{m}^{(\ell)}%
-\omega_{\ell}\left(  z\right)  } \label{eq:m(z)_det_equiv}%
\end{equation}
where the complex function $\omega_{\ell}\left(  z\right)  $ is defined as the
unique solution in $\mathbb{C}^{\mathbb{+}}$ of the polynomial equation
\begin{equation}
z=\omega_{\ell}\left(  z\right)  \left(  1-\frac{1}{N}\sum_{m=0}^{\bar
{M}_{\ell}}K_{m}^{(\ell)}\frac{\gamma_{m}^{(\ell)}}{\gamma_{m}^{(\ell)}%
-\omega_{\ell}\left(  z\right)  }\right)  . \label{eq:definition_w(z)}%
\end{equation}

\end{theorem}

By using the analycity and boundedness of the function $\left\vert \hat
{m}_{\ell}(z)-\bar{m}_{\ell}(z)\right\vert $ on the set $\mathbb{C}%
\backslash\mathcal{S}_{\ell}\mathcal{\cup}\left\{  0\right\}  $ one may invoke
Montel's theorem to establish uniform convergence on compact sets in
$\mathbb{C}\backslash\mathcal{S}_{\ell}\mathcal{\cup}\left\{  0\right\}  $. In
this case, the definition of the function $\omega_{\ell}\left(  z\right)  $ is
extended by conventional analytical continuation. In particular, one can
establish that $\sup_{z\in\mathcal{C}_{\ell}}\left\vert \hat{m}_{\ell}%
(z)-\bar{m}_{\ell}(z)\right\vert \rightarrow0$ with probability one. Let us
now define $\bar{\eta}_{C}(\theta_{M}^{(\ell)})$ and $\bar{\eta}_{U}%
(\theta_{M}^{(\ell)}) $ as in (\ref{eq:eta_CML_int})-(\ref{eq:eta_UML_int})
but replacing $\hat{m}_{\ell}(z)$ by $\bar{m}_{\ell}(z)$, $\ell\geq0$. By the
Dominated Convergence Theorem (DCT) together with
Theorem\ \ref{theorem:first_order_general} we can readily see that $|\hat
{\eta}_{C}(\theta_{M}^{(\ell)})-\bar{\eta}_{C}(\theta_{M}^{(\ell
)})|\rightarrow0$. Regarding the UML cost function, we can write
\begin{equation}
\hat{\eta}_{U}\left(  \theta_{M}^{(\ell)}\right)  -\bar{\eta}_{U}\left(
\theta_{M}^{(\ell)}\right)  =\frac{1}{2\pi\operatorname*{j}}%
{\displaystyle\oint\nolimits_{\mathcal{C}_{\ell}^{-}}}
\log z~\left[  \hat{m}_{\ell}(z)-\bar{m}_{\ell}(z)\right]  dz+\frac
{M-\widetilde{K}}{M}\log\left[  1+\frac{\hat{\eta}_{C}\left(  \theta
_{M}^{(\ell)}\right)  -\bar{\eta}_{C}\left(  \theta_{M}^{(\ell)}\right)
}{\bar{\eta}_{C}\left(  \theta_{M}^{(\ell)}\right)  }\right]  .
\label{eq:difer_uml}%
\end{equation}
The first term of the above equation converges almost surely to zero by the
DCT and Theorem\ \ref{theorem:first_order_general}. Regarding the second term,
one can clearly see that by $\mathbf{(As4)}$ $\inf_{M}\bar{\eta}_{C}%
(\theta_{M}^{(\ell)})>0$ so that by convergence of $\hat{\eta}_{C}(\theta
_{M}^{(\ell)})-\bar{\eta}_{C}(\theta_{M}^{(\ell)})$ to zero and the bound
$\log\left(  1+x\right)  <x$ for $\left\vert x\right\vert <1$ we can establish
the same result.

At this point, it only remains to prove that the CML and UML\ deterministic
cost functions defined by direct substitution of $\hat{m}_{\ell}(z)$ by
$\bar{m}_{\ell}(z)$, $\ell\geq0$, in (\ref{eq:eta_CML_int}%
)-(\ref{eq:eta_UML_int}) correspond to those in the statement of the theorem.
In other words, it remains to solve the corresponding integrals. To do this,
we consider the change of variables $z\mapsto\omega=\omega_{\ell}\left(
z\right)  $, where $\omega_{\ell}\left(  z\right)  $ is as defined in
(\ref{eq:definition_w(z)}), and observe that
\begin{equation}
\omega_{\ell}^{\prime}\left(  z\right)  =\left(  1-\frac{1}{N}\sum_{m=1}%
^{\bar{M}_{\ell}}K_{m}^{(\ell)}\left(  \frac{\gamma_{m}^{(\ell)}}{\gamma
_{m}^{(\ell)}-\omega_{\ell}\left(  z\right)  }\right)  ^{2}\right)  ^{-1}.
\label{eq:wprime(z)}%
\end{equation}
If we denote $\mathcal{C}_{\omega_{\ell}}^{-}=\omega_{\ell}\left(
\mathcal{C}_{\ell}^{-}\right)  $, we can express $\bar{\eta}_{C}\left(
\theta_{M}^{(\ell)}\right)  $ as
\[
\bar{\eta}_{C}\left(  \theta_{M}^{(\ell)}\right)  =\frac{1}{M}%
\operatorname*{tr}\left[  \mathbf{R}\right]  -\frac{1}{2\pi\operatorname*{j}%
}{\displaystyle\oint\nolimits_{\mathcal{C}_{\omega_{\ell}}^{-}}} \frac{1}%
{M}\sum_{m=0}^{\bar{M}_{\ell}}K_{m}^{(\ell)}\frac{\omega}{\gamma_{m}^{(\ell
)}-\omega}\frac{1}{\omega_{\ell}^{\prime}}d\omega
\]
where we have used the definition of $\bar{m}_{\ell}(z)$ in
(\ref{eq:m(z)_det_equiv}) together with the $z\mapsto\omega$ change of
variables, and where $\omega_{\ell}^{\prime}$ should be understood as the
derivative defined in (\ref{eq:wprime(z)}) as a function of $\omega$. It was
shown in \cite{mestreeigsp08} that, for $\ell=1,\ldots,L$, the contour
$\mathcal{C}_{\omega_{\ell}}^{-}$ encloses all the positive eigenvalues of
$\mathbf{R}_{A}^{(\ell)}$, namely $\gamma_{1}^{(\ell)}<\ldots<\gamma_{\bar
{M}_{\ell}}^{(\ell)}$, which are the unique singularities of the above
integrand ($\gamma_{0}^{(\ell)}=0$ is not a singularity). Hence, by
conventional Cauchy integration, we obtain $\bar{\eta}_{C}\left(  \theta
_{M}^{(\ell)}\right)  =M^{-1}\operatorname*{tr}\left[  \mathbf{R}\right]
-M^{-1}\operatorname*{tr}\left[  \mathbf{R}_{A}^{(\ell)}\right]  $ as we
wanted to show. Regarding $\bar{\eta}_{U}\left(  \theta\right)  $, we only
need to solve the integral corresponding to the second term in
(\ref{eq:eta_UML_int}), namely
\begin{align*}
\mathcal{I}_{\ell}  &  \triangleq\frac{1}{2\pi\operatorname*{j}}%
{\displaystyle\oint\nolimits_{\mathcal{C}_{\ell}^{-}}} \log z\bar{m}_{\ell
}(z)dz\\
&  =\frac{1}{2\pi\operatorname*{j}}{\displaystyle\oint\nolimits_{\mathcal{C}%
_{\ell}^{-}}} \frac{\log z}{z}\frac{1}{M}\sum_{m=1}^{\bar{M}_{\ell}}%
\frac{K_{m}^{(\ell)}\omega_{\ell}\left(  z\right)  }{\gamma_{m}^{(\ell
)}-\omega_{\ell}\left(  z\right)  }dz\\
&  =\frac{1}{2\pi\operatorname*{j}}{\displaystyle\oint\nolimits_{\mathcal{C}%
_{\omega_{\ell}}^{-}}} \frac{\mathcal{L}_{\ell}(\omega)\frac{1}{M}\sum
_{m=1}^{\bar{M}_{\ell}}K_{m}^{(\ell)}\frac{1}{\gamma_{m}^{(\ell)}-\omega}%
}{1-\frac{1}{N}\sum_{m=1}^{\bar{M}_{\ell}}K_{m}^{(\ell)}\frac{\gamma
_{m}^{(\ell)}}{\gamma_{m}^{(\ell)}-\omega}}\frac{d\omega}{\omega_{\ell
}^{\prime}}%
\end{align*}
for $\ell=1,\ldots,L$, where in the first identity we used the definition of
$\bar{m}_{\ell}(z)$ in (\ref{eq:m(z)_det_equiv}) together with the fact that
$z=0$ is not enclosed by $\mathcal{C}_{\ell}^{-}$ (so we can drop the sum term
in $m=0$), and where the second identity follows from the change of variables
$z\mapsto\omega$ and the definition
\begin{equation}
\mathcal{L}_{\ell}(\omega)=\log\left[  \omega\left(  1-\frac{1}{N}\sum
_{m=1}^{\bar{M}_{\ell}}K_{m}^{(\ell)}\frac{\gamma_{m}^{(\ell)}}{\gamma
_{m}^{(\ell)}-\omega}\right)  \right]  . \label{eq:definition_L(w)}%
\end{equation}
The following proposition establishes the value of this integral.

\begin{proposition}
\label{prop:integrals_log1} For every $\ell\geq1$, the integral $\mathcal{I}%
_{\ell}$ takes the value
\[
\mathcal{I}_{\ell}=\frac{1}{M}\sum_{k=1}^{\bar{M}_{\ell}}K_{k}^{(\ell)}%
\log\gamma_{k}^{(\ell)}+\frac{N-K}{M}\log\frac{N}{N-K}-\frac{K}{M}%
\]
when $N>K$ (oversampled regime), whereas%
\[
\mathcal{I}_{\ell}=\frac{1}{M}\sum_{k=1}^{\bar{M}_{\ell}}K_{k}^{(\ell)}%
\log\left(  \gamma_{k}^{(\ell)}-\phi_{0}^{(\ell)}\right)  +\frac{N-K}{M}%
\log\left\vert \phi_{0}^{(\ell)}\right\vert -\frac{N}{M}%
\]
when $N<K$ (undersampled regime), where $\phi_{0}^{(\ell)}$ is the only
negative solution to the equation in (\ref{eq:negative_sol}) for
$\theta=\theta_{M}^{(\ell)}$.
\end{proposition}

\begin{IEEEproof}%
See Appendix \ref{app:integrals_log}.%
\end{IEEEproof}

In order to conclude the proof of Theorem \ref{theorem:first_order}, we need
to ensure that the above integral makes sense even in $N\rightarrow\infty$. In
particular, we need to show that $\inf_{N,\ell}|\phi_{0}^{(\ell)}|>0$ in the
undersampled regime, so that the logarithm in the above expression is well
defined for all large $N$. From the definition of $\phi_{0}^{(\ell)}$, we can
write
\[
\frac{K}{N}-1=\frac{\phi_{0}^{(\ell)}}{N}\sum_{m=1}^{\bar{M}_{\ell}}%
K_{m}^{(\ell)}\frac{1}{\phi_{0}^{(\ell)}-\gamma_{m}^{(\ell)}}%
\]
from where it follows that
\[
\left\vert \phi_{0}^{(\ell)}\right\vert \geq\left(  \frac{1}{N}\sum
_{m=1}^{\bar{M}_{\ell}}K_{m}^{(\ell)}\left(  \gamma_{m}^{(\ell)}\right)
^{-1}\right)  ^{-1}\left\vert \frac{K}{N}-1\right\vert
\]
and consequently $\inf_{N,\ell}|\phi_{0}^{(\ell)}|>0$ as a consequence of
($\mathbf{As3}$) and ($\mathbf{As4}$).

\section{\label{sec:proof_theorem_CLT}Proof of Theorem
\ref{theorem:second_order}}

Following the same approach as in the proof of Theorem
\ref{theorem:first_order}, we see that we are able to express
\[
M\left(  \hat{\eta}_{C}\left(  \theta_{M}^{(\ell)}\right)  -\bar{\eta}%
_{C}\left(  \theta_{M}^{(\ell)}\right)  \right)  =\frac{1}{2\pi
\operatorname*{j}}{\displaystyle\oint\nolimits_{\mathcal{C}_{0}^{-}}}zM\left(
\hat{m}_{0}(z)-\bar{m}_{0}(z)\right)  dz-\frac{1}{2\pi\operatorname*{j}%
}{\displaystyle\oint\nolimits_{\mathcal{C}_{\ell}^{-}}}zM\left(  \hat{m}%
_{\ell}(z)-\bar{m}_{\ell}(z)\right)  dz
\]
where $\mathcal{C}_{0}^{-}$ and $\mathcal{C}_{\ell}^{-}$ respectively enclose
all the positive eigenvalues of $\mathbf{R}$ and $\mathbf{R}_{A}^{(\ell)}$,
and not zero. On the other hand, using (\ref{eq:difer_uml}), we can also
write
\begin{equation}
M\left(  \hat{\eta}_{U}\left(  \theta_{M}^{(\ell)}\right)  -\bar{\eta}%
_{U}\left(  \theta_{M}^{(\ell)}\right)  \right)  =\frac{M-\widetilde{K}}%
{\bar{\eta}_{C}\left(  \theta_{M}\right)  }\left(  \hat{\eta}_{C}\left(
\theta_{M}^{(\ell)}\right)  -\bar{\eta}_{C}\left(  \theta_{M}^{(\ell)}\right)
\right)  +\frac{1}{2\pi\operatorname*{j}}{\displaystyle\oint
\nolimits_{\mathcal{C}^{-}}}\log z~M\left[  \hat{m}_{\ell}(z)-\bar{m}_{\ell
}(z)\right]  dz+\epsilon_{M}^{(\ell)} \label{eq:K_eta_UML}%
\end{equation}
where we have defined the error term
\begin{equation}
\epsilon_{M}^{(\ell)}=\left(  M-\widetilde{K}\right)  \log\left[  1+\frac
{\hat{\eta}_{C}\left(  \theta_{M}^{(\ell)}\right)  -\bar{\eta}_{C}\left(
\theta_{M}^{(\ell)}\right)  }{\bar{\eta}_{C}\left(  \theta_{M}^{(\ell
)}\right)  }\right]  -\left(  M-\widetilde{K}\right)  \frac{\hat{\eta}%
_{C}\left(  \theta_{M}^{(\ell)}\right)  -\bar{\eta}_{C}\left(  \theta
_{M}^{(\ell)}\right)  }{\bar{\eta}_{C}\left(  \theta_{M}^{(\ell)}\right)  }.
\label{eq:error_quants_prob0}%
\end{equation}
Using the fact that $\left\vert \log\left(  1-x\right)  -x\right\vert
<\left\vert x\right\vert ^{2}\left(  1-\left\vert x\right\vert \right)  ^{-1}$
for $\left\vert x\right\vert <1$ one can readily see that $\epsilon_{M}%
^{(\ell)}\rightarrow0$ in probability, so that we can disregard this term in
the asymptotic analysis. Now, observe that we can express all the remaining
terms in the form
\begin{equation}
\frac{1}{2\pi\operatorname*{j}}{\displaystyle\oint\nolimits_{\mathcal{C}%
_{\ell}^{-}}}f_{\ell}\left(  z\right)  M\left(  \hat{m}_{\ell}(z)-\bar
{m}_{\ell}(z)\right)  dz \label{eq:quantities_error_CLT}%
\end{equation}
where $f_{\ell}\left(  z\right)  $ is a certain complex function that is
holomorphic on the positive real axis. For every fixed $\ell$, it was shown in
\cite{bai04} that the above statistic asymptotically fluctuates as a Gaussian
random variable with zero mean and positive variance. However, in this paper
we need a different result, that describes the asymptotic joint distribution
of collections of variables with the form above.

\begin{theorem}
\label{theorem:CLT_general}Let $\left\{  \mathbf{R}_{1},\ldots,\mathbf{R}%
_{R}\right\}  \ $denote a collection of $R$ Hermitian positive semidefinite
matrices of dimension $M\times M$ with bounded spectral norm. For
$r=1,\ldots,R$, $\mathbf{\hat{R}}_{r}=N^{-1}\mathbf{R}_{r}^{1/2}%
\mathbf{XX}^{H}\mathbf{R}_{r}^{H/2}$, where $\mathbf{R}_{r}^{1/2}$ is a
non-necessarily Hermitian $M\times M$\ matrix such that $\mathbf{R}%
_{r}=\mathbf{R}_{r}^{1/2}\mathbf{R}_{r}^{H/2}$, and $\mathbf{X}$ is an
$M\times N$ matrix of i.i.d. Gaussian random variables with law $\mathcal{CN}%
(0,1)$. Let $\hat{m}_{r}(z)$, $\bar{m}_{r}(z)$ and $\omega_{r}(z)$ be defined
as in (\ref{eq:Stieltjes_def}), (\ref{eq:m(z)_det_equiv}) and
(\ref{eq:definition_w(z)}) respectively. Let $\left\{  f_{1}(z),\ldots
,f_{R}(z)\right\}  $ be complex functions that are holomorphic on the positive
real axis, $\mathbb{R}_{\star}^{+}$ and define $\underline{\hat{\eta}}%
=[\hat{\eta}^{\left(  1\right)  },\ldots,\hat{\eta}^{\left(  R\right)  }]^{T}$
where
\begin{equation}
\hat{\eta}^{\left(  r\right)  }=\frac{1}{2\pi\operatorname*{j}}\oint
\nolimits_{\mathcal{C}_{r}^{-}}f_{r}(z)\hat{m}_{r}(z)dz \label{eq_corr_fl}%
\end{equation}
where $\mathcal{C}_{r}^{-}$ is a clockwise oriented contour enclosing all the
positive eigenvalues of $\mathbf{\hat{R}}_{r}$\ and not $\left\{  0\right\}
$. Let $\underline{\bar{\eta}}=[\bar{\eta}^{\left(  1\right)  },\ldots
,\bar{\eta}^{\left(  R\right)  }]^{T}$, where $\bar{\eta}^{(r)}$ is defined as
$\hat{\eta}^{(r)}$ replacing $\hat{m}_{r}(z)$\ with $\bar{m}_{r}(z)$, and
consider an $R\times R$ matrix $\Gamma$ with entries%
\begin{equation}
\left\{  \Gamma\right\}  _{r,m}=\frac{-1}{4\pi^{2}}\oint\nolimits_{\mathcal{C}%
_{\omega_{r}}^{+}}\oint\nolimits_{\mathcal{C}_{\omega_{m}}^{+}}g_{r}\left(
\omega_{1}\right)  g_{m}\left(  \omega_{2}\right)  \Phi_{r,m}\left(
\omega_{1},\omega_{2}\right)  d\omega_{1}d\omega_{2} \label{eq_Gamma_generic}%
\end{equation}
where $C_{\omega_{r}}=\omega_{r}(C_{r})$,
\begin{gather*}
g_{r}(\omega)=f_{r}\left(  \omega\left(  1-\frac{1}{N}\operatorname*{tr}%
\left[  \mathbf{R}_{r}\left(  \mathbf{R}_{r}-\omega\right)  ^{-1}\right]
\right)  \right) \\
\Phi_{r,m}\left(  \omega_{1},\omega_{2}\right)  =\frac{-\partial^{2}%
\log\left(  1-\Psi_{r,m}\left(  \omega_{1},\omega_{2}\right)  \right)
}{\partial\omega_{1}\partial\omega_{2}}%
\end{gather*}
and where
\[
\Psi_{r,m}\left(  \omega_{1},\omega_{2}\right)  =\frac{1}{N}\operatorname*{tr}%
\Bigg[\left(  \mathbf{R}_{r}^{1/2}\right)  ^{H}\left(  \mathbf{R}_{r}%
-\omega_{1}\mathbf{I}_{M}\right)  ^{-1}\mathbf{R}_{r}^{1/2}\mathbf{R}%
_{m}^{H/2}\left(  \mathbf{R}_{m}-\omega_{2}\mathbf{I}_{M}\right)
^{-1}\mathbf{R}_{m}^{1/2}\Bigg].
\]
Consider an $L\times R$ complex transformation matrix $\mathbf{\Xi}$ and
assume that $\mathbf{\Xi}\Gamma\mathbf{\Xi}^{H}$ is invertible and that the
spectral norm of $\left(  \mathbf{\Xi}\Gamma\mathbf{\Xi}^{H}\right)  ^{-1}$ is
bounded in $M$. Assume, finally that $M\rightarrow\infty$ when $N\rightarrow
\infty$ so that the quotient $M/N$ is enclosed by a compact of the positive
real axis. Under these conditions, the column vector $M\left(  \mathbf{\Xi
}\Gamma\mathbf{\Xi}^{H}\right)  ^{-1/2}\mathbf{\Xi}\left(  \underline
{\hat{\eta}}-\underline{\bar{\eta}}\right)  $ converges in law to a
multivariate standardized Gaussian random vector.
\end{theorem}

\begin{IEEEproof}%
See Appendix \ref{sec:proofTheorem4}.%
\end{IEEEproof}

We can readily particularize Theorem \ref{theorem:CLT_general} to the problem
at hand. For the CML\ cost function, we can use the above theorem with $R=L$,
$\mathbf{\Xi=I}_{L}$, $f_{r}(z)=z$ and $\mathbf{R}_{r}^{1/2}=\mathbf{P}
_{A}\left(  \theta_{M}^{(r)}\right)  \mathbf{R}^{1/2}$, $r=1,\ldots,L$.
Computing the integrals in (\ref{eq_Gamma_generic}) or directly using the
formula for the expectation of four Gaussian random variables we can establish
that
\[
\left\{  \Gamma_{C}\right\}  _{r,m}=\frac{1}{N}\operatorname*{tr}\left[
\mathbf{P}_{A}^{\perp}\left(  \theta_{M}^{(r)}\right)  \mathbf{RP}_{A}^{\perp
}\left(  \theta_{M}^{(m)}\right)  \mathbf{R}\right]  .
\]
The matrix $\Gamma_{C}$ has its minimum eigenvalue bounded away from zero.
Indeed, observe that we can write%
\[
\Gamma_{C}=\mathbf{W}^{H}\left(  \mathbf{R}^{T}\mathbf{\otimes R}\right)
\mathbf{W}%
\]
where $\mathbf{W}$ is defined as $M^{2}\times L$ matrix%
\begin{equation}
\mathbf{W=}\frac{1}{\sqrt{N}}\left[  \operatorname*{vec}\left(  \mathbf{P}%
_{A}^{\perp}\left(  \theta_{M}^{(1)}\right)  \right)  ,\ldots
,\operatorname*{vec}\left(  \mathbf{P}_{A}^{\perp}\left(  \theta_{M}%
^{(L)}\right)  \right)  \right]  . \label{eq:definition_W}%
\end{equation}
Therefore, for any unit norm vector $\mathbf{u}\in\mathbb{C}^{L\times1}$ we
have $\mathbf{u}^{H}\Gamma_{C}\mathbf{u}\geq\lambda_{\min}^{2}(\mathbf{R}%
)\mathbf{u}^{H}\mathbf{W}^{H}\mathbf{Wu}\geq\lambda_{\min}^{2}(\mathbf{R}%
)\lambda_{\min}(\mathbf{W}^{H}\mathbf{W})=\lambda_{\min}^{2}(\mathbf{R}%
)\lambda_{\min}(\mathbf{W}_{P})$, where $\lambda_{\min}\left(
\text{\textperiodcentered}\right)  $ is the minimum eigenvalue of a matrix and
$\mathbf{W}_{P}$ is defined in (\ref{eq:defWp}). Therefore, taking infimum
with respect to $\mathbf{u}$ and invoking $\mathbf{(As2)}$ and $\mathbf{(As4)}%
$ we see that the minimum eigenvalue of $\Gamma_{C}$ is bounded away from zero
uniformly in $M$. Therefore, we can apply Theorem \ref{theorem:CLT_general} to
conclude that that $M\Gamma_{C}^{-1}\left(  \underline{\hat{\eta}}%
_{C}-\underline{\bar{\eta}}_{C}\right)  $ converges to a standardized Gaussian distribution.

Regarding the UML\ cost function, we see from (\ref{eq:K_eta_UML}) that we can
express%
\[
M\left(  \underline{\hat{\eta}}_{U}-\underline{\bar{\eta}}_{U}\right)
=\Xi\mathbf{\xi}+\epsilon_{M}%
\]
where $\Xi\in\mathbb{C}^{L\times2L+1}$ is a deterministic transformation
matrix as in the statement of Theorem \ref{theorem:CLT_general}, $\mathbf{\xi
}\in\mathbb{C}^{2L+1\times1}$ is a random column vector with quantities of the
form in (\ref{eq:quantities_error_CLT}) and where $\epsilon_{M}\in
\mathbb{C}^{L\times1}$ contains the error quantities $\epsilon_{M}^{(\ell)}$
defined in (\ref{eq:error_quants_prob0}), which converge to zero in
probability. More specifically, the column vector $\mathbf{\xi}$ can be
written as $\underline{\hat{\eta}}=[\mathbf{\xi}^{T}(1)\mathbf{,\xi}%
^{T}(2)]^{T}$ where $\mathbf{\xi}^{T}\left(  1\right)  \in\mathbb{C}%
^{L+1\times1}$ has its $\left(  \ell+1\right)  $th entry equal to
\[
\left\{  \mathbf{\xi}\left(  1\right)  \right\}  _{\ell+1}=\frac{1}%
{2\pi\operatorname*{j}}\oint\nolimits_{\mathcal{C}_{\ell}^{-}}zM\left(
\hat{m}_{\ell}(z)-\bar{m}_{\ell}(z)\right)  dz
\]
for $\ell\geq0$, whereas $\mathbf{\xi}\left(  2\right)  \in\mathbb{C}%
^{L\times1}$ has its $\ell$th entry equal to
\[
\left\{  \mathbf{\xi}_{M}\left(  2\right)  \right\}  _{\ell}=\frac{1}%
{2\pi\operatorname*{j}}\oint\nolimits_{\mathcal{C}_{\ell}^{-}}\log zM\left(
\hat{m}_{\ell}(z)-\bar{m}_{\ell}(z)\right)  dz.
\]
On the other hand, the $\ell$th row of matrix $\Xi$ has zeros everywhere
except for the $1$st, the $\left(  \ell+1\right)  $th and the $\left(
\ell+L+1\right)  $th positions, which take the values%
\begin{gather*}
\left\{  \Xi\right\}  _{\ell,1}=\frac{M-\widetilde{K}}{M\bar{\eta}_{C}\left(
\theta_{M}^{(\ell)}\right)  },\\
\left\{  \Xi\right\}  _{\ell,\ell+1}=-\left\{  \Xi\right\}  _{\ell,1}%
,\quad\left\{  \Xi\right\}  _{\ell,\ell+L+1}=1
\end{gather*}
respectively. We can now invoke Theorem \ref{theorem:CLT_general} in order to
establish a CLT\ on the\ random vector $\Xi\mathbf{\xi}$. Let $\mathbf{C}%
_{\xi}\in\mathbb{C}^{2L+1\times2L+1}$ denote the asymptotic covariance matrix
of $\mathbf{\xi}$, equivalent to $\Gamma$ in the statement of Theorem
\ref{theorem:CLT_general}. Consider first the upper left entries of the matrix
$\mathbf{C}_{\xi}$. In this case, we may directly establish that
$\{\mathbf{C}_{\xi}\}_{\ell+1,m+1}=N^{-1}\mathrm{tr}[\mathbf{P}_{A}^{(\ell
)}\mathbf{R\mathbf{P}}_{A}^{(m)}\mathbf{R]}$ for $0\leq\ell,m\leq L$, where we
have used the short hand notation%
\[
\mathbf{P}_{A}^{(\ell)}=\mathbf{P}_{A}\left(  \theta_{M}^{(\ell)}\right)
,\ell\geq1\text{, and }\mathbf{P}_{A}^{(0)}=\mathbf{I}_{M}.
\]
Now, in order to compute the other elements of the matrix, namely
$\{\mathbf{C}_{\xi}\}_{M+\ell,:}$ for $\ell\geq1$ and $\{\mathbf{C}_{\xi
}\}_{:,M+m}$ for $m\geq1$, we need to solve the integrals%
\begin{align}
\Gamma_{1}^{(\ell,m)}  &  =\frac{1}{2\pi\operatorname*{i}}\frac{1}%
{2\pi\operatorname*{i}}\oint\nolimits_{\mathcal{C}_{\omega_{\ell}}^{+}}%
\oint\nolimits_{\mathcal{C}_{\omega_{m}}^{+}}z_{1}\log z_{2}\Phi_{\ell
,m}\left(  \omega_{1},\omega_{2}\right)  d\omega_{1}d\omega_{2}%
\label{eq:Gamma1_def}\\
\Gamma_{2}^{(\ell,m)}  &  =\frac{1}{2\pi\operatorname*{i}}\frac{1}%
{2\pi\operatorname*{i}}\oint\nolimits_{\mathcal{C}_{\omega_{\ell}}^{+}}%
\oint\nolimits_{\mathcal{C}_{\omega_{m}}^{+}}\log z_{1}\log z_{2}\Phi_{\ell
,m}\left(  \omega_{1},\omega_{2}\right)  d\omega_{1}d\omega_{2}
\label{eq:Gamma2_def}%
\end{align}
for $\ell\geq0,m\geq1$, with $z_{1}$ (resp. $z_{2}$) replaced by the right
hand side of (\ref{eq:definition_w(z)}) as a function of $\omega_{1}$ (resp.
$\omega_{2}$). It is shown in Appendix \ref{app:integral_covariances} that%
\begin{equation}
\Gamma_{1}^{(\ell,m)}=\frac{1}{N}\operatorname*{tr}\left[  \mathbf{R}%
^{1/2}\mathbf{P}_{A}^{(\ell)}\mathbf{R}^{1/2}\mathcal{Q}_{m}\right]
\label{eq:result_Gamma1}%
\end{equation}
whereas
\begin{equation}
\Gamma_{2}^{(\ell,m)}=-\log\left\vert 1-\frac{1}{N}\operatorname*{tr}\left[
\mathcal{Q}_{\ell}\mathcal{Q}_{m}\right]  \right\vert
\label{eq:solution_Gamma2}%
\end{equation}
where $\mathcal{Q}_{l}$ is defined in (\ref{eq:def_Qcal}). An easy computation
shows that $\Xi_{M}\mathbf{C}_{\xi}\Xi_{M}^{T}=\Gamma_{U}$ as given in the
statement of Theorem \ref{theorem:second_order}. However, we need to check
that the expression in (\ref{eq:solution_Gamma2}) makes sense even for
asymptotically large $N$. By the Cauchy-Schwarz inequality, it is enough to
see that $\sup_{N,\ell}N^{-1}\operatorname*{tr}[\mathcal{Q}_{\ell}^{2}]<1 $.
Observe that we can express
\[
\frac{1}{N}\operatorname*{tr}\left[  \mathcal{Q}_{\ell}^{2}\right]  =\frac
{1}{N}\sum_{m=1}^{\bar{M}_{\ell}}K_{m}^{(\ell)}\left(  \frac{\gamma_{m}%
^{(\ell)}}{\gamma_{m}^{(\ell)}-\phi_{0}^{(\ell)}}\right)  ^{2}.
\]
In the undersampled case we have $\phi_{0}^{(\ell)}=0$ and the above quantity
is equal to $K/N$, which is obviously bounded away from $1$ according to
($\mathbf{As3}$). For the undersampled case, we use the fact that
$\inf_{N,\ell}|\phi_{0}^{(\ell)}|>0$, which is proven at the end of Section
\ref{sec_prooftheorem}, so that
\[
\frac{1}{N}\operatorname*{tr}\left[  \mathcal{Q}_{\ell}^{2}\right]
=1-\left\vert \phi_{0}^{(\ell)}\right\vert \frac{1}{N}\sum_{m=1}^{\bar
{M}_{\ell}}K_{m}^{(\ell)}\frac{\gamma_{m}^{(\ell)}}{\left(  \gamma_{m}%
^{(\ell)}-\phi_{0}^{(\ell)}\right)  ^{2}}<1
\]
uniformly in $M,\ell$. Finally, it remains to show that the minimum eigenvalue
of $\Gamma_{U}$ is bounded away from zero. To see this, we observe that we can
write $\Gamma_{U}=\Gamma_{U}^{(1)}+\Gamma_{U}^{(2)}$ where
\begin{align*}
\left\{  \Gamma_{U}^{(1)}\right\}  _{\ell,m}  &  =\frac{1}{N}%
\operatorname*{tr}\left[  \left(  \frac{1}{\sigma_{m}^{2}}\mathcal{P}%
_{m}^{\perp}+\mathcal{Q}_{m}\right)  \left(  \frac{1}{\sigma_{\ell}^{2}%
}\mathcal{P}_{\ell}^{\perp}+\mathcal{Q}_{\ell}\right)  \right] \\
\left\{  \Gamma_{U}^{(2)}\right\}  _{\ell,m}  &  =-\frac{1}{N}%
\operatorname*{tr}\left[  \mathcal{Q}_{\ell}\mathcal{Q}_{m}\right]
-\log\left\vert 1-\frac{1}{N}\operatorname*{tr}\left[  \mathcal{Q}_{\ell
}\mathcal{Q}_{m}\right]  \right\vert .
\end{align*}
We can readily see that $\Gamma_{U}^{(1)}\geq0$, so it suffices to show that
$\Gamma_{U}^{(2)}>0$ uniformly in $M$. Indeed, observe that $\Gamma_{U}^{(2)}
$ is a Hadamard matrix function \cite[p.449]{hornjohnson2} of the $L\times L$
matrix $\mathbf{W}_{Q}$ defined in (\ref{eq:defWq}). Indeed, we can express
$\Gamma_{U}^{(2)}$ as
\[
\Gamma_{U}^{(2)}=\sum_{n\geq2}\frac{1}{n}\mathbf{W}_{Q}^{(n)}%
\]
where $\mathbf{W}_{Q}^{(n)}$ is the $n$th Hadamard product of $\mathbf{W}_{Q}
$ with itself. This matrix is positive definite uniformly in $M$ by
$(\mathbf{As4})$, implying that $\Gamma_{U}^{(2)}>\frac{1}{2}\mathbf{W}%
_{Q}^{(2)}>0$ uniformly in $M$, as we wanted to show.

\section{\label{sec:NumericalEval}Numerical Validation}

\begin{figure}[tbh]
\includegraphics[width=\columnwidth]{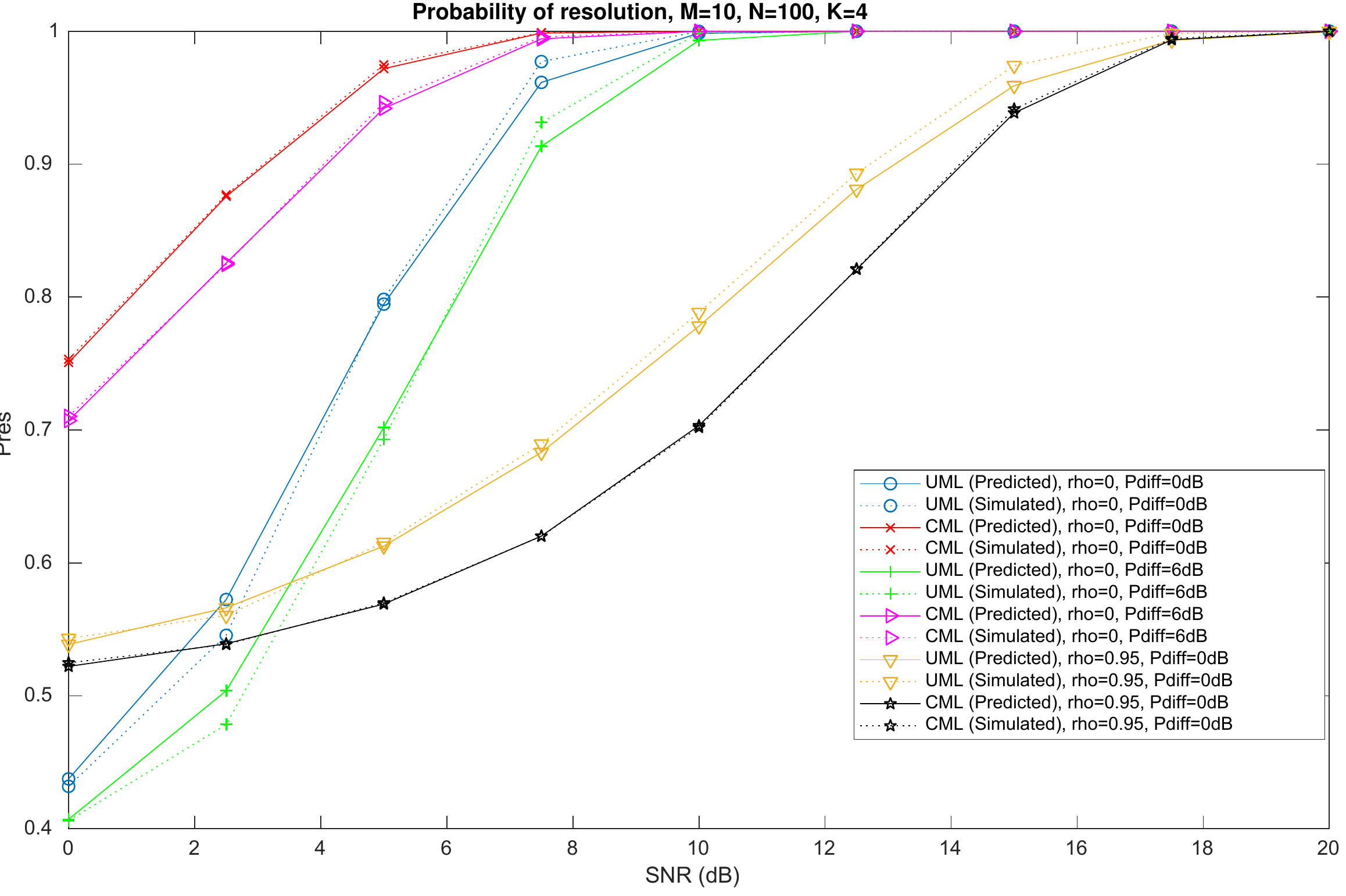}\newline\caption{Simulated vs.
asymptotic resolution probability of the UML and CML methods when $N=100$.}%
\label{fig:PresN100}%
\end{figure}

\begin{figure}[tbh]
\includegraphics[width=\columnwidth]{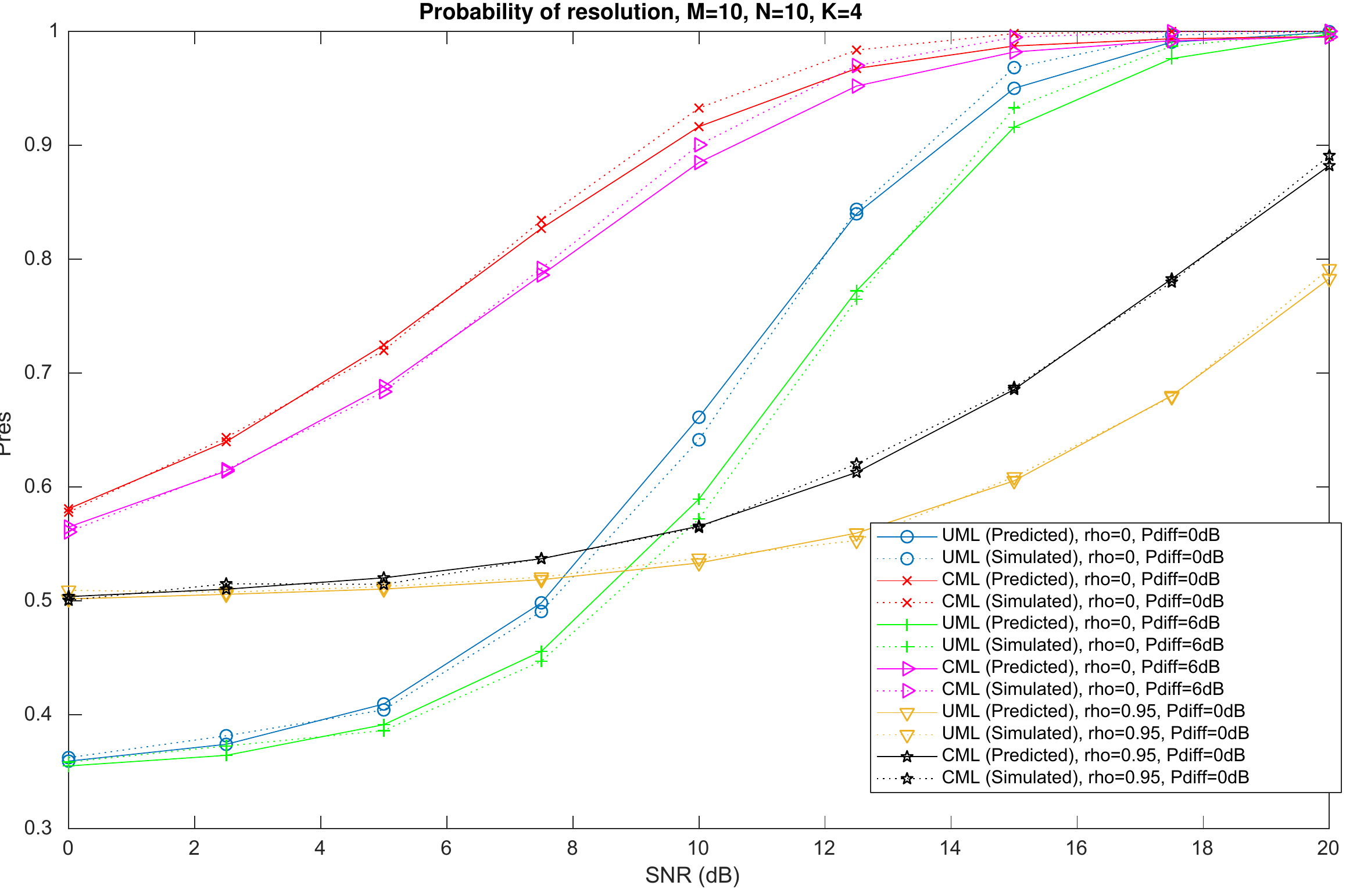}\caption{Simulated vs. asymptotic
resolution probability of the UML and CML methods when $N=10$.}%
\label{fig:PresN10}%
\end{figure}

\begin{figure}[tbh]
\includegraphics[width=\columnwidth]{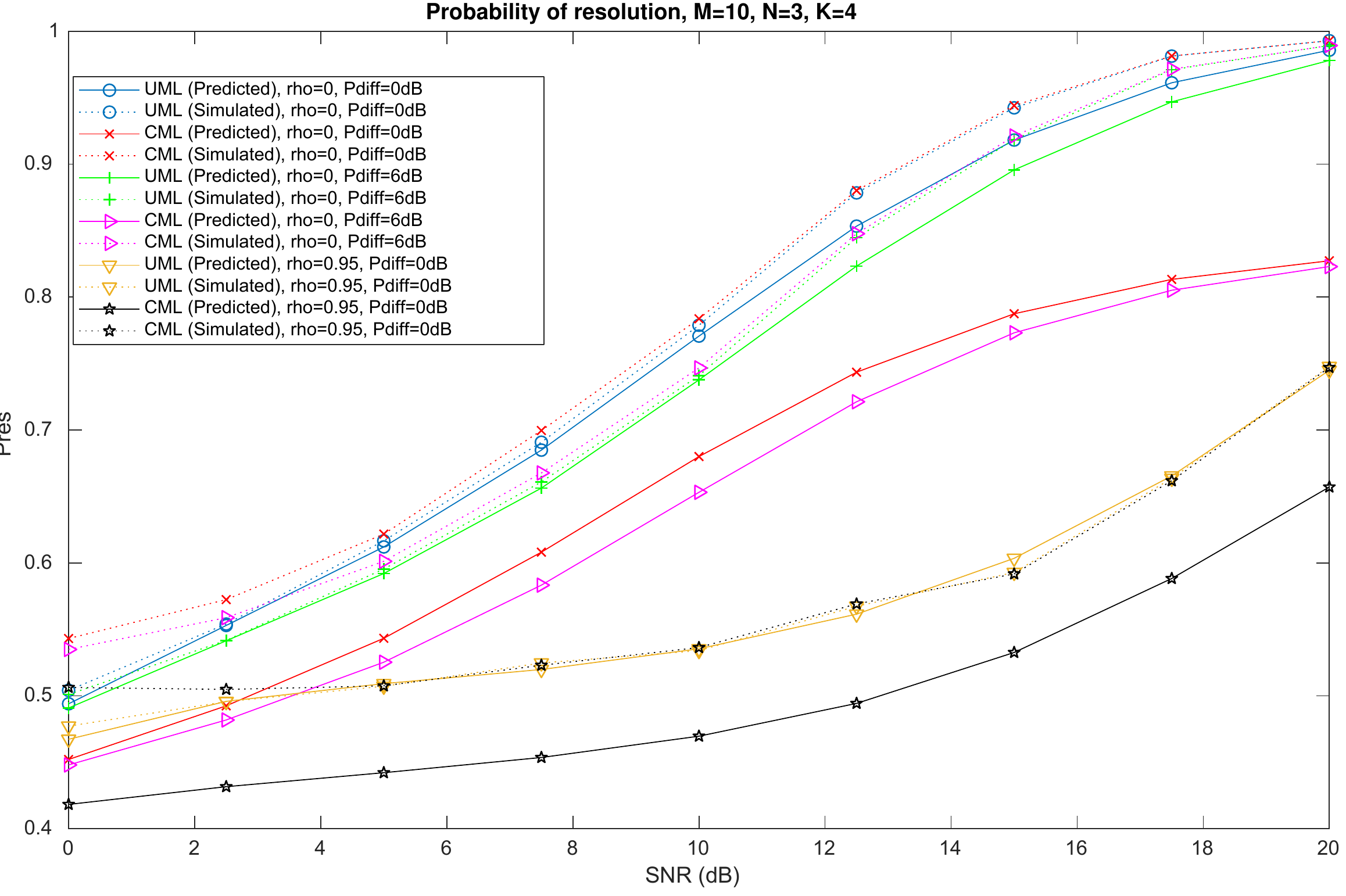}\caption{Simulated vs. asymptotic
resolution probability of the UML and CML methods when $N=3$.}%
\label{fig:PresN3}%
\end{figure}

In this section, we provide a numerical validation of the main results
established in Section \ref{sec:signal_model}. We considered a uniformly
spaced antenna array of $M=10\,\ $elements separated a quarter of wavelength
apart, receiving $K=4$ sources coming from DoAs: $16^{\circ},18^{\circ
},60^{\circ},-50^{\circ}$. The four signals were received with the same power
and were all uncorrelated, except for the first two (closely spaced), which
presented a correlation coefficient $\rho$. Figures \ref{fig:PresN100},
\ref{fig:PresN10} and \ref{fig:PresN3} represent the resolution probability of
the UML and\ CML methods when the sample size consisted of $N=100,10$ and $3$
snapshots respectively. In these figures, dotted lines represent the
resolution probability simulated with $10^{4}$ sample realizations, whereas
solid lines represent the behavior predicted by the asymptotic formulas in
Theorem \ref{theorem:second_order}. The values of the local minima where
obtained by running an accelerated proximal gradient search method
\cite[Section 4.3]{Parikh13} over the electrical angle space\footnote{That is,
over $\theta=\pi\sin\beta$, with $\beta$ the physical DoA.}, taking as initial
search values a collection of $331$ points uniformly distributed among the
feasibility set in (\ref{eq:feasiblity_set}). At each iteration, this
algorithm performs a gradient update followed by an orthogonal projection onto
the feasibility set in (\ref{eq:feasiblity_set}) with $\epsilon=0.0262$
radians, which corresponds to $1/4$ of the minimum separation between sources
in the scenario. The resulting values where then grouped into clusters using
an agglomerative hierarchical clustering algorithm based on single linkage
(distance between clusters defined as the minimum of the distance among any
pair of elements). Clusters where formed by ensuring a minimum distance of at
least $0.6$ radians between clusters, whose centroids were used as
representatives for the position of the local minima. We considered three
different scenarios:\ (i) uncorrelated and equi-powered sources; (ii)
uncorrelated sources with powers $\{2,0.5,1,1\}$ (so that the closely spaced
sources are received with a $6$dB power difference); and (iii) almost coherent
equi-powered sources (where the closely spaced sources were received with a
correlation coefficient $\rho=0.95$). The predicted probability of resolution
is evaluated by assuming that the cost functions follow a multi-dimensional
Gaussian distribution, i.e. $\underline{\hat{\eta}}_{C}\sim\mathcal{N}%
(\underline{\bar{\eta}}_{C},M^{-2}\Gamma_{C})$ and $\underline{\hat{\eta}}%
_{U}\sim\mathcal{N}(\underline{\bar{\eta}}_{U},M^{-2}\Gamma_{U})$, where we
used the Matlab$^{\copyright }$ implementation of the multidimensional
cumulative distribution function (mvncdf.m), a quasi-Monte Carlo integration
algorithm based on methods developed by Genz and Bretz \cite{Genz99}.

Observe that in general terms the asymptotic expressions are very good
approximations of the actual resolution probability even for relatively low
values of the sample size, the only exception being the performance of the CML
method in the undersampled regime ($N=3$). Surprisingly enough, the asymptotic
resolution probability of the UML method provides an accurate prediction even
for extremely small vales of $M,N$. Regarding the actual performance of the
methods, the CML method generally provides better performance in terms of
resolution probability. The only exceptions appear to be situations where
sources are very highly correlated and the sample size is relatively high
(Figure \ref{fig:PresN100}).

\begin{figure}[tbh]
\includegraphics[width=\columnwidth]{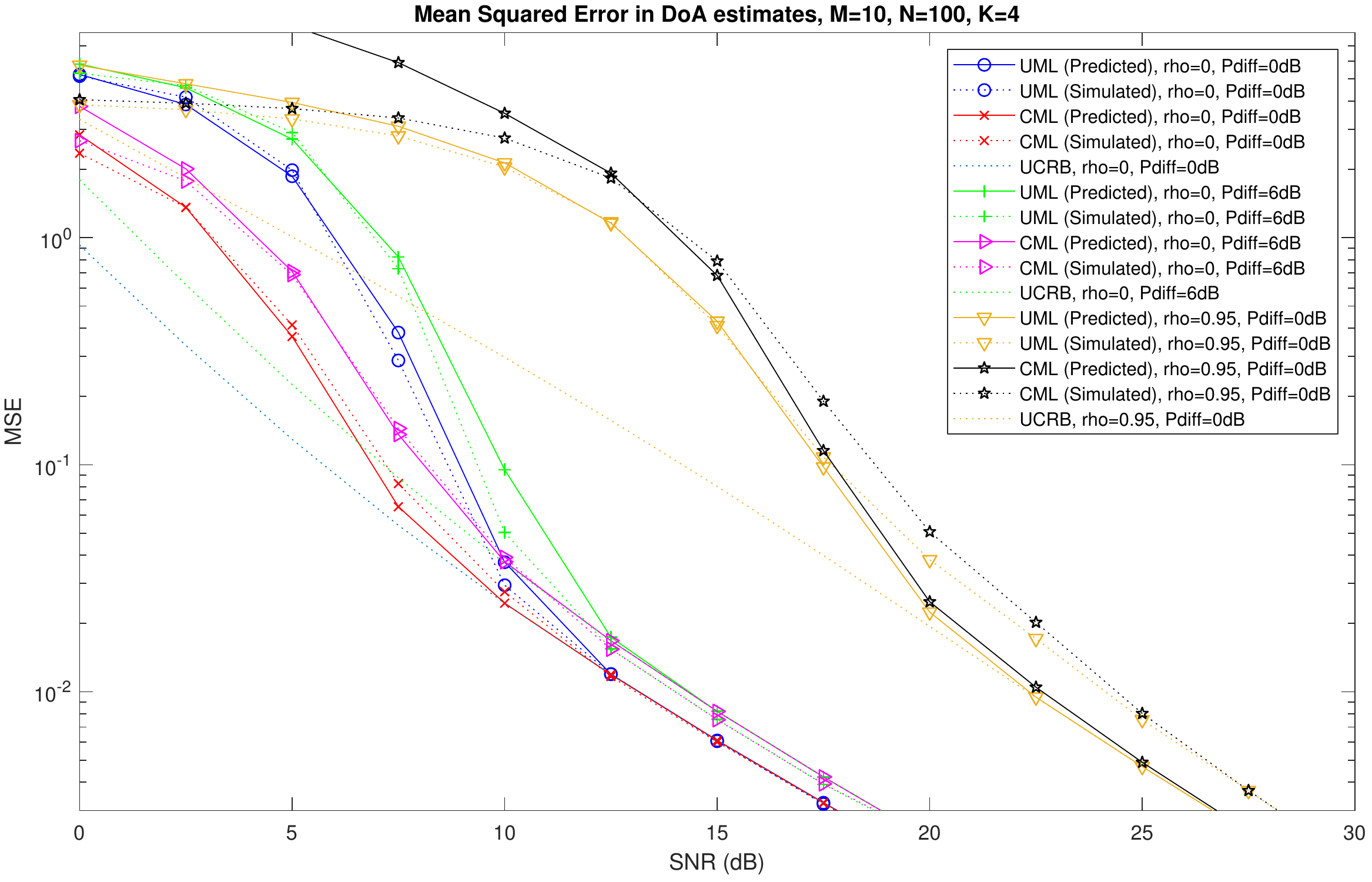}\caption{Simulated vs. asymptotic
mean squared error (MSE) of the UML and CML methods when $N=100$.}%
\label{fig:MSEN100}%
\end{figure}

In order to justify the usefulness of the resolution probability, in Figure
\ref{fig:MSEN100}\ we represent the mean squared error (MSE) of the CML and
UML estimates as a function of the SNR for the same scenario as above, taking
$N=100$. The predicted MSE\ is obtained using the expression in
(\ref{eq:Pred_MSE}), where the $MSE_{\text{small}}$ is as given in
\cite{STOI90A} and where $MSE_{\text{large}}$ is the MSE that is obtained by
choosing uniformly at random a point in the feasibility region $\Theta_{K}$.
Using \cite[Integral 4.631]{Gradshteyn2000}, one can readily see that
(assuming $\varepsilon\rightarrow0$ in the feasibility region of
(\ref{eq:feasiblity_set}))
\[
MSE_{\text{large}}=\left(  2\pi\right)  ^{2}\left[  \frac{K}{6\left(
K+1\right)  }+\sum_{m=1}^{K}\left(  \frac{m}{K+1}-\frac{\pi+\bar{\theta
}\left(  m\right)  }{2\pi}\right)  ^{2}\right]
\]
where we recall that $\bar{\theta}\left(  1\right)  <\ldots<\bar{\theta
}\left(  K\right)  $ are the true DoAs, assumed to take values on the interval
$\left(  -\pi,\pi\right)  $. Observe that the aysmptotic expressions of the
resolution probability are extremely useful in order to characterize the
threshold SNR at which outliers begin to appear. In fact, the main limitation
of the MSE\ prediction based on (\ref{eq:Pred_MSE}) is the fact that the small
error term $MSE_{\text{small}}$ is only valid as an approximation when
$N\rightarrow\infty$ and does not fully characterize the finite $N$ scenario,
especially in the presence of highly correlated sources (similar results are
reported in \cite{abramovich11, Abramovich12Asilomar}).

\section{\label{sec:Conclusions}Conclusions}

The use of random matrix theory techniques has proven to be a very good tool
to characterize the behavior of conditional and unconditional ML algorithms
for DoA estimation in the threshold region. By studying the first and second
order behavior of these two ML cost functions, it has been shown that the
finite dimensional distributions of these two cost functions asymptotically
fluctuate as multivariate Gaussian random vectors with a certain mean and
covariance matrix, which have been derived in closed form. These results have
been used in order to approximate and characterize the resolution probability
of both methods in threshold region. These studies corroborate the fact that,
in general terms, the CML method provides better resolution probabilities than
UML, except for very specific cases with highly correlated source signals and
relatively large sample size.

\appendices

\section{\label{sec:proof_Lemma1}Proof of Lemma \ref{lemma:global_max}}

The fact that $\bar{\eta}_{C}\left(  \theta\right)  $ achieves its maximum at
$\theta=\bar{\theta}$ is well known in the literature, as it is the case for
$\bar{\eta}_{U}\left(  \theta\right)  $ in the oversampled regime. Thus, we
will only prove the result for $\bar{\eta}_{U}\left(  \theta\right)  $ in the
undersampled regime. Let us recall that $\mathbf{U}_{A}\left(  \theta\right)
=\mathbf{A}(\theta)\left(  \mathbf{A}^{H}(\theta)\mathbf{A}(\theta)\right)
^{-1/2}$. Observe first that we can re-write
\[
\bar{\eta}_{U}\left(  \theta\right)  =\Upsilon\left(  \mathbf{U}_{A}\left(
\theta\right)  \right)  +\frac{M-N}{M}\log\left(  \frac{1}{M-N}%
\operatorname*{tr}\left[  \mathbf{R}\right]  \right)
\]
where%
\[
\Upsilon\left(  \mathbf{U}\right)  =-\frac{N}{M}+\frac{1}{M}\log\det\left[
\mathbf{U}^{H}\mathbf{RU}-\xi\left(  \mathbf{U}\right)  \mathbf{I}_{K}\right]
+\frac{M-N}{M}\log\left(  1-\frac{\operatorname*{tr}\left[  \mathbf{U}%
^{H}\mathbf{RU}\right]  }{\operatorname*{tr}\left[  \mathbf{R}\right]
}\right)  -\frac{K-N}{M}\log\left\vert \xi\left(  \mathbf{U}\right)
\right\vert
\]
and where $\xi\left(  \mathbf{U}_{A}\left(  \theta\right)  \right)  =\phi
_{0}\left(  \theta\right)  $, so that $\xi\left(  \mathbf{U}\right)  $ is the
unique negative solution of the equation
\begin{equation}
\frac{1}{N}\operatorname*{tr}\left[  \mathbf{U}^{H}\mathbf{RU}\left(
\mathbf{U}^{H}\mathbf{RU}-\xi\mathbf{I}_{K}\right)  ^{-1}\right]  =1.
\label{eq:solution_psi}%
\end{equation}
Consider the unconstrained maximization of the function $\Upsilon\left(
\mathbf{U}\right)  $ with respect to the set of $M\times K$ matrices with
orthogonal columns. Let $\mathbf{\Sigma}\in\mathbb{C}^{K\times K}$ be a
complex Hermitian matrix that contains the Lagrange multipliers corresponding
to the set of constraints $\mathbf{U}^{H}\mathbf{U=I}_{K}$. By constructing
the Lagrangian associated to this optimization problem and forcing the
derivatives with respect to the entries of $\mathbf{U}$ to zero we obtain the
optimality conditions
\begin{equation}
\frac{1}{\bar{\sigma}_{N}^{2}\left(  \mathbf{U}\right)  }\mathbf{RU}%
-\mathbf{RU}\left(  \mathbf{U}^{H}\mathbf{RU-}\xi\left(  \mathbf{U}\right)
\mathbf{I}_{K}\right)  ^{-1}=M\mathbf{U\Sigma} \label{eq:deriv=0}%
\end{equation}
where, with some abuse of notation, we have defined%
\[
\bar{\sigma}_{N}^{2}\left(  \mathbf{U}\right)  =\frac{1}{M-K}%
\operatorname*{tr}\left[  \mathbf{R}\left(  \mathbf{I}_{M}-\mathbf{UU}%
^{H}\right)  \right]  .
\]
Imposing the orthogonality constraint we obtain an expression for
$\mathbf{\Sigma}$, which inserted into (\ref{eq:deriv=0}) leads to%
\begin{equation}
\left[  \mathbf{RU}-\mathbf{U}\left(  \mathbf{U}^{H}\mathbf{RU}\right)
\right]  \left[  \mathbf{I}_{K}-\bar{\sigma}_{N}^{2}\left(  \mathbf{U}\right)
\left(  \mathbf{U}^{H}\mathbf{RU-}\xi\left(  \mathbf{U}\right)  \mathbf{I}%
_{K}\right)  ^{-1}\right]  =\mathbf{0.} \label{eq:identity_solutions}%
\end{equation}
Hence, all local minima of $\Upsilon\left(  \mathbf{U}\right)  $ must be zeros
of one of the two factors above.

Let us first analyze the zeros of the first term, namely the solutions to
$\mathbf{RU=U}(\mathbf{U}^{H}\mathbf{RU)}$. Noting that $\mathbf{U}%
^{H}\mathbf{RU}$ is Hermitian and positive definite, so we can consider its
eigenvalue decomposition, namely $\mathbf{Q}\Lambda\mathbf{Q}^{H}$. Using
this, the above equation can be rewritten as $\mathbf{RUQ=UQ}\Lambda$ and we
see that the columns of matrix $\mathbf{UQ}$ are equal to $K$ different
eigenvectors of $\mathbf{R}$. Since the value of the cost function is
invariant to the choice of $\mathbf{Q}$, we have determined all the set of
solutions corresponding to the first part of the identity in
(\ref{eq:identity_solutions}). Regarding the second term in
(\ref{eq:identity_solutions}), we need to have%
\[
\mathbf{U}^{H}\mathbf{RU}=\left(  \xi\left(  \mathbf{U}\right)  +\bar{\sigma
}_{N}^{2}\left(  \mathbf{U}\right)  \right)  \mathbf{I}_{K}.
\]
This means that $\mathbf{U}$ must belong to the kernel of $\mathbf{R-}\left(
\xi\left(  \mathbf{U}\right)  +\bar{\sigma}_{N}^{2}\left(  \mathbf{U}\right)
\right)  \mathbf{I}_{M},$\ so that here again the columns of $\mathbf{U}$ must
be formed by distinct eigenvectors of $\mathbf{R}$ up to right multiplication
by an orthogonal matrix. Let $\lambda_{1}\leq\ldots\leq\lambda_{M}$ denote the
eigenvalues of $\mathbf{R}$ (some of which may be repeated), and let
$\mathcal{I}$ denote the index set of the $K$ eigenvectors of $\mathbf{R}$ in
the selected matrix $\mathbf{U}$ and $\overline{\mathcal{I}}$ its
complementary. At the local extrema, we see that the cost function
$\Upsilon\left(  \mathbf{U}\right)  $ can be expressed as%
\begin{multline*}
\Upsilon\left(  \mathbf{U}\right)  =-\frac{N}{M}+\frac{1}{M}\log\det
\mathbf{R}-\frac{M-N}{M}\log\frac{\operatorname*{tr}\left[  \mathbf{R}\right]
}{M-K}\\
+\frac{M-N}{M}\left(  \log\left(  \frac{1}{M-K}\sum_{k\in\overline
{\mathcal{I}}}\lambda_{k}\right)  -\frac{1}{M-K}\sum_{k\in\overline
{\mathcal{I}}}\log\lambda_{k}\right)  +\frac{1}{M}\frac{K-N}{M-K}\sum
_{k\in\overline{\mathcal{I}}}\log\lambda_{k}\\
-\frac{1}{M}\sum_{k\in\mathcal{I}}\log\frac{\lambda_{k}}{\lambda_{k}%
-\xi_{\mathcal{I}}}-\frac{K-N}{M}\log\left\vert \xi_{\mathcal{I}}\right\vert
\end{multline*}
where $\xi_{\mathcal{I}}$ is the negative solution to
\[
\frac{1}{N}\sum_{i\in\mathcal{I}}\frac{\lambda_{i}}{\lambda_{i}-\xi
_{\mathcal{I}}}=1.
\]
The first term is independent of $\mathcal{I}$ and can be obviated. The second
term is non-negative by the arithmetic-geometric mean inequality and reaches
its minimum when the eigenvalues $\left(  \lambda_{k}\right)  _{k\in
\overline{\mathcal{I}}}$ are all equal, that is when $\overline{\mathcal{I}%
}=\left\{  1,\ldots,M-K\right\}  $ so that all these eigenvalues are equal to
the noise power. Since $K>N$, the third term is minimum under the same
circumstances. Regarding the final term, it can readily be seen that it
decreasing in $\left(  \lambda_{k}\right)  _{k\in\mathcal{I}}$ because its
derivative with respect to $\lambda_{j}$ is equal to
\[
-\frac{1}{M}\frac{\left(  -\xi_{\mathcal{I}}\right)  }{\lambda_{j}\left(
\lambda_{j}-\xi_{\mathcal{I}}\right)  }<0
\]
implying that the minimum is achieved when the selected eigenvalues are
maximum, that is$\ \mathcal{I}=\left\{  M-K+1,\ldots,M\right\}  $.

From all the above, we conclude that $\Upsilon\left(  \mathbf{U}\right)  $ is
minimized when the columns of $\mathbf{U}$ are selected as the eigenvectors
associated with the $K$ largest eigenvalues of $\mathbf{R}$. However, these
eigenvectors span the column space of $\mathbf{A}\left(  \bar{\theta}\right)
$ and by the manifold regularity condition one must necessarily have
$\bar{\theta}=\arg\min_{\theta}\bar{\eta}_{U}\left(  \theta\right)  $ as we
wanted to show.

\section{\label{app:integrals_log}Proof of Proposition
\ref{prop:integrals_log1}}

In this appendix, we drop from the notation the dependence on $\ell$ (point
sequence index) to simplify the exposition. Since $\ell\geq1$, we will have
$0=\gamma_{0}<\gamma_{1}<\ldots<\gamma_{\bar{M}}$. We note that we can rewrite
$\mathcal{L}(\omega)$ in (\ref{eq:definition_L(w)}) as $\mathcal{L}%
(\omega)=\log\left[  \omega\left(  1-\Phi\left(  \omega\right)  \right)
\right]  $, where we have defined
\begin{equation}
\Phi\left(  \omega\right)  =\frac{1}{N}\sum_{m=1}^{\bar{M}}K_{m}\frac
{\gamma_{m}}{\gamma_{m}-\omega}.\, \label{eq:definition_Phiconv}%
\end{equation}
Let $\phi_{0}<\phi_{1}<\ldots<\phi_{\bar{M}}$ denote the $\bar{M}$ different
roots of the equation $\phi\left(  1-\Phi\left(  \phi\right)  \right)  =0$. It
is not difficult to see \cite{mestreeigsp08}\ that only the roots $\phi
_{1},\ldots,\phi_{\bar{M}}$ are enclosed by $\mathcal{C}_{\omega}$, whereas
$\phi_{0}$ is always outside the contour $\mathcal{C}_{\omega}$. Furthermore,
when $K<N$ we have $\phi_{0}=0$ and $\phi_{1}>0$ whereas when $K>N$ we have
$\phi_{0}<0$ and $\phi_{1}=0$. In particular, we see that $\{0\}$ is enclosed
by $\mathcal{C}_{\omega}$ only in the undersampled regime ($K>N$).

First of all, note that, using the fact that $\phi_{0}\left(  1-\Phi\left(
\phi_{0}\right)  \right)  =0$ we may re-write $\mathcal{L}(\omega)$ in
(\ref{eq:definition_L(w)})\ as%
\[
\mathcal{L}(\omega)=\log\left[  \left(  \omega-\phi_{0}\right)  \left(
1-\widetilde{\Phi}\left(  \omega\right)  \right)  \right]
\]
where we have defined
\begin{equation}
\widetilde{\Phi}\left(  \omega\right)  =\frac{1}{N}\sum_{m=1}^{\bar{M}_{m}%
}\widetilde{K}_{m}\frac{\gamma_{m}}{\gamma_{m}-\omega}\quad\widetilde{K}%
_{m}=K_{m}\frac{\gamma_{m}}{\gamma_{m}-\phi_{0}}. \label{eq:defPsiTilde}%
\end{equation}
Consider the function $\mathcal{I}(x):[0,1]\rightarrow\mathbb{C}$, defined as
\begin{equation}
\mathcal{I}(x)=\frac{1}{2\pi\operatorname*{j}}{\displaystyle\oint
\nolimits_{\mathcal{C}_{\omega}^{-}}} \frac{\mathcal{L}(\omega,x)}%
{1-\Phi\left(  \omega\right)  }\left(  \frac{1}{M}\sum_{m=1}^{\bar{M}}%
\frac{K_{m}}{\gamma_{m}-\omega}\right)  \frac{1}{\omega^{\prime}}%
d\omega\label{eq:def_I(x)}%
\end{equation}
where $\mathcal{L}(\omega,x)=\log[(\omega-\phi_{0})(1-x\widetilde{\Phi}\left(
\omega\right)  )]$ and where $\omega^{\prime}$ should be understood as the
derivative defined in (\ref{eq:wprime(z)}) as a function of $\omega$. Note
that the above integral we want to solve can be expressed as $\mathcal{I}(1)$.
It follows from the DCT\ (cf. \cite[Lemma 4]{Mestre16coherence}) that
$\mathcal{I}(x)$ is a differentiable function of $x$, and therefore we can
compute $\mathcal{I}(1)$ by first computing the derivative $\mathcal{I}%
^{\prime}(x)$ and its primitive and then using $\mathcal{I}(0)$ to fix the
value of the undetermined constant.

We consider first the integral $\mathcal{I}(0)$, which can be expressed as
\[
\mathcal{I}(0)=\frac{1}{2\pi\operatorname*{j}}{\displaystyle\oint
\nolimits_{\mathcal{C}_{\omega}^{-}}}\frac{\log\left(  \omega-\phi_{0}\right)
}{1-\Phi\left(  \omega\right)  }\left(  \frac{1}{M}\sum_{m=1}^{\bar{M}}%
K_{m}\frac{1}{\gamma_{m}-\omega}\right)  \frac{1}{\omega^{\prime}}d\omega.
\]
Noting that $\Phi\left(  0\right)  \neq1$, we see that $\phi_{k}=0$ does not
constitute a singularity of the above integrand, so the poles that contribute
to the above integral are $\left\{  \gamma_{k},k\geq1\right\}  $
and\ $\left\{  \phi_{k},k\geq1,\phi_{k}\neq0\right\}  $. By simple residue
calculus, we can establish that \
\[
\mathcal{I}(0)=\frac{1}{M}\sum_{k=1}^{\bar{M}}K_{k}\log\left(  \gamma_{k}%
-\phi_{0}\right)  -\frac{N}{M}\Phi\left(  \phi_{0}\right)  +\frac{N-K}{M}%
\log\frac{{\displaystyle\prod\nolimits_{k=1}^{\bar{M}}}\left(  \gamma_{k}%
-\phi_{0}\right)  }{{\displaystyle\prod\nolimits_{\substack{k=1 \\\phi_{k}%
\neq0}}^{\bar{M}}}\left(  \phi_{k}-\phi_{0}\right)  }%
\]

Next, we focus on the derivative $\mathcal{I}^{\prime}(x)$, which takes the
form%
\[
\mathcal{I}^{\prime}(x)=\frac{1}{2\pi\operatorname*{j}}{\displaystyle\oint
\nolimits_{\mathcal{C}_{\omega}^{+}}} \frac{\widetilde{\Phi}\left(
\omega\right)  \left(  \frac{1}{M}\sum_{m=1}^{\bar{M}}K_{m}\frac{1}{\gamma
_{m}-\omega}\right)  }{\left(  1-x\widetilde{\Phi}\left(  \omega\right)
\right)  \left(  1-\Phi\left(  \omega\right)  \right)  }\frac{1}%
{\omega^{\prime}}d\omega.
\]
Let us consider $x\in\left(  0,1\right)  $ and let $\widetilde{\phi}%
_{1}\left(  x\right)  <\ldots<\widetilde{\phi}_{\bar{M}}\left(  x\right)  $
denote the $\bar{M}$ distinct solutions to the equation $x\widetilde{\Phi
}\left(  \widetilde{\nu}\right)  =1$. It can readily be seen following the
approach in \cite{mestreeigsp08} that when $x\in\left(  0,1\right)  $ all the
roots $\widetilde{\phi}_{1}\left(  x\right)  ,\ldots,\widetilde{\phi}_{\bar
{M}}\left(  x\right)  $ are located inside $\mathcal{C}_{\omega}$. Observing
that $\omega=0$ is not a singularity of the above integrand, we see that the
poles inside the contour are $\{\gamma_{k},k\geq1\}$, $\{\phi_{k},k\geq
1,\phi_{k}\neq0\}$ and $\{\widetilde{\phi}_{k}\left(  x\right)  ,k\geq1\}$.
Computing the corresponding residues, we find that%
\begin{multline*}
\mathcal{I}^{\prime}(x)=\frac{1}{x}\left(  \frac{K}{M}+\frac{N-K}{M}\bar
{M}\right)  +\frac{1}{x^{2}}\frac{N}{M}\sum_{k=1}^{\bar{M}}\frac{K_{k}%
}{\widetilde{K}_{k}}\\
+\frac{1}{M}\sum_{\substack{k=1 \\\widetilde{\phi}_{k}\left(  x\right)  \neq
0}}^{\bar{M}}\sum_{m=1}^{\bar{M}}K_{m}\left(  1+\frac{\gamma_{m}}{\gamma
_{m}-\widetilde{\phi}_{k}\left(  x\right)  }\right)  \frac{\widetilde{\phi
}_{k}^{\prime}\left(  x\right)  }{\gamma_{m}-\widetilde{\phi}_{k}\left(
x\right)  }\\
+\frac{N-K}{M}\left(  \sum_{\substack{k=1 \\\phi_{k}\neq0}}^{\bar{M}}%
\frac{\widetilde{\Phi}\left(  \phi_{k}\right)  }{1-x\widetilde{\Phi}\left(
\phi_{k}\right)  }-\sum_{\substack{k=1 \\\widetilde{\phi}_{k}\left(  x\right)
\neq0}}^{\bar{M}}\frac{\Phi^{\prime}\left(  \widetilde{\phi}_{k}\left(
x\right)  \right)  \widetilde{\phi}_{k}^{\prime}\left(  x\right)  }%
{1-\Phi\left(  \widetilde{\phi}_{k}\left(  x\right)  \right)  }\right)
\end{multline*}
where we have used the fact that, for $k>1$, we have $x\widetilde{\Phi
}(\widetilde{\phi}_{k}(x))=\Phi(\phi_{k})=1$ and that $\widetilde{\phi}%
_{k}\left(  x\right)  $ is a differentiable function of $x$ with derivative%
\[
\widetilde{\phi}_{k}^{\prime}\left(  x\right)  =\left(  \frac{-x^{2}}{N}%
\sum_{m=1}^{\bar{M}}\widetilde{K}_{m}\frac{\gamma_{m}}{\left(  \gamma
_{m}-\widetilde{\phi}_{k}\left(  x\right)  \right)  ^{2}}\right)  ^{-1}.
\]
One can easily find a primitive of $\mathcal{I}^{\prime}(x)$ and fix the
undetermined constraint according to the value of $\mathcal{I}(0)$. In this
process, one needs to use the fact that, for $k\geq1$, when $x\rightarrow0$ we
have $\widetilde{\phi}_{k}\left(  x\right)  \rightarrow\gamma_{k}$,
$x^{-1}(\gamma_{k}-\widetilde{\phi}_{k}\left(  x\right)  )\rightarrow
\frac{\gamma_{k}}{N}\widetilde{K}_{k}$ and
\[
\frac{1}{x}\left(  \frac{\gamma_{k}-\widetilde{\phi}_{k}\left(  x\right)  }%
{x}-\frac{\gamma_{k}}{N}\widetilde{K}_{k}\right)  \rightarrow\frac{\gamma_{k}%
}{N}\widetilde{K}_{k}\frac{1}{N}\sum_{\substack{m=1 \\m\neq k}}^{\bar{M}%
}\widetilde{K}_{m}\frac{\gamma_{m}}{\gamma_{m}-\gamma_{k}}.
\]

Once we have fixed the undetermined constant, we can obtain%
\begin{align*}
\mathcal{I}(x)  &  =\frac{1}{M}\sum_{k=1}^{\bar{M}}K_{k}\log\left(  \gamma
_{k}-\phi_{0}\right)  -\frac{N}{M}\Phi\left(  \phi_{0}\right) \\
&  +\frac{N-K}{M}\log\left\vert \left(  1-x\widetilde{\Phi}\left(  \phi
_{0}\right)  1_{\phi_{0}\neq0}\right)  \frac{{\displaystyle\prod
\nolimits_{k=1}^{\bar{M}}} \left(  \gamma_{k}-\phi_{0}\right)  }%
{{\displaystyle\prod\nolimits_{\substack{k=1 \\\phi_{k}\neq0}}^{\bar{M}}}
\left(  \phi_{k}-\phi_{0}\right)  }\right\vert \\
&  +\xi_{1}(x)+\xi_{2}(x)+\xi_{3}(x)
\end{align*}
where $1_{\left\{  \text{\textperiodcentered}\right\}  }$ is the indicator
function and where we have defined
\begin{align*}
\xi_{1}(x)  &  =\frac{N-K}{M}\log\left\vert \frac{\prod\nolimits_{k=1}%
^{\bar{M}}\left(  1-\Phi\left(  \widetilde{\phi}_{k}\left(  x\right)  \right)
\right)  }{\prod\nolimits_{\substack{k=0 \\\phi_{k}\neq0}}^{\bar{M}}\left(
1-x\widetilde{\Phi}\left(  \phi_{k}\right)  \right)  }\prod\limits_{k=1}%
^{\bar{M}}\left(  \frac{\widetilde{K}_{k}}{K_{k}}x\right)  \right\vert \\
\xi_{2}(x)  &  =\frac{1}{M}\sum_{m=1}^{\bar{M}}K_{m}\log\left\vert
\frac{x\frac{\gamma_{m}}{N}\widetilde{K}_{m}\prod\nolimits_{\substack{k=1
\\k\neq m}}^{\bar{M}}\left(  \gamma_{m}-\gamma_{k}\right)  }{\prod
\nolimits_{k=1}^{\bar{M}}\left(  \gamma_{m}-\widetilde{\phi}_{k}\left(
x\right)  \right)  }\right\vert \\
\xi_{3}(x)  &  =\frac{1}{M}\sum_{m=1}^{\bar{M}}\sum_{k=1}^{\bar{M}}\frac
{K_{m}\gamma_{m}}{\gamma_{m}-\widetilde{\phi}_{k}\left(  x\right)  }+\frac
{1}{M}\sum_{k=1}^{\bar{M}}\sum_{\substack{m=1 \\m\neq k}}^{\bar{M}}\frac
{K_{k}}{\widetilde{K}_{k}}\frac{\widetilde{K}_{m}\gamma_{m}}{\gamma_{m}%
-\gamma_{k}}\\
&  +\frac{1}{M}\sum_{k=1}^{\bar{M}}\sum_{\substack{m=1 \\m\neq k}}^{\bar{M}%
}\frac{K_{m}\gamma_{m}}{\gamma_{k}-\gamma_{m}}-\frac{1}{x}\frac{N}{M}%
\sum_{k=1}^{\bar{M}}\frac{K_{k}}{\widetilde{K}_{k}}.
\end{align*}

\begin{lemma}
\label{lemma:polyident}For any $x\in\left(  0,1\right)  $, it holds that
\[
\xi_{1}(x)=\xi_{2}(x)=\xi_{3}(x)=0.
\]
Furthermore,
\begin{equation}
\frac{\prod\nolimits_{\substack{k=1 \\\phi_{k}\neq0}}^{\bar{M}}\left(
\phi_{k}-\phi_{0}\right)  }{\prod\nolimits_{k=1}^{\bar{M}}\left(  \gamma
_{k}-\phi_{0}\right)  }=\left\{
\begin{array}
[c]{ccc}%
\frac{1}{N}\sum_{m=1}^{\bar{M}}K_{m}\frac{\gamma_{m}}{\left(  \gamma_{m}%
-\phi_{0}\right)  ^{2}} &  & \phi_{0}\neq0\\
\frac{N-K}{N} &  & \phi_{0}=0.
\end{array}
\right.  \label{eq:identyphis}%
\end{equation}

\end{lemma}

\begin{IEEEproof}%
See Appendix \ref{sec:proofLemma2}.%
\end{IEEEproof}

From all the above, we see that when $K<N$ the function $\mathcal{I}(x)$ is
constant in $x$ and equal to the value in statement of the proposition.
Finally when $K>N$ we will have
\begin{align*}
\mathcal{I}(x)  &  =\frac{1}{M}\sum_{k=1}^{\bar{M}}K_{k}\log\left(  \gamma
_{k}-\phi_{0}\right)  -\frac{N}{M}+\frac{N-K}{M}\log\left\vert \phi
_{0}\right\vert \\
&  +\frac{N-K}{M}\log\left\vert \frac{1-x\widetilde{\Phi}\left(  \phi
_{0}\right)  }{1-\widetilde{\Phi}\left(  \phi_{0}\right)  }\right\vert
\end{align*}
and therefore $\mathcal{I}(1)$ is equal to the value in the statement of the proposition.

\section{\label{app:integral_covariances}Solution to the covariance integrals
$\Gamma_{1}^{(\ell,m)},\Gamma_{2}^{(\ell,m)}$}

In this appendix, we provide an alternative solution to the covariance
integrals in (\ref{eq:Gamma1_def})-(\ref{eq:Gamma2_def}). We will consider the
case $\ell\geq1,m\geq1$, since the cases $\ell=0$ and/or $m=0$ can be obtained
by a simple renumbering of the eigenvalues. Both integrals are particular
instances of the generic covariance integral given in (\ref{eq_Gamma_generic}%
), where we recall that we can express
\[
\Phi_{\ell,m}\left(  \omega_{1},\omega_{2}\right)  =-\frac{\partial^{2}%
\log\left(  1-\Psi_{\ell,m}\left(  \omega_{1},\omega_{2}\right)  \right)
}{\partial\omega_{1}\partial\omega_{2}}%
\]
where the partial derivatives are well defined because at the contours we have
$\left\vert \Psi_{\ell,m}\left(  \omega_{1},\omega_{2}\right)  \right\vert <1$
(simply use Cauchy-Schwarz and apply the results in \cite[Appendix
I]{mestreeigsp08}). Therefore, using the integration by parts formula we can
alternatively express (\ref{eq_Gamma_generic}) as%
\[
\left\{  \Gamma\right\}  _{\ell,m}=-\frac{1}{2\pi\operatorname*{i}}\frac
{1}{2\pi\operatorname*{i}}\oint\nolimits_{\mathcal{C}_{\omega_{\ell}}^{+}%
}\oint\nolimits_{\mathcal{C}_{\omega_{m}}^{+}}\frac{dg_{\ell}\left(
\omega_{1}\right)  }{d\omega_{1}}\frac{dg_{m}\left(  \omega_{2}\right)
}{d\omega_{2}}\log\left(  1-\Psi_{\ell,m}\left(  \omega_{1},\omega_{2}\right)
\right)  d\omega_{1}d\omega_{2}.
\]
This means that we can re-write $\Gamma_{1}^{(\ell,m)}$ and $\Gamma_{2}%
^{(\ell,m)}$ as%

\begin{align*}
\Gamma_{1}^{(\ell,m)}  &  =-\frac{1}{2\pi\operatorname*{i}}\oint
\nolimits_{\mathcal{C}_{\omega_{\ell}}^{+}}\frac{dz_{1}}{d\omega_{1}}%
I(\omega_{1})d\omega_{1}\\
\Gamma_{2}^{(\ell,m)}  &  =-\frac{1}{2\pi\operatorname*{i}}\oint
\nolimits_{\mathcal{C}_{\omega_{\ell}}^{+}}\frac{d\mathcal{L}_{\ell}%
(\omega_{1})}{d\omega_{1}}I(\omega_{1})d\omega_{1}%
\end{align*}
where $\mathcal{L}_{m}(\omega)$ is defined in (\ref{eq:definition_L(w)}) and
where we have defined the integral
\[
I(\omega_{1})=\frac{1}{2\pi\operatorname*{i}}\oint\nolimits_{\mathcal{C}%
_{\omega_{m}}^{+}}\frac{d\mathcal{L}_{m}(\omega_{2})}{d\omega_{2}}\log\left(
1-\Psi_{\ell,m}\left(  \omega_{1},\omega_{2}\right)  \right)  d\omega_{2}.
\]
In order to solve this integral, we consider the singular value decomposition%
\[
\mathbf{R}_{\ell}^{1/2}=\mathbf{P}_{A}^{(\ell)}\mathbf{R}^{1/2}=\sum
_{r=0}^{\bar{M}_{\ell}}\sqrt{\gamma_{r}^{(\ell)}}\mathbf{U}_{r}^{(\ell
)}\left(  \mathbf{V}_{r}^{(\ell)}\right)  ^{H}%
\]
where $\bar{M}_{\ell}+1$ is the total number of singular values of
$\mathbf{R}_{\ell}^{1/2}$ (including $\{0\}$ when $\ell>0$). With all these
definitions, we may express the function $\Psi_{\ell,m}$ as
\begin{gather*}
\Psi_{\ell,m}\left(  \omega_{1},\omega_{2}\right)  =\sum_{r=0}^{\bar{M}_{\ell
}}\sum_{k=0}^{\bar{M}_{m}}\kappa_{rk}^{\ell m}\frac{\gamma_{r}^{(\ell)}%
\gamma_{k}^{(m)}}{\left(  \gamma_{r}^{(\ell)}-\omega_{1}\right)  \left(
\gamma_{k}^{(m)}-\omega_{2}\right)  }\\
\kappa_{rk}^{\ell m}=\frac{1}{N}\operatorname*{tr}\left[  \mathbf{V}%
_{r}^{(\ell)}\left(  \mathbf{V}_{r}^{(\ell)}\right)  ^{H}\mathbf{V}_{k}%
^{(m)}\left(  \mathbf{V}_{k}^{(m)}\right)  ^{H}\right]  .
\end{gather*}
and note that this notation is valid for $\ell,m\geq0$. Using the definition
of $\phi_{0}^{(m)}$, we can find an equivalent definition of the function
$\mathcal{L}_{m}(\omega_{2})$ that will be more convenient from now on, that
is%
\begin{equation}
\mathcal{L}_{m}(\omega_{2})=\log\left[  \left(  \omega_{2}-\phi_{0}%
^{(m)}\right)  \left(  1-\frac{1}{N}\sum_{k=1}^{\bar{M}_{m}}\widetilde{K}%
_{k}^{(m)}\frac{\gamma_{k}^{(m)}}{\gamma_{k}^{(m)}-\omega_{2}}\right)
\right]  \label{eq:definition_Log}%
\end{equation}
where
\[
\widetilde{K}_{k}^{(m)}=K_{k}^{(m)}\frac{\gamma_{k}^{(m)}}{\gamma_{k}%
^{(m)}-\phi_{0}^{(m)}}.
\]
In order to be able to handle both the undersampled and oversampled cases, we
let $0\leq\widetilde{\phi}_{1}^{(m)}<\ldots<\widetilde{\phi}_{\bar{M}_{m}%
}^{(m)}$ denote the $\bar{M}_{m}$ roots of the equation $\widetilde{\Phi
}^{(m)}(\widetilde{\phi})=1$, where $\widetilde{\Phi}^{(m)}\left(
\phi\right)  $ is defined as in (\ref{eq:defPsiTilde}) but replacing all
quantities associated with $\mathbf{R}$ by the equivalent ones associated with
$\mathbf{R}_{m}$.

\subsection{Computation of $I(\omega_{1})$}

In order to find a close form expression for $I(\omega_{1})$, we will follow
the same approach as in the proof of Proposition \ref{prop:integrals_log1}
above. More specifically, we express $I(\omega_{1})=I(\omega_{1},1)$ where
$I(\omega_{1},x)$ is a differentiable function of $x\in\left[  0,1\right]  $
defined as
\[
I(\omega_{1},x)=\frac{1}{2\pi\operatorname*{i}}\oint\nolimits_{\mathcal{C}%
_{\omega_{m}}^{+}}\frac{d\mathcal{L}_{m}(\omega_{2})}{d\omega_{2}}\log\left(
1-x\Psi_{\ell,m}\right)  d\omega_{2}.
\]
Obviously, we will have $I(\omega_{1},0)=0$. On the other hand, using the DCT
one can see that $I(\omega_{1},x)$ is a differentiable function of
$x\in\lbrack0,1]$ with derivative
\[
\frac{\partial I(\omega_{1},x)}{\partial x}=\frac{-1}{2\pi\operatorname*{i}%
}\oint\nolimits_{\mathcal{C}_{\omega_{m}}^{+}}\frac{d\mathcal{L}_{m}%
(\omega_{2})}{d\omega_{2}}\frac{\Psi_{\ell,m}}{1-x\Psi_{\ell,m}}d\omega_{2}.
\]
One can trivially find a closed form expression for the above integral using
conventional residue calculus. One only needs to realize that, since $\phi
_{0}^{(m)}$ is never enclosed by $\mathcal{C}_{\omega_{m}}$ (the proof is
similar to that of Lemma \ref{lemma:locnroots} below), the above integrand
only has singularities at the points $\{\gamma_{k}^{(m)},\widetilde{\phi}%
_{k}^{(m)},k=1,\ldots,\bar{M}_{m}\}$ plus the solutions to the equation
$x\Psi_{\ell,m}\left(  \omega_{1},\omega_{2}\right)  =1$ as a function of
$\omega_{1}$. The following lemma establishes that there exist exactly
$\bar{M}_{m}$ solutions to this equation enclosed by $\mathcal{C}_{\omega_{m}%
}$, which will be denoted as $\zeta_{j}^{(\ell,m)}\left(  \omega_{1},x\right)
$, $j=1,\ldots,\bar{M}_{m}$.

\begin{lemma}
\label{lemma:locnroots}Define, for $\ell\geq0$, the region
\[
\Omega_{\ell}=\left\{  \omega\in\mathbb{C}:\frac{1}{N}\sum_{r=0}^{\bar
{M}_{\ell}}K_{r}^{(\ell)}\left\vert \frac{\gamma_{r}^{(\ell)}}{\gamma
_{r}^{(\ell)}-\omega}\right\vert ^{2}<1\right\}
\]
and let $\omega_{1}\in\Omega_{\ell}$. Then, when $m\geq1$ and for $x\in(0,1]$,
the following equation in the complex variable $\zeta$
\begin{equation}
x\Psi_{\ell,m}\left(  \omega_{1},\zeta\right)  =1. \label{eq:rootsPsilm}%
\end{equation}
has exactly $\bar{M}_{m}$ roots inside $\partial\Omega_{m}$.
\end{lemma}

\begin{IEEEproof}%
See Appendix \ref{sec:proofLemma3}.%
\end{IEEEproof}

It can easily be seen that $\{\widetilde{\phi}_{k}^{(m)},k=1,\ldots,\bar
{M}_{m}\}$ are always enclosed by the contour $\partial\Omega_{m}$. Since the
contour $\mathcal{C}_{\omega_{m}}^{+}$ is always enclosing the region
$\Omega_{m}$ (cf. \cite{mestreeigsp08}) we can readily compute the value of
$\partial I(\omega_{1},x)/\partial x$ using conventional Cauchy calculus,%
\begin{multline*}
\frac{\partial I(\omega_{1},x)}{\partial x}=-\frac{\bar{M}_{m}}{x}-\sum
_{j=1}^{\bar{M}_{m}}\frac{\Psi_{\ell,m}\left(  \omega_{1},\widetilde{\phi}%
_{j}^{(m)}\right)  }{1-x\Psi_{\ell,m}\left(  \omega_{1},\widetilde{\phi}%
_{j}^{(m)}\right)  }+\\
+\sum_{j=1}^{\bar{M}_{m}}\left(  \frac{\widetilde{\Phi}^{(m)\prime}\left(
\zeta_{j}^{(\ell,m)}\left(  \omega_{1,}x\right)  \right)  }{1-\widetilde{\Phi
}^{(m)}\left(  \zeta_{j}^{(\ell,m)}\left(  \omega_{1,}x\right)  \right)
}-\frac{1}{\zeta_{j}^{(\ell,m)}\left(  \omega_{1},x\right)  -\phi_{0}^{(m)}%
}\right)  \frac{\partial\zeta_{j}^{(\ell,m)}\left(  \omega_{1},x\right)
}{\partial x}%
\end{multline*}
where $\widetilde{\Phi}^{(m)\prime}$ denotes the derivative of $\widetilde
{\Phi}^{(m)}$ and where we have used the fact that $\zeta_{j}^{(\ell
,m)}\left(  \omega_{1},x\right)  $ is a differentiable function of $x$ with
derivative%
\[
\frac{\partial\zeta_{j}^{(\ell,m)}\left(  \omega_{1},x\right)  }{\partial
x}=\left(  -x^{2}\left.  \frac{\partial\Psi_{\ell,m}\left(  \omega_{1}%
,\omega_{2}\right)  }{\partial\omega_{2}}\right\vert _{\omega_{2}=\zeta
_{j}^{(\ell,m)}\left(  \omega_{1},x\right)  }\right)  ^{-1}.
\]
We can therefore find a primitive of $\partial I(\omega_{1},x)/\partial x$ and
force the value of $I(\omega_{1},0)=0$ to fix the undetermined constraint. In
order to do this, we need to use the fact that $\lim_{x\rightarrow0}\zeta
_{j}^{(\ell,m)}\left(  \omega_{1,}x\right)  =\gamma_{j}^{(m)}$ and%
\[
\lim_{x\rightarrow0}x^{-1}\left(  \gamma_{j}^{(m)}-\zeta_{j}^{(\ell,m)}\left(
\omega_{1,}x\right)  \right)  =\sum_{r=1}^{\bar{M}_{\ell}}\kappa_{rj}^{\ell
m}\frac{\gamma_{r}^{(\ell)}\gamma_{j}^{(m)}}{\gamma_{r}^{(\ell)}-\omega_{1}}.
\]
All this leads to%
\[
I(\omega_{1},x)=-\log\left(  1-x\Psi_{\ell,m}\left(  \omega_{1},\phi_{0}%
^{(m)}\right)  \right)  +\xi_{1}(\omega_{1},x)+\xi_{2}(\omega_{1},x)
\]
where we have defined
\begin{align*}
\xi_{1}(\omega_{1},x)=\sum_{j=1}^{\bar{M}_{m}}\log\left(  \frac{1-x\Psi
_{\ell,m}\left(  \omega_{1},\widetilde{\phi}_{j}^{(m)}\right)  }%
{1-\widetilde{\Phi}^{(m)}\left(  \zeta_{j}^{(\ell,m)}\left(  \omega
_{1},x\right)  \right)  }\right)   &  -\sum_{j=1}^{\bar{M}_{m}}\log\left(
\frac{xN}{\widetilde{K}_{j}^{(m)}}\sum_{r=1}^{\bar{M}_{\ell}}\kappa_{rj}^{\ell
m}\frac{\gamma_{r}^{(\ell)}}{\omega_{1}-\gamma_{r}^{(\ell)}}\right) \\
\xi_{2}(\omega_{1},x)=\log\left(  1-x\Psi_{\ell,m}\left(  \omega_{1},\phi
_{0}^{(m)}\right)  \right)   &  +\sum_{j=1}^{\bar{M}_{m}}\log\left(
\frac{\gamma_{i}^{(m)}-\phi_{0}^{(m)}}{\zeta_{i}^{(\ell,m)}\left(  \omega
_{1},x\right)  -\phi_{0}^{(m)}}\right)
\end{align*}

\begin{lemma}
\label{lemma:ups1ups2}For any $x\in\left(  0,1\right]  $, we have $\xi
_{1}(\omega_{1},x)=\xi_{2}(\omega_{1},x)=0$.
\end{lemma}

\begin{IEEEproof}%
See Appendix \ref{sec:proofLemma4}.%
\end{IEEEproof}

By simply taking \thinspace$x=1$ in the resulting expression for $I(\omega
_{1},x)$ we get $I(\omega_{1})=-\log(1-\Psi_{\ell,m}(\omega_{1},\phi_{0}%
^{(m)}))$.

\subsection{Computation of $\Gamma_{1}^{(\ell,m)}$}

By applying the integration by parts formula we can write
\[
\Gamma_{1}^{(\ell,m)}=\frac{1}{2\pi\operatorname*{i}}\oint
\nolimits_{\mathcal{C}_{\omega_{\ell}}^{+}}z_{1}\frac{\sum_{r=1}^{\bar
{M}_{\ell}}\widetilde{\kappa}_{r}^{\ell m}\frac{\gamma_{r}^{(\ell)}}{\left(
\gamma_{r}^{(\ell)}-\omega_{1}\right)  ^{2}}}{1-\sum_{r=1}^{\bar{M}_{\ell}%
}\widetilde{\kappa}_{r}^{\ell m}\frac{\gamma_{r}^{(\ell)}}{\gamma_{r}^{(\ell
)}-\omega_{1}}}d\omega_{1}%
\]
where we have defined
\[
\widetilde{\kappa}_{r}^{\ell m}=\sum_{k=1}^{\bar{M}_{m}}\kappa_{rk}^{\ell
m}\frac{\gamma_{k}^{(m)}}{\gamma_{k}^{(m)}-\phi_{0}^{(m)}}.
\]
Let $\nu_{j}^{(\ell,m)}$ denote the solutions to the equation
\begin{equation}
1=\sum_{r=0}^{\bar{M}_{\ell}}\widetilde{\kappa}_{r}^{\ell m}\frac{\gamma
_{r}^{(\ell)}}{\gamma_{r}^{(\ell)}-\nu_{j}^{(\ell,m)}}.
\label{eq:equation_mus}%
\end{equation}

\begin{lemma}
\label{lemma:possols2}This equation has $\bar{M}_{\ell}$ distinct solutions,
namely \ $\nu_{1}^{(\ell,m)}<\ldots<\nu_{\bar{M}_{\ell}}^{(\ell,m)}$, all of
which are located inside $\mathcal{C}_{\omega_{\ell}}^{+}$.
\end{lemma}

\begin{IEEEproof}%
See Appendix \ref{sec:prooflemma5}.%
\end{IEEEproof}

Hence, we can obtain the value of $\Gamma_{1}^{(\ell,m)}$ by using classical
residue calculus on the poles $\{\gamma_{j}^{(\ell)},\nu_{j}^{(\ell,m)}\}$
that are enclosed by $\mathcal{C}_{\omega_{\ell}}^{+}$. Close examination of
the integrand of $\Gamma_{1}^{(\ell,m)}$ leads to the conclusion that the only
poles enclosed by $\mathcal{C}_{\omega_{\ell}}^{+}$ are $\{\gamma_{j}^{(\ell
)},\nu_{j}^{(\ell,m)},j=1,\ldots,\bar{M}_{\ell}\}$, and direct computation of
the residues allows us to conclude that
\begin{multline*}
\Gamma_{1}^{(\ell,m)}=\sum_{j=1}^{\bar{M}_{\ell}}K_{j}^{(\ell)}\gamma
_{j}^{(\ell)}+\sum_{j=1}^{\bar{M}_{\ell}}\gamma_{j}^{(\ell)}\left(
1-\sum_{\substack{r=1 \\r\neq j}}^{\bar{M}_{\ell}}K_{r}^{(\ell)}\frac
{\gamma_{r}^{(\ell)}}{\gamma_{r}^{(\ell)}-\gamma_{j}^{(\ell)}}\right)  +\\
-\sum_{j=1}^{\bar{M}_{\ell}}\nu_{j}^{(\ell,m)}\left(  1-\sum_{r=1}^{\bar
{M}_{\ell}}K_{r}^{(\ell)}\frac{\gamma_{r}^{(\ell)}}{\gamma_{r}^{(\ell)}%
-\nu_{j}^{(\ell,m)}}\right)  -\sum_{j=1}^{\bar{M}_{\ell}}\frac{K_{j}^{(\ell)}%
}{\widetilde{\kappa}_{j}^{\ell m}}\gamma_{j}^{(\ell)}\left(  1-\sum
_{\substack{r=1 \\r\neq j}}^{\bar{M}_{\ell}}\widetilde{\kappa}_{r}^{\ell
m}\frac{\gamma_{r}^{(\ell)}}{\gamma_{r}^{(\ell)}-\gamma_{j}^{(\ell)}}\right)
\end{multline*}
In order to simplify this, we introduce the following lemma.

\begin{lemma}
\label{lemma_sum_mus}Let $\ell\geq1$. For $j=1,\ldots,\bar{M}_{\ell}$ we have%
\begin{align}
\sum_{r=1}^{\bar{M}_{\ell}}\nu_{r}^{(\ell,m)}  &  =\sum_{r=1}^{\bar{M}_{\ell}%
}\gamma_{r}^{(\ell)}-\sum_{r=1}^{\bar{M}_{\ell}}\widetilde{\kappa}_{r}^{\ell
m}\gamma_{r}^{(\ell)}\label{eq:sum_gamma_nus}\\
\sum_{i=1}^{\bar{M}_{\ell}}\frac{\widetilde{\kappa}_{j}^{\ell m}\gamma
_{j}^{(\ell)}}{\nu_{i}^{(\ell,m)}-\gamma_{j}^{(\ell)}}  &  =\sum
_{\substack{r=1 \\r\neq j}}^{\bar{M}_{\ell}}\widetilde{\kappa}_{r}^{\ell
m}\frac{\gamma_{r}^{(\ell)}}{\gamma_{r}^{(\ell)}-\gamma_{j}^{(\ell)}}%
+\sum_{\substack{i=1 \\i\neq j}}^{\bar{M}_{\ell}}\widetilde{\kappa}_{j}^{\ell
m}\frac{\gamma_{j}^{(\ell)}}{\gamma_{i}^{(\ell)}-\gamma_{j}^{(\ell)}}-1
\label{eq:sum_inv_nu_minus_gamma}%
\end{align}

\end{lemma}

\begin{IEEEproof}%
See Appendix \ref{sec:prooflemma6}.%
\end{IEEEproof}

Now, using (\ref{eq:sum_gamma_nus})-(\ref{eq:sum_inv_nu_minus_gamma}) in the
expression of $\Gamma_{1}^{(\ell,m)}$ we directly obtain
\[
\Gamma_{1}^{(\ell,m)}=\sum_{r=1}^{\bar{M}_{\ell}}\widetilde{\kappa}_{r}^{\ell
m}\gamma_{r}^{(\ell)}=\frac{1}{N}\operatorname*{tr}\left[  \mathbf{R}%
^{1/2}\mathbf{P}_{A}^{(\ell)}\mathbf{R}^{1/2}\sum_{k=1}^{\bar{M}_{m}}%
\frac{\gamma_{k}^{(m)}}{\gamma_{k}^{(m)}-\phi_{0}^{(m)}}\mathbf{V}_{k}%
^{(m)}\left(  \mathbf{V}_{k}^{(m)}\right)  ^{H}\right]
\]
which can alternatively be expressed as in (\ref{eq:result_Gamma1}).

\subsection{Computation of $\Gamma_{2}^{(\ell,m)}$}

Using the expression of $I(\omega_{1})$ and the analycity of $\mathcal{L}%
_{\ell}(\omega_{1})$, we may express $\Gamma_{2}^{(\ell,m)}$ as $\Gamma
_{2}^{(\ell,m)}(1)$ where%
\[
\Gamma_{2}^{(\ell,m)}\left(  x\right)  =\frac{-1}{2\pi\operatorname*{i}}%
\oint\nolimits_{\mathcal{C}_{\omega_{\ell}}^{+}}\frac{d\mathcal{L}_{\ell
}(\omega_{1})}{d\omega_{1}}\log\left(  1-x\Psi_{\ell,m}\left(  \omega_{1}%
,\phi_{0}^{(m)}\right)  \right)  d\omega_{1}%
\]
is a differentiable function of $x$ in the unit interval. Taking the
definition of $\mathcal{L}_{\ell}(\omega_{1})$ in (\ref{eq:definition_Log}),
we can readily see that $\Gamma_{2}^{(\ell,m)}\left(  0\right)  =0$. On the
other hand, using the dominated convergence theorem we can compute the
derivative as
\[
\frac{d\Gamma_{2}^{(\ell,m)}\left(  x\right)  }{dx}=\frac{1}{2\pi
\operatorname*{i}}\oint\nolimits_{\mathcal{C}_{\omega_{\ell}}^{+}}%
\frac{d\mathcal{L}_{\ell}(\omega_{1})}{d\omega_{1}}\frac{\Psi_{\ell,m}\left(
\omega_{1},\phi_{0}^{(m)}\right)  }{1-x\Psi_{\ell,m}\left(  \omega_{1}%
,\phi_{0}^{(m)}\right)  }d\omega_{1}.
\]
This integral can be computed via conventional Cauchy integration, using the
fact that the singularities are located at the points $\{\gamma_{r}^{(\ell
)},\widetilde{\phi}_{j}^{(\ell)},\nu_{j}^{(\ell,m)}\left(  x\right)
,j=1,\ldots,\bar{M}_{\ell}\}$ where we recall that $\widetilde{\phi}%
_{j}^{(\ell)}$ are the solutions to $\widetilde{\Phi}^{(\ell)}\left(
\omega_{1}\right)  =1$ and where we have introduced the quantities $\nu
_{j}^{(\ell,m)}\left(  x\right)  $, defined as the solutions to the equation
$x\Psi_{\ell,m}(\omega_{1},\phi_{0}^{(m)})=1$. It can be shown that all these
points are inside $\mathcal{C}_{\omega_{\ell}}$. Computing the corresponding
residues, we obtain%
\[
\frac{d\Gamma_{2}^{(\ell,m)}\left(  x\right)  }{dx}=\frac{1}{x}\bar{M}_{\ell
}+\sum_{j=1}^{\bar{M}_{\ell}}\frac{\sum_{r=1}^{\bar{M}_{\ell}}\widetilde
{\kappa}_{r}^{\ell m}\frac{\gamma_{r}^{(\ell)}}{\gamma_{r}^{(\ell)}%
-\widetilde{\phi}_{j}^{(\ell)}}}{1-x\sum_{r=1}^{\bar{M}_{\ell}}\widetilde
{\kappa}_{r}^{\ell m}\frac{\gamma_{r}^{(\ell)}}{\gamma_{r}^{(\ell)}%
-\widetilde{\phi}_{j}^{(\ell)}}}+\sum_{j=1}^{\bar{M}_{\ell}}\left(  \frac
{1}{\nu_{j}^{(\ell,m)}\left(  x\right)  -\phi_{0}^{(\ell)}}-\frac{\Phi
^{(\ell)\prime}(\nu_{j}^{(\ell,m)}\left(  x\right)  )}{1-\Phi^{(\ell)}(\nu
_{j}^{(\ell,m)}\left(  x\right)  )}\right)  \frac{d\nu_{j}^{(\ell,m)}\left(
x\right)  }{dx}%
\]
where $\Phi^{(\ell)}(\omega_{1})$ is defined as in (\ref{eq:defPsiTilde}) and
where $\Phi^{(\ell)\prime}(\omega_{1})$ denotes its derivative. We can readily
find a primitive of the above equation and finally obtain
\begin{multline*}
\Gamma_{2}^{(\ell,m)}=-\sum_{j=1}^{\bar{M}_{\ell}}\log\left\vert 1-\sum
_{r=1}^{\bar{M}_{\ell}}\widetilde{\kappa}_{r}^{\ell m}\frac{\gamma_{r}%
^{(\ell)}}{\gamma_{r}^{(\ell)}-\widetilde{\phi}_{j}^{(\ell)}}\right\vert \\
+\sum_{j=1}^{\bar{M}_{\ell}}\log\left\vert N\frac{\widetilde{\kappa}_{j}^{\ell
m}}{\widetilde{K}_{j}^{(\ell)}}\frac{\nu_{j}^{(\ell,m)}-\phi_{0}^{(\ell)}%
}{\gamma_{j}^{(\ell)}-\phi_{0}^{(\ell)}}\right\vert +\sum_{j=1}^{\bar{M}%
_{\ell}}\log\left\vert 1-\frac{1}{N}\sum_{r=1}^{\bar{M}_{\ell}}\widetilde
{K}_{r}^{(\ell)}\frac{\gamma_{r}^{(\ell)}}{\gamma_{r}^{(\ell)}-\nu_{j}%
^{(\ell,m)}}\right\vert
\end{multline*}
where we have used the fact that $\Gamma_{2}^{(\ell,m)}(0)=0$ to fix the
indeterminate constraint. We can now simplify this expression using the
following result.

\begin{lemma}
\label{lemma:transform_final}The following identities hold for any $j\geq1$,
\begin{align*}
\prod\limits_{r=1}^{\bar{M}_{\ell}}\frac{\nu_{r}^{(\ell,m)}-\widetilde{\phi
}_{j}^{(\ell)}}{\gamma_{r}^{(\ell)}-\widetilde{\phi}_{j}^{(\ell)}}  &
=1-\sum_{r=1}^{\bar{M}_{\ell}}\widetilde{\kappa}_{r}^{\ell m}\frac{\gamma
_{r}^{(\ell)}}{\gamma_{r}^{(\ell)}-\widetilde{\phi}_{j}^{(\ell)}}\\
\prod\limits_{r=1}^{\bar{M}_{\ell}}\left(  \nu_{r}^{(\ell,m)}-\gamma
_{j}^{(\ell)}\right)   &  =-\widetilde{\kappa}_{j}^{\ell m}\gamma_{j}^{(\ell
)}\prod\limits_{\substack{k=1 \\k\neq j}}^{\bar{M}_{\ell}}\left(  \gamma
_{k}^{(\ell)}-\gamma_{j}^{(\ell)}\right) \\
{\displaystyle\prod\limits_{k=1}^{\bar{M}_{\ell}}} \frac{\widetilde{\phi}%
_{k}^{(\ell)}-\nu_{j}^{(\ell,m)}}{\gamma_{k}^{(\ell)}-\nu_{j}^{(\ell,m)}}  &
=1-\frac{1}{N}\sum_{r=1}^{\bar{M}_{\ell}}\widetilde{K}_{r}^{(\ell)}%
\frac{\gamma_{r}^{(\ell)}}{\gamma_{r}^{(\ell)}-\nu_{j}^{(\ell,m)}}\\
{\displaystyle\prod\limits_{k=1}^{\bar{M}_{\ell}}} \left(  \widetilde{\phi
}_{k}^{(\ell)}-\gamma_{j}^{(\ell)}\right)   &  =-\frac{1}{N}\widetilde{K}%
_{j}^{(\ell)}\gamma_{j}^{(\ell)}{\displaystyle\prod\limits_{\substack{k=1
\\k\neq j}}^{\bar{M}_{\ell}}} \left(  \gamma_{k}^{(\ell)}-\gamma_{j}^{(\ell
)}\right) \\
\prod\limits_{r=1}^{\bar{M}_{\ell}}\frac{\nu_{r}^{(\ell,m)}-\phi_{0}^{(\ell)}%
}{\gamma_{k}^{(\ell)}-\phi_{0}^{(\ell)}}  &  =1-\sum_{r=1}^{\bar{M}_{\ell}%
}\widetilde{\kappa}_{r}^{\ell m}\frac{\gamma_{r}^{(\ell)}}{\gamma_{r}^{(\ell
)}-\phi_{0}^{(\ell)}}.
\end{align*}

\end{lemma}

\begin{IEEEproof}%
See Appendix \ref{sec:prooflemma7}.%
\end{IEEEproof}

By direct application of the identities in Lemma \ref{lemma:transform_final}
below, we can simplify the expression of $\Gamma_{2}^{(\ell,m)}$ as%
\[
\Gamma_{2}^{(\ell,m)}=\log\left\vert 1-\sum_{r=1}^{\bar{M}_{\ell}}%
\widetilde{\kappa}_{r}^{\ell m}\frac{\gamma_{r}^{(\ell)}}{\gamma_{r}^{(\ell
)}-\phi_{0}^{(\ell)}}\right\vert
\]
which can alternatively be expressed as in (\ref{eq:solution_Gamma2}).

\section{\label{sec:proofTheorem4}Proof of Theorem \ref{theorem:CLT_general}}

Let us consider the following real-valued random variable $\eta_{\ell}%
=\frac{M}{2\pi\operatorname*{i}}\oint\nolimits_{\mathcal{C}_{\ell}}f_{\ell
}(z)\left(  \hat{m}_{\ell}(z)-\bar{m}_{\ell}(z)\right)  dz$ where
\begin{gather*}
\hat{m}_{\ell}(z)=\frac{1}{M}\operatorname*{tr}\left[  \mathbf{\hat{Q}}_{\ell
}(z)\right]  ,\quad\bar{m}_{\ell}(z)=\frac{1}{M}\operatorname*{tr}\left[
\mathbf{\bar{Q}}_{\ell}(z)\right] \\
\mathbf{\hat{Q}}_{\ell}(z)=\left(  \mathbf{\hat{R}}_{\ell}-z\mathbf{I}%
_{M}\right)  ^{-1}\quad\mathbf{\bar{Q}}_{\ell}(z)=\frac{\omega_{\ell}\left(
z\right)  }{z}\left(  \mathbf{R}_{\ell}-\omega_{\ell}\left(  z\right)
\mathbf{I}_{M}\right)  ^{-1}%
\end{gather*}
with $\omega_{\ell}\left(  z\right)  $ being the unique solution to the
following equation
\[
z=\omega_{\ell}\left(  z\right)  \left(  1-\frac{1}{N}\operatorname*{tr}%
\left[  \mathbf{R}_{\ell}\left(  \mathbf{R}_{\ell}-\omega_{\ell}\left(
z\right)  \mathbf{I}_{M}\right)  ^{-1}\right]  \right)
\]
that belongs to $\mathbb{C}^{\mathbb{+}}$ for $z\in\mathbb{C}^{\mathbb{+}}$.
Let $a_{\ell}$, $\ell=1\ldots L$, denote real-valued bounded quantities and
consider%
\begin{equation}
\eta=\sum_{\ell=1}^{L}a_{\ell}\eta_{\ell}=\sum_{\ell=1}^{L}a_{\ell}\frac
{1}{2\pi\operatorname*{i}}\oint\nolimits_{\mathcal{C}_{\ell}}f_{\ell
}(z)M\left(  \hat{m}_{\ell}(z)-\bar{m}_{\ell}(z)\right)  dz
\label{eq:definition_eta}%
\end{equation}
Let $\Psi(u)=\exp\left(  \operatorname*{i}u\eta\right)  $ and consider the
characteristic function $\mathbb{E}\left[  \Psi(u)\right]  $. The objective is
to show that, in the limit when $M,N\rightarrow\infty$, we have
\begin{equation}
\mathbb{E}\left[  \Psi\left(  u\right)  \right]  -\exp\left(  -\frac
{\mathbf{a}^{T}\mathbf{\Gamma a}}{2}u^{2}\right)  \rightarrow0
\label{eq:pointwise_conv_charact}%
\end{equation}
pointwise in $u$, where $\mathbf{a}=\left[  a_{1},\ldots,a_{L}\right]  ^{T}$
and where $\mathbf{\Gamma}$ is defined in the statement of Theorem
\ref{theorem:CLT_general}. Given the boundedness assumptions in the statement
of the theorem, the result will follow from a trivial modification of
\cite[Proposition 6]{hachem08}. The rest of the section is therefore devoted
to showing (\ref{eq:pointwise_conv_charact}).

The first step of the proof consists in replacing the original contour
$\mathcal{C}_{\ell}$, which depends on $\mathbf{\hat{R}}_{\ell}$ (it encloses
all the eigenvalues of $\mathbf{\hat{R}}_{\ell}$ except zero)\ by a fixed
deterministic contour that only depends on $\mathbf{R}_{\ell}$, which will be
denoted by $\mathcal{\bar{C}}_{\ell}$. To do this, we use the fact that, for
all large $M,N$, the eigenvalues of $\mathbf{\hat{R}}_{\ell}$ are all located
inside a certain support, $\mathcal{S}_{\ell}$ \cite{silverstein98}.
Unfortunately, the random variable $\eta$\ in (\ref{eq:definition_eta}) does
not need to have a characteristic function for all $M$\ when $\mathcal{C}%
_{\ell}$ is replaced with $\mathcal{\bar{C}}_{\ell}$. This is because there
might exist realizations for which the eigenvalues of $\mathbf{\hat{R}}_{\ell
}$ become dangerously close the contour $\mathcal{C}_{\ell}$ or even on
$\mathcal{C}_{\ell}$. In order to overcome this difficulty, we will follow the
approach in \cite{Pastur2011, Hachem12} and consider an equivalent (large-$M$)
representation of $\eta$ that is guaranteed to have characteristic function
for all $M$. Indeed, let us define $\mathcal{S}_{\ell}^{\mathcal{\epsilon}%
}=\left\{  x\in\mathbb{R}:\operatorname*{dist}\left(  x,\mathcal{S}_{\ell
}\right)  \leq\epsilon\right\}  $ for $\epsilon>0$. Assume that $\epsilon$ is
small enough such that $\mathcal{S}_{\ell}^{2\mathcal{\epsilon}}$ does not
contain $\{0\}$. Let $\phi_{\ell}$ denote a smooth function $\phi_{\ell
}:\mathbb{R}\rightarrow\lbrack0,1]$ such that $\phi_{\ell}(x)=1$ for
$x\in\mathcal{S}_{\ell}^{\mathcal{\epsilon}}$ and $\phi_{\ell}(x)=0$ for
$x\in\mathbb{R}\backslash\mathcal{S}_{\ell}^{2\mathcal{\epsilon}}$. We will
write $\phi_{\ell}=\det\phi_{\ell}\left(  \mathbf{\hat{R}}_{\ell}\right)  $.
By \cite{silverstein98}, we know that $\phi_{M}=1$ with probability one for
all $M$ sufficiently large. Therefore, we may represent $\eta$ as
\begin{equation}
\eta=\sum_{\ell=1}^{L}a_{\ell}\frac{1}{2\pi\operatorname*{i}}\oint
\nolimits_{\mathcal{\bar{C}}_{\ell}}f_{\ell}(z)M\left(  \hat{m}_{\ell}%
(z)-\bar{m}_{\ell}(z)\right)  \phi_{\ell}dz \label{eq:eta_regularized}%
\end{equation}
almost surely for all $M$ sufficiently large. The characteristic function of
(\ref{eq:eta_regularized}) exists for every realization and every possible $M
$.\ Having introduced this regularization parameter $\phi_{\ell}$ and the
deterministic contours, we are now in the position of introducing the main
technical tools that will be used in the proof of this theorem. Following the
approach in \cite{hachem08}, our derivations will be based on the partial
integration formula for Gaussian functionals, together with the
Poincar\'{e}-Nash inequality. We introduce these tools in the following proposition.

\begin{remark}
\label{remark_ezNn}In what follows, the symbol $\mathcal{O}(N^{-k})$ will
denote a general bivariate complex function that is bounded in magnitude by
$\epsilon\left(  z_{1},z_{2}\right)  N^{-k}$, where $\epsilon\left(
z_{1},z_{2}\right)  $ does not depend on $N$ and is such that
\begin{equation}
\max_{m,n}\sup_{\left(  z_{1},z_{2}\right)  \in\mathcal{C}_{m}\mathcal{\times
C}_{n}}\left\Vert \epsilon\left(  z_{1},z_{2}\right)  \right\Vert <+\infty.
\label{eq:supremum_abs(ez)}%
\end{equation}
The function itself\ may be different from one line to another, and it may be
matrix valued, in which case (\ref{eq:supremum_abs(ez)}) is understood as the
spectral norm. On the other hand, $\mathcal{O}(N^{-\mathbb{N}})$ should be
understood as a bivariate complex function that can be written as
$\mathcal{O}(N^{-\ell})$ for every $\ell\in\mathbb{N}$.
\end{remark}

\begin{proposition}
\label{lemma:ipp_regularization}Assume that, for each fixed $z\in\mathbb{C}$,
the function $\Omega\left(  \mathbf{X},\mathbf{X}^{\ast},z\right)  :$
$\mathbb{R}^{2MN}\rightarrow\mathbb{C}$ is continuously differentiable and
such that both itself and its partial derivatives are polynomically bounded.
If $\mathbf{X}$ is real valued, simply consider $\Gamma$ as a function on
$\mathbb{R}^{MN}$, with the same properties. Than, under $\mathbf{(As1)}$ we
can write
\begin{equation}
\mathbb{E}\left[  X_{ij}\Omega\left(  \mathbf{X},\mathbf{X}^{\ast},z\right)
\right]  =\mathbb{E}\left[  \frac{\partial\Omega\left(  \mathbf{X}%
,\mathbf{X}^{\ast},z\right)  }{\partial X_{ij}^{\ast}}\right]
\label{eq:ipp_reg1}%
\end{equation}
where
\[
\frac{\partial}{\partial X_{ij}^{\ast}}=\frac{1}{2}\frac{\partial}%
{\partial\operatorname{Re}\left[  X_{ij}\right]  }+\operatorname*{j}\frac
{1}{2}\frac{\partial}{\partial\operatorname{Im}\left[  X_{ij}\right]  }.
\]
On the other hand, we can also write
\begin{equation}
\operatorname*{var}\left[  \Omega\left(  \mathbf{X},\mathbf{X}^{\ast
},z\right)  \right]  \leq\sum_{i=1}^{M}\sum_{j=1}^{N}\mathbb{E}\left[
\left\vert \frac{\partial\Omega\left(  \mathbf{X},\mathbf{X}^{\ast},z\right)
}{\partial X_{ij}}\right\vert ^{2}+\left\vert \frac{\partial\Omega\left(
\mathbf{X},\mathbf{X}^{\ast},z\right)  }{\partial X_{ij}^{\ast}}\right\vert
^{2}\right]  \label{eq:Nash-Poincare}%
\end{equation}
where now%
\[
\frac{\partial}{\partial X_{ij}}=\frac{1}{2}\frac{\partial}{\partial
\operatorname{Re}\left[  X_{ij}\right]  }-\operatorname*{j}\frac{1}{2}%
\frac{\partial}{\partial\operatorname{Im}\left[  X_{ij}\right]  }.
\]
\ The function $\phi_{\ell}$ is continuously differentiable (on $\mathbb{R}%
^{2MN}$) with polynomically bounded partial derivatives. If, in addition,
$\sup_{z\in\mathcal{C}}\mathbb{E}\left(  \left\vert \Omega\left(
\mathbf{X},\mathbf{X}^{\ast},z\right)  \phi_{\ell}\right\vert ^{2}\right)  <C$
for some positive deterministic $C$ independent of $M$, then under
$(\mathbf{As1})$,%
\begin{equation}
\mathbb{E}\left[  \Omega\left(  \mathbf{X},\mathbf{X}^{\ast},z\right)
\phi_{M}^{r}\right]  =\mathbb{E}\left[  \Omega\left(  \mathbf{X}%
,\mathbf{X}^{\ast},z\right)  \phi_{\ell}\right]  +\mathcal{O}\left(
N^{-\mathbb{N}}\right)  \label{eq:getridofphi}%
\end{equation}
for any $r\in\mathbb{N}$, and also
\begin{equation}
\mathbb{E}\left[  \Omega\left(  \mathbf{X},\mathbf{X}^{\ast},z\right)
\frac{\partial\phi_{\ell}}{\partial X_{ij}}\right]  =\mathcal{O}\left(
N^{-\mathbb{N}}\right)  \label{eq:getridofphider}%
\end{equation}
where the term $\mathcal{O}\left(  N^{-\mathbb{N}}\right)  $ should be
understood as in Remark \ref{remark_ezNn} above.
\end{proposition}

The above results are well known in the random matrix literature and the proof
is therefore omitted. One of the conclusions of Proposition
\ref{lemma:ipp_regularization} is the fact that we can basically ignore the
presence of the regularization term $\phi_{\ell}$ up to an error of order
$\mathcal{O}\left(  N^{-m}\right)  $ for any $m\in\mathbb{N}$, which will be
irrelevant for the purposes of our derivations.

From now on we will therefore consider the definition of $\eta$ in
(\ref{eq:eta_regularized}) and apply the above tools to investigate the
asymptotic behavior of its characteristic function. Using the dominated
convergence theorem, we can establish that $\mathbb{E}\left[  \Psi(u)\right]
$ is a differentiable function of $u\,\ $with derivative%
\[
\frac{d\mathbb{E}\left[  \Psi(u)\right]  }{du}=\operatorname*{i}\sum_{\ell
=1}^{L}a_{\ell}\frac{M}{2\pi\operatorname*{i}}\oint\nolimits_{\mathcal{C}%
_{\ell}}f_{\ell}(z_{1})\mathbb{E}\left[  \left(  \hat{m}_{\ell}(z_{1})-\bar
{m}_{\ell}(z_{1})\right)  \Psi(u)\phi_{\ell}\right]  dz_{1}.
\]
We can study the quantity inside the expectation using the resolvent identity,
which states that
\[
z_{1}\mathbf{\hat{Q}}_{\ell}(z_{1})=\mathbf{\hat{Q}}_{\ell}(z_{1}%
)\mathbf{\hat{R}}_{\ell}-\mathbf{I}_{M}.
\]
We start by noting that we can develop the term
\[
\mathbb{E}\left[  \mathbf{\hat{Q}}_{\ell}(z_{1})\mathbf{\hat{R}}_{\ell}%
\Psi(u)\phi_{\ell}\right]  =\sum_{i=1}^{M}\sum_{j=1}^{N}\mathbb{E}\left[
X_{ij}\mathbf{\hat{Q}}_{\ell}(z_{1})\mathbf{R}_{\ell}^{1/2}\frac
{\mathbf{e}_{i}\mathbf{x}_{j}^{H}}{N}\mathbf{R}_{\ell}^{H/2}\Psi(u)\phi_{\ell
}\right]
\]
which can be further developed using the integration by parts formula in
(\ref{eq:ipp_reg1}), together with the identity
\[
\frac{\partial}{\partial\overline{X_{ij}}}\mathbf{\hat{Q}}_{\ell}%
(z_{1})=-\mathbf{\hat{Q}}_{\ell}(z_{1})\mathbf{R}_{\ell}^{1/2}\frac
{\mathbf{x}_{j}\mathbf{e}_{i}^{H}}{N}\mathbf{R}_{\ell}^{H/2}\mathbf{\hat{Q}%
}_{\ell}(z_{1}).
\]
A direct application of these techniques and the resolvent identity allows us
to write
\begin{align}
\mathbb{E}\left[  \mathbf{\hat{Q}}_{\ell}(z_{1})\phi_{\ell}\Psi(u)\right]   &
=-z_{1}^{-1}\mathbf{I}_{M}\mathbb{E}\left[  \Psi(u)\phi_{\ell}\right]
+\frac{1}{\omega_{\ell}\left(  z_{1}\right)  }\mathbb{E}\left[  \mathbf{\hat
{Q}}_{\ell}(z_{1})\mathbf{R}_{\ell}\Psi(u)\phi_{\ell}\right]  -\frac{1}%
{\omega_{\ell}\left(  z_{1}\right)  }\mathbb{E}\left[  \mathbf{\hat{Q}}_{\ell
}(z_{1})\mathbf{\hat{R}}_{\ell}\Psi(u)\alpha_{\ell}(z_{1})\phi_{\ell}\right]
\label{eq:tempEQzPsi}\\
&  -\operatorname*{i}u\frac{1}{\omega_{\ell}\left(  z_{1}\right)  }\sum
_{m=1}^{L}\frac{a_{m}}{2\pi\operatorname*{i}}\oint\nolimits_{\mathcal{C}_{m}%
}f_{m}(z_{2})\frac{1}{N}\mathbb{E}\left[  \mathbf{\hat{Q}}_{\ell}%
(z_{1})\mathbf{R}_{\ell,m}\mathbf{\hat{Q}}_{m}^{2}(z_{2})\mathbf{\hat{R}%
}_{m,\ell}\Psi(u)\phi_{\ell}\right]  dz_{2}\nonumber\\
&  +\mathcal{O}\left(  N^{-\mathbb{N}}\right) \nonumber
\end{align}
with the definitions%
\begin{align*}
\alpha_{\ell}(z_{1})  &  =\frac{1}{N}\operatorname*{tr}\left[  \mathbf{R}%
_{\ell}\mathbf{\hat{Q}}_{\ell}(z_{1})\right]  \phi_{\ell}-\frac{1}%
{N}\operatorname*{tr}\left[  \mathbf{R}_{\ell}\mathbf{\bar{Q}}_{\ell}%
(z_{1})\right] \\
\mathbf{R}_{\ell,m}  &  =\mathbf{R}_{\ell}^{1/2}\mathbf{R}_{m}^{H/2}\\
\mathbf{\hat{R}}_{\ell,m}  &  =\mathbf{R}_{\ell}^{1/2}\frac{\mathbf{XX}^{H}%
}{N}\mathbf{R}_{m}^{H/2}%
\end{align*}
and where we have used the identity%
\begin{equation}
1+\frac{1}{N}\operatorname*{tr}\left[  \mathbf{R}_{\ell}\mathbf{\bar{Q}}%
_{\ell}(z_{1})\right]  =1+\frac{\omega_{\ell}\left(  z_{1}\right)  }{z_{1}%
}\frac{1}{N}\operatorname*{tr}\left[  \mathbf{R}_{\ell}\left(  \mathbf{R}%
_{\ell}-\omega_{\ell}\left(  z_{1}\right)  \mathbf{I}_{M}\right)
^{-1}\right]  =\frac{\omega_{\ell}\left(  z_{1}\right)  }{z_{1}}
\label{eq:identity_w/z}%
\end{equation}
which follows directly from the definition of $\omega_{\ell}\left(
z_{1}\right)  $. In order to investigate the asymptotic behavior of
$\mathbb{E}\left[  \hat{m}_{\ell}(z_{1})\Psi(u)\phi_{\ell}\right]  $ from
(\ref{eq:tempEQzPsi}), we move the second term on the right hand side of
(\ref{eq:tempEQzPsi}) to the left hand side and multiply both sides by
$\mathbf{\bar{Q}}_{\ell}(z_{1})$ so that we obtain
\begin{multline}
\mathbb{E}\left[  \left(  \mathbf{\hat{Q}}_{\ell}(z_{1})-\mathbf{\bar{Q}%
}_{\ell}(z_{1})\right)  \Psi(u)\phi_{\ell}\right]  =\frac{z_{1}}{\omega_{\ell
}\left(  z_{1}\right)  }\mathbb{E}\left[  \mathbf{\hat{Q}}_{\ell}%
(z_{1})\mathbf{\hat{R}}_{\ell}\mathbf{\bar{Q}}_{\ell}(z_{1})\Psi
(u)\alpha_{\ell}(z_{1})\phi_{\ell}\right] \label{eq:diff_resolvents}\\
+\operatorname*{i}u\frac{z_{1}}{\omega_{\ell}\left(  z_{1}\right)  }\sum
_{m=1}^{L}\frac{a_{m}}{2\pi\operatorname*{i}}\oint\nolimits_{\mathcal{C}_{m}%
}f_{m}(z_{2})\frac{1}{N}\mathbb{E}\left[  \mathbf{\hat{Q}}_{\ell}%
(z_{1})\mathbf{R}_{\ell,m}\mathbf{\hat{Q}}_{m}^{2}(z_{2})\mathbf{\hat{R}%
}_{m,\ell}\mathbf{\bar{Q}}_{\ell}(z_{1})\Psi(u)\phi_{\ell}\phi_{m}\right]
dz_{2}+\mathcal{O}\left(  N^{-\mathbb{N}}\right)  .
\end{multline}
Observe that the trace of the left hand side of the above equation is directly
equal to $M\mathbb{E}\left[  \left(  \hat{m}_{\ell}(z_{1})-\bar{m}_{\ell
}(z_{1})\right)  \Psi(u)\phi_{\ell}\right]  $, which is the original quantity
that we want to analyze.

Let us now analyze the behavior of the quantity $\mathbb{E}\left[
\alpha_{\ell}(z_{1})\Psi(u)\phi_{\ell}\right]  .$ To to this, we multiply both
sides of (\ref{eq:diff_resolvents}) by $\mathbf{R}_{\ell}$ and take traces of
the result, so that we can write
\begin{multline}
\mathbb{E}\left[  \alpha_{\ell}(z_{1})\Psi(u)\phi_{\ell}\right]  =\frac
{1}{1-\gamma_{\ell,\ell}\left(  z_{1},z_{1}\right)  }\mathbb{E}\left[
\alpha_{\ell}(z_{1})\beta_{\ell}(z_{1})\Psi(u)\phi_{\ell}\right]
\label{eq:Ealpha1PSi}\\
+\frac{1}{N}\frac{\operatorname*{i}u}{1-\gamma_{\ell,\ell}\left(  z_{1}%
,z_{1}\right)  }\frac{z_{1}}{\omega_{\ell}\left(  z_{1}\right)  }\sum
_{m=1}^{L}\frac{a_{m}}{2\pi\operatorname*{i}}\oint\nolimits_{\mathcal{C}_{m}%
}f_{m}(z_{2})\frac{1}{N}\mathbb{E}\operatorname*{tr}\left[  \mathbf{\bar{Q}%
}_{\ell}(z_{1})\mathbf{R}_{\ell}\mathbf{\hat{Q}}_{\ell}(z_{1})\mathbf{R}%
_{\ell,m}\mathbf{\hat{Q}}_{m}^{2}(z_{2})\mathbf{\hat{R}}_{m,\ell}\Psi
(u)\phi_{\ell}\phi_{m}\right]  dz_{2}+\mathcal{O}\left(  N^{-\mathbb{N}%
}\right)
\end{multline}
where we have defined
\begin{align*}
\beta_{\ell}(z_{1})  &  =\frac{z_{1}}{\omega_{\ell}\left(  z_{1}\right)
}\frac{1}{N}\operatorname*{tr}\left[  \mathbf{\bar{Q}}_{\ell}(z_{1}%
)\mathbf{R}_{\ell}\mathbf{\hat{Q}}_{\ell}(z_{1})\mathbf{\hat{R}}_{\ell
}\right]  \phi_{\ell}-\gamma_{\ell,\ell}\left(  z_{1},z_{1}\right) \\
\gamma_{\ell,m}\left(  z_{1},z_{2}\right)   &  =\frac{z_{1}z_{2}}{\omega
_{\ell}\left(  z_{1}\right)  \omega_{m}\left(  z_{2}\right)  }\frac{1}%
{N}\operatorname*{tr}\left[  \mathbf{R}_{m\ell}\mathbf{\bar{Q}}_{\ell}%
(z_{1})\mathbf{\newline R}_{\ell m}\mathbf{\bar{Q}}_{m}(z_{2})\right]
\end{align*}
and where we have used the well known fact that%
\[
\sup_{M}\sup_{z_{1}\in\mathcal{C}_{\ell},z_{2}\in\mathcal{C}_{m}}\left\vert
\gamma_{\ell,m}\left(  z_{1},z_{2}\right)  \right\vert <1
\]
so that the quantity $1-\gamma_{\ell,\ell}\left(  z_{1},z_{1}\right)  $ is
always invertible.

Consider the first term on the right hand side of (\ref{eq:Ealpha1PSi}).
According to Lemma \ref{lemma:auxiliar_1resolvent} presented below, the
expectation of $\alpha_{\ell}(z_{1}),\beta_{\ell}(z_{1})$ is $\mathcal{O}%
(N^{-1})$, and its variance decays as $\mathcal{O}(N^{-2})$. Hence, we can
write (by Cauchy-Schwarz inequality)%
\[
\left\vert \mathbb{E}\left[  \alpha_{\ell}(z_{1})\beta_{\ell}(z_{1}%
)\Psi(u)\right]  \right\vert ^{2}\leq\mathbb{E}\left[  \left\vert \alpha
_{\ell}(z_{1})\right\vert ^{2}\right]  \mathbb{E}\left[  \left\vert
\beta_{\ell}(z_{1})\right\vert ^{2}\right]  =\mathcal{O}(N^{-2})
\]
so that the first term on the right hand side\ of (\ref{eq:Ealpha1PSi}) decays
as $\mathcal{O}(N^{-1})$. Regarding the second term, we can use a similar
argument together with Lemma \ref{lemma:auxiliar_3resolvents} presented below
to establish that
\begin{multline*}
\frac{1}{N}\mathbb{E}\operatorname*{tr}\left[  \mathbf{\bar{Q}}_{\ell}%
(z_{1})\mathbf{R}_{\ell}\mathbf{\hat{Q}}_{\ell}(z_{1})\mathbf{R}_{\ell
,m}\mathbf{\hat{Q}}_{m}^{2}(z_{2})\mathbf{\hat{R}}_{m,\ell}\Psi(u)\phi_{\ell
}\phi_{m}\right]  =\\
=\frac{1}{N}\mathbb{E}\operatorname*{tr}\left[  \mathbf{\bar{Q}}_{\ell}%
(z_{1})\mathbf{R}_{\ell}\mathbf{\hat{Q}}_{\ell}(z_{1})\mathbf{R}_{\ell
,m}\mathbf{\hat{Q}}_{m}^{2}(z_{2})\mathbf{\hat{R}}_{m,\ell}\phi_{\ell}\phi
_{m}\right]  \mathbb{E}\left[  \Psi(u)\right]  +\mathcal{O}(N^{-1}).
\end{multline*}
Furthermore, the approximations in Lemma \ref{lemma:auxiliar_3resolvents}
allow to express
\begin{multline*}
\frac{1}{N}\mathbb{E}\operatorname*{tr}\left[  \mathbf{\bar{Q}}_{\ell}%
(z_{1})\mathbf{R}_{\ell}\mathbf{\hat{Q}}_{\ell}(z_{1})\mathbf{\newline%
\mathbf{R}}_{\ell,m}\mathbf{\hat{Q}}_{m}^{2}(z_{2})\mathbf{\hat{R}}_{m,\ell
}\phi_{\ell}\phi_{m}\right]  =\\
=\left(  \frac{\omega_{\ell}\left(  z_{1}\right)  }{z_{1}}\right)  ^{2}%
\frac{\omega_{\ell}\left(  z_{1}\right)  }{\left(  1-\gamma_{\ell,m}\left(
z_{1},z_{2}\right)  \right)  ^{2}}\frac{\gamma_{\ell,m}^{(1,2)}\left(
z_{1},z_{2}\right)  \gamma_{\ell,m}^{(2,1)}\left(  z_{1},z_{2}\right)
}{1-\gamma_{m,m}\left(  z_{2},z_{2}\right)  }+\left(  \frac{\omega_{\ell
}\left(  z_{1}\right)  }{z_{1}}\right)  ^{2}\frac{\omega_{\ell}\left(
z_{1}\right)  }{1-\gamma_{\ell,m}\left(  z_{1},z_{2}\right)  }\frac
{\gamma_{\ell,m}^{(2,2)}\left(  z_{1},z_{2}\right)  }{1-\gamma_{m,m}\left(
z_{2},z_{2}\right)  }\\
+\left(  \frac{\omega_{\ell}\left(  z_{1}\right)  }{z_{1}}\right)  ^{2}%
\frac{\gamma_{\ell,m}^{(1,2)}\left(  z_{1},z_{2}\right)  }{\left(
1-\gamma_{\ell,m}\left(  z_{1},z_{2}\right)  \right)  ^{2}}\frac
{1-\gamma_{\ell,\ell}\left(  z_{1},z_{1}\right)  }{1-\gamma_{m,m}\left(
z_{2},z_{2}\right)  }+\mathcal{O}(N^{-1})
\end{multline*}
where, for $r,s\in\mathbb{N}$, we have defined
\[
\gamma_{\ell,m}^{(r,s)}\left(  z_{1},z_{2}\right)  =\left(  \frac{z_{1}%
}{\omega_{\ell}\left(  z_{1}\right)  }\right)  ^{r}\left(  \frac{z_{2}}%
{\omega_{m}\left(  z_{2}\right)  }\right)  ^{s}\frac{1}{N}\operatorname*{tr}%
\left[  \mathbf{R}_{m\ell}\mathbf{\bar{Q}}_{\ell}^{r}(z_{1})\mathbf{\newline
R}_{\ell m}\mathbf{\bar{Q}}_{m}^{s}(z_{2})\right]  .
\]
so that in particular $\gamma_{\ell,m}\left(  z_{1},z_{2}\right)
=\gamma_{\ell,m}^{(1,1)}\left(  z_{1},z_{2}\right)  ,$\ and where we have used
(\ref{eq:identity_w/z}) together with the identity
\[
\frac{1}{N}\operatorname*{tr}\left[  \mathbf{R_{\ell}\bar{Q}}_{\ell}^{2}%
(z_{2})\right]  =\frac{1}{z_{1}}-\frac{\omega_{\ell}\left(  z_{1}\right)
}{z_{1}^{2}}\left(  1-\gamma_{\ell,\ell}\left(  z_{1},z_{1}\right)  \right)
.
\]

Let us now go back to the expression in (\ref{eq:diff_resolvents}). Taking
traces and applying the same decorrelation technique together with Lemma
\ref{lemma:auxiliar_3resolvents}, we can write
\begin{multline*}
M\mathbb{E}\left[  \left(  \hat{m}_{\ell}(z_{1})-\bar{m}_{\ell}(z_{1})\right)
\Psi(u)\phi_{\ell}\right]  =\\
=\frac{z_{1}}{\omega_{\ell}\left(  z_{1}\right)  }N\mathbb{E}\left[  \xi
(z_{1})\alpha_{\ell}(z_{1})\Psi(u)\right]  +\left(  \frac{z_{1}}{\omega_{\ell
}\left(  z_{1}\right)  }\right)  ^{2}\operatorname*{tr}\left[  \mathbf{R}%
_{\ell}\mathbf{\bar{Q}}_{\ell}^{2}(z_{1})\right]  \mathbb{E}\left[
\alpha_{\ell}(z_{1})\Psi(u)\right] \\
+\operatorname*{i}u\frac{z_{1}}{\omega_{\ell}\left(  z_{1}\right)  }\sum
_{m=1}^{L}\frac{a_{m}}{2\pi\operatorname*{i}}\oint\nolimits_{\mathcal{C}_{m}%
}f_{m}(z_{2})\frac{1}{N}\mathbb{E}\operatorname*{tr}\left[  \mathbf{\bar{Q}%
}_{\ell}(z_{1})\mathbf{\hat{Q}}_{\ell}(z_{1})\mathbf{R}_{\ell,m}%
\mathbf{\hat{Q}}_{m}^{2}(z_{2})\mathbf{\hat{R}}_{m,\ell}\phi_{\ell}\phi
_{m}\right]  dz_{2}\mathbb{E}\Psi(u)+\mathcal{O}(N^{-1}).
\end{multline*}
where we have defined
\[
\xi(z_{1})=\frac{1}{N}\operatorname*{tr}\left[  \mathbf{\hat{Q}}_{\ell}%
(z_{1})\mathbf{\hat{R}}_{\ell}\mathbf{\bar{Q}}_{\ell}(z_{1})\right]
\phi_{\ell}-\frac{z_{1}}{\omega_{\ell}\left(  z_{1}\right)  }\frac{1}%
{N}\operatorname*{tr}\left[  \mathbf{R}_{\ell}\mathbf{\bar{Q}}_{\ell}%
^{2}(z_{1})\right]  .
\]
Now, the first term is of order $\mathcal{O}(N^{-1})$ because by
Cauchy-Schwarz
\[
N\left\vert \mathbb{E}\left[  \xi(z_{1})\alpha_{\ell}(z_{1})\Psi(u)\right]
\right\vert \leq N\sqrt{\mathbb{E}\left[  \left\vert \xi(z_{1})\right\vert
^{2}\right]  \mathbb{E}\left[  \left\vert \alpha_{\ell}(z_{1})\right\vert
^{2}\right]  }=N\mathcal{O}(N^{-2})=\mathcal{O}(N^{-1}).
\]
As for the second term, we see that it corresponds to the one derived in
(\ref{eq:Ealpha1PSi}). Hence, inserting the derived expression for
$\mathbb{E}\left[  \alpha_{\ell}(z_{1})\Psi(u)\right]  $ derived above and
using the approximations in Lemma \ref{lemma:auxiliar_3resolvents} we obtain
\[
M\mathbb{E}\left[  \left(  \hat{m}_{\ell}(z_{1})-\bar{m}_{\ell}(z_{1})\right)
\Psi(u)\right]  =\operatorname*{i}u\sum_{m=1}^{L}\frac{a_{m}}{2\pi
\operatorname*{i}}\oint\nolimits_{\mathcal{C}_{m}}f_{m}(z_{2})\sigma_{\ell
,m}^{2}\left(  z_{1},z_{2}\right)  dz_{2}\mathbb{E}\Psi(u)+\mathcal{O}%
(N^{-1})
\]
where
\[
\sigma_{\ell,m}^{2}\left(  z_{1},z_{2}\right)  =\frac{1}{1-\gamma_{\ell,\ell
}\left(  z_{1},z_{1}\right)  }\frac{1}{1-\gamma_{m,m}\left(  z_{2}%
,z_{2}\right)  }\left(  \frac{\gamma_{\ell,m}^{(1,2)}\left(  z_{1}%
,z_{2}\right)  \gamma_{\ell,m}^{(2,1)}\left(  z_{1},z_{2}\right)  }{\left(
1-\gamma_{\ell,m}\left(  z_{1},z_{2}\right)  \right)  ^{2}}+\frac{\gamma
_{\ell,m}^{(2,2)}\left(  z_{1},z_{2}\right)  }{1-\gamma_{\ell,m}\left(
z_{1},z_{2}\right)  }\right)  .
\]
We can conclude the proof by noting that we can alternatively express
$\gamma_{\ell,m}^{(r,s)}\left(  z_{1},z_{2}\right)  $ as
\[
\gamma_{\ell,m}^{(r,s)}\left(  z_{1},z_{2}\right)  =\Gamma_{\ell,m}%
^{(r,s)}\left(  \omega_{\ell}\left(  z_{1}\right)  ,\omega_{m}\left(
z_{2}\right)  \right)
\]
where%
\[
\Gamma_{\ell,m}^{(r,s)}\left(  \omega_{1},\omega_{2}\right)  =\frac{1}%
{N}\operatorname*{tr}\left[  \mathbf{R}_{m\ell}\left(  \mathbf{R}_{\ell
}-\omega_{\ell}\left(  z_{1}\right)  \mathbf{I}_{M}\right)  ^{-r}%
\mathbf{R}_{\ell m}\left(  \mathbf{R}_{m}-\omega_{m}\left(  z_{2}\right)
\mathbf{I}_{M}\right)  ^{-s}\right]
\]
and
\[
\left(  1-\gamma_{\ell,\ell}\left(  z_{1},z_{1}\right)  \right)  ^{-1}%
=\omega_{\ell}^{\prime}\left(  z_{1}\right)
\]
so that
\[
\sigma_{\ell,m}^{2}\left(  z_{1},z_{2}\right)  =\left.  \frac{\Gamma_{\ell
,m}^{(1,2)}\left(  \omega_{1},\omega_{2}\right)  \Gamma_{\ell,m}%
^{(2,1)}\left(  \omega_{1},\omega_{2}\right)  }{\left(  1-\Gamma_{\ell
,m}\left(  \omega_{1},\omega_{2}\right)  \right)  ^{2}}+\frac{\Gamma_{\ell
,m}^{(2,2)}\left(  \omega_{1},\omega_{2}\right)  }{1-\Gamma_{\ell,m}\left(
\omega_{1},\omega_{2}\right)  }\right\vert _{\left(  \omega_{1},\omega
_{2}\right)  =\left(  \omega_{\ell}\left(  z_{1}\right)  ,\omega_{m}\left(
z_{2}\right)  \right)  }\omega_{\ell}^{\prime}\left(  z_{1}\right)  \omega
_{m}^{\prime}\left(  z_{2}\right)  .
\]
or, alternatively,
\[
\sigma_{\ell,m}^{2}\left(  z_{1},z_{2}\right)  =-\left.  \frac{\partial^{2}%
}{\partial\omega_{1}\partial\omega_{2}}\log\left(  1-\Gamma_{\ell,m}\left(
\omega_{1},\omega_{2}\right)  \right)  \right\vert _{\left(  \omega_{1}%
,\omega_{2}\right)  =\left(  \omega_{\ell}\left(  z_{1}\right)  ,\omega
_{m}\left(  z_{2}\right)  \right)  }\omega_{\ell}^{\prime}\left(
z_{1}\right)  \omega_{m}^{\prime}\left(  z_{2}\right)  .
\]
Therefore, applying the change of variables $z_{1}\mapsto\omega_{\ell}\left(
z_{1}\right)  $, $z_{2}\mapsto\omega_{m}\left(  z_{2}\right)  $ we arrive at
the result of the theorem.

\subsection{Auxiliary Lemmas}

In this appendix, we provide some bounds on expectations and variances of
different random functions of complex variable. The notation $\mathcal{O}%
(N^{-\ell})$ should be understood as a deterministic term whose magnitude is
upper bounded by a quantity of the form $\varepsilon\left(  z_{1}%
,z_{2}\right)  N^{-\ell}$, where $\varepsilon\left(  z_{1},z_{2}\right)  $ is
a bivariate real valued positive function independent of $N$ such that
$\sup_{\left(  z_{1},z_{2}\right)  \in\mathcal{C}_{\ell}\times\mathcal{C}_{m}%
}\varepsilon\left(  z_{1},z_{2}\right)  <\infty$.

\begin{lemma}
\label{lemma:auxiliar_1resolvent}Let $\mathbf{A}$ denote an $M\times M$
deterministic matrix with bounded spectral norm. Then, we can write
\begin{align*}
\frac{1}{N}\mathbb{E}\operatorname*{tr}\left[  \mathbf{A\hat{Q}}_{\ell}%
(z_{1})\phi_{\ell}\mathbf{\newline}\right]   &  =\frac{1}{N}\operatorname*{tr}%
\left[  \mathbf{A\bar{Q}}_{\ell}(z_{1})\mathbf{\newline}\right]
+\mathcal{O}(N^{-1})\\
\frac{1}{N}\mathbb{E}\operatorname*{tr}\left[  \mathbf{A\hat{Q}}_{\ell}%
(z_{1})\mathbf{\newline\mathbf{R}}_{\ell}^{1/2}\frac{\mathbf{XX}^{H}}{N}%
\phi_{\ell}\right]   &  =\frac{z_{1}}{\omega_{\ell}\left(  z_{1}\right)
}\frac{1}{N}\operatorname*{tr}\left[  \mathbf{A\bar{Q}}_{\ell}(z_{1}%
)\mathbf{\newline\mathbf{R}}_{\ell}^{1/2}\right]  +\mathcal{O}(N^{-1})
\end{align*}
and also
\begin{align*}
\operatorname*{var}\left(  \frac{1}{N}\operatorname*{tr}\left[  \mathbf{A\hat
{Q}}_{\ell}(z_{1})\mathbf{\newline}\right]  \phi_{\ell}\right)   &
=\mathcal{O}(N^{-2})\\
\operatorname*{var}\left(  \frac{1}{N}\operatorname*{tr}\left[  \mathbf{A\hat
{Q}}_{\ell}(z_{1})\mathbf{\mathbf{R}}_{\ell}^{1/2}\frac{\mathbf{XX}^{H}}%
{N}\right]  \phi_{\ell}\right)   &  =\mathcal{O}(N^{-2}).
\end{align*}

\end{lemma}

\begin{IEEEproof}%
It follows from the application of the integration by parts formula in
(\ref{eq:ipp_reg1}) together with the Nash-Poincar\'{e} variance inequality in
(\ref{eq:Nash-Poincare}).%
\end{IEEEproof}

\begin{lemma}
\label{lemma:auxiliar_2resolvents}Let $\mathbf{A}$ denote an $M\times M$
deterministic matrix with bounded spectral norm. Then, we can write%
\begin{align}
\frac{1}{N}\mathbb{E}\operatorname*{tr}\left[  \mathbf{\mathbf{A}\hat{Q}%
}_{\ell}(z_{1})\mathbf{\newline\mathbf{R}}_{\ell,m}\mathbf{\hat{Q}}_{m}%
(z_{2})\phi_{\ell}\phi_{m}\right]   &  =\frac{1}{1-\gamma_{\ell,m}\left(
z_{1},z_{2}\right)  }\frac{1}{N}\operatorname*{tr}\left[  \mathbf{\mathbf{A}%
\bar{Q}}_{\ell}(z_{1})\mathbf{R}_{\ell,m}\mathbf{\bar{Q}}_{m}(z_{2})\right]
+\mathcal{O}(N^{-1})\label{eq:lemma2resolv_1}\\
\frac{1}{N}\mathbb{E}\operatorname*{tr}\left[  \mathbf{A\hat{Q}}_{\ell}%
(z_{1})\mathbf{\newline\mathbf{R}}_{\ell,m}\mathbf{\hat{Q}}_{m}(z_{2}%
)\mathbf{\mathbf{R}}_{m}^{1/2}\frac{\mathbf{XX}^{H}}{N}\phi_{\ell}\phi
_{m}\right]   &  =\frac{z_{2}}{\omega_{m}\left(  z_{2}\right)  }\frac
{1}{1-\gamma_{\ell,m}\left(  z_{1},z_{2}\right)  }\frac{1}{N}%
\operatorname*{tr}\left[  \mathbf{\mathbf{A}\bar{Q}}_{\ell}(z_{1}%
)\mathbf{R}_{\ell,m}\mathbf{\bar{Q}}_{m}(z_{2})\mathbf{R}_{m}^{1/2}\right]
\label{eq:lemma2resolv_2}\\
&  -\frac{\gamma_{\ell,m}\left(  z_{1},z_{2}\right)  }{1-\gamma_{\ell
,m}\left(  z_{1},z_{2}\right)  }\frac{1}{N}\operatorname*{tr}\left[
\mathbf{A\bar{Q}}_{\ell}(z_{1})\mathbf{R}_{\ell}^{1/2}\right]  +\mathcal{O}%
(N^{-1})\nonumber
\end{align}
and also
\begin{align*}
\operatorname*{var}\frac{1}{N}\operatorname*{tr}\left[  \mathbf{A\hat{Q}%
}_{\ell}(z_{1})\mathbf{\newline\mathbf{R}}_{\ell,m}\mathbf{\hat{Q}}_{m}%
(z_{2})\phi_{\ell}\phi_{m}\right]   &  =\mathcal{O}(N^{-2})\\
\operatorname*{var}\frac{1}{N}\operatorname*{tr}\left[  \mathbf{A\hat{Q}%
}_{\ell}(z_{1})\mathbf{\newline\mathbf{R}}_{\ell,m}\mathbf{\hat{Q}}_{m}%
(z_{2})\mathbf{\mathbf{\hat{R}}}_{m,\ell}\phi_{\ell}\phi_{m}\right]   &
=\mathcal{O}(N^{-2})
\end{align*}

\end{lemma}

\begin{IEEEproof}%
The proof of the variances follows directly from the Nash-Poincar\'{e}
variance inequality in (\ref{eq:Nash-Poincare}), so that we will only prove
the first two identities. Let us first consider on the first identity. Using
the resolvent's identity on $\mathbf{\hat{Q}}_{m}(z_{2})$, developing with
respect to $\mathbf{X}$ and applying the integration by parts formula in
(\ref{eq:ipp_reg1}), we obtain
\begin{align*}
&  \frac{1}{N}\mathbb{E}\operatorname*{tr}\left[  \mathbf{A\hat{Q}}_{\ell
}(z_{1})\mathbf{\newline\mathbf{R}}_{\ell,m}\mathbf{\hat{Q}}_{m}(z_{2}%
)\phi_{\ell}\phi_{m}\right] \\
&  =\mathbf{-}z_{2}^{-1}\frac{z_{1}}{\omega_{\ell}\left(  z_{1}\right)  }%
\frac{1}{N}\mathbb{E}\operatorname*{tr}\left[  \mathbf{A\bar{Q}}_{\ell}%
(z_{1})\mathbf{R}_{\ell,m}\right]  \frac{1}{N}\mathbb{E}\operatorname*{tr}%
\left[  \mathbf{\hat{Q}}_{\ell}(z_{1})\mathbf{\newline\mathbf{R}}_{\ell
,m}\mathbf{\hat{Q}}_{m}(z_{2})\mathbf{\mathbf{R}}_{m,\ell}\phi_{\ell}\phi
_{m}\right] \\
&  -z_{2}^{-1}\frac{1}{N}\mathbb{E}\operatorname*{tr}\left[  \mathbf{A\hat{Q}%
}_{\ell}(z_{1})\mathbf{\newline\mathbf{R}}_{\ell,m}\mathbf{\hat{Q}}_{m}%
(z_{2})\mathbf{\hat{R}}_{m}\phi_{\ell}\phi_{m}\right]  \frac{1}{N}%
\mathbb{E}\operatorname*{tr}\left[  \mathbf{\bar{Q}}_{m}(z_{2}%
)\mathbf{\mathbf{R}}_{m}\right] \\
&  +z_{2}^{-1}\frac{1}{N}\mathbb{E}\operatorname*{tr}\left[  \mathbf{A\hat{Q}%
}_{\ell}(z_{1})\mathbf{\newline\mathbf{R}}_{\ell,m}\mathbf{\hat{Q}}_{m}%
(z_{2})\mathbf{\mathbf{R}}_{m}\phi_{\ell}\phi_{m}\right]  -z_{2}^{-1}\frac
{1}{N}\mathbb{E}\operatorname*{tr}\left[  \mathbf{A\hat{Q}}_{\ell}%
(z_{1})\mathbf{\newline\mathbf{R}}_{\ell,m}\phi_{\ell}\right]  +\mathcal{O}%
(N^{-1})
\end{align*}
where we have additionally decorrelated the expectations of two terms and used
the approximations in Lemma \ref{lemma:auxiliar_1resolvent} above. We can also
apply the integration by parts formula on the quantity%
\begin{align}
&  \frac{1}{N}\mathbb{E}\operatorname*{tr}\left[  \mathbf{A\hat{Q}}_{\ell
}(z_{1})\mathbf{\newline\mathbf{R}}_{\ell,m}\mathbf{\hat{Q}}_{m}%
(z_{2})\mathbf{R}_{m}^{1/2}\frac{\mathbf{XX}^{H}}{N}\phi_{\ell}\phi_{m}\right]
\label{eq:tempAQRQRXX}\\
&  =-\frac{z_{1}z_{2}}{\omega_{\ell}\left(  z_{1}\right)  \omega_{m}\left(
z_{2}\right)  }\frac{1}{N}\operatorname*{tr}\left[  \mathbf{A\bar{Q}}_{\ell
}(z_{1})\mathbf{R}_{\ell}^{1/2}\right]  \frac{1}{N}\mathbb{E}%
\operatorname*{tr}\left[  \mathbf{\hat{Q}}_{\ell}(z_{1})\mathbf{\mathbf{R}%
}_{\ell,m}\mathbf{\hat{Q}}_{m}(z_{2})\mathbf{\mathbf{R}}_{m,\ell}\phi_{\ell
}\phi_{m}\right] \nonumber\\
&  +\frac{z_{2}}{\omega_{m}\left(  z_{2}\right)  }\frac{1}{N}\mathbb{E}%
\operatorname*{tr}\left[  \mathbf{A\hat{Q}}_{\ell}(z_{1})\mathbf{\newline%
\mathbf{R}}_{\ell,m}\mathbf{\hat{Q}}_{m}(z_{2})\mathbf{R}_{m}^{1/2}\phi_{\ell
}\phi_{m}\right]  +\mathcal{O}(N^{-1})\nonumber
\end{align}
where we have also decorrelated terms of expectations using the fact that all
variances decay as $\mathcal{O}(N^{-2})$ and subsequently applied the
approximations in Lemma \ref{lemma:auxiliar_1resolvent}. Inserting this last
equation with $\mathbf{A}$ replaced by $\mathbf{R}_{m}^{H/2}\mathbf{A}$ into
the first one we obtain
\begin{align}
&  \frac{1}{N}\mathbb{E}\operatorname*{tr}\left[  \mathbf{A\hat{Q}}_{\ell
}(z_{1})\mathbf{\newline\mathbf{R}}_{\ell,m}\mathbf{\hat{Q}}_{m}(z_{2}%
)\phi_{\ell}\phi_{m}\right] \label{eq:tempAQRQ}\\
&  =-\frac{z_{1}}{\omega_{\ell}\left(  z_{1}\right)  \omega_{m}\left(
z_{2}\right)  }\frac{1}{N}\operatorname*{tr}\left[  \mathbf{A\bar{Q}}_{\ell
}(z_{1})\mathbf{R}_{\ell,m}\right]  \frac{1}{N}\mathbb{E}\operatorname*{tr}%
\left[  \mathbf{\hat{Q}}_{\ell}(z_{1})\mathbf{\newline\mathbf{R}}_{\ell
,m}\mathbf{\hat{Q}}_{m}(z_{2})\mathbf{\mathbf{R}}_{m,\ell}\phi_{\ell}\phi
_{m}\right] \nonumber\\
&  +\frac{1}{\omega_{m}\left(  z_{2}\right)  }\frac{1}{N}\mathbb{E}%
\operatorname*{tr}\left[  \mathbf{A\hat{Q}}_{\ell}(z_{1})\mathbf{\newline%
\mathbf{R}}_{\ell,m}\mathbf{\hat{Q}}_{m}(z_{2})\mathbf{\mathbf{R}}_{m}%
\phi_{\ell}\phi_{m}\right]  -z_{2}^{-1}\frac{1}{N}\mathbb{E}\operatorname*{tr}%
\left[  \mathbf{A\hat{Q}}_{\ell}(z_{1})\mathbf{\newline\mathbf{R}}_{\ell
,m}\phi_{\ell}\right]  +\mathcal{O}(N^{-1})\nonumber
\end{align}
Particularizing this expression for $\mathbf{A}$ replaced with $\mathbf{\bar
{Q}}_{m}(z_{2})\mathbf{\mathbf{R}}_{m,\ell}$ and using Lemma
\ref{lemma:auxiliar_1resolvent} we obtain
\[
\frac{1}{N}\mathbb{E}\operatorname*{tr}\left[  \mathbf{\mathbf{R}}_{m,\ell
}\mathbf{\hat{Q}}_{\ell}(z_{1})\mathbf{\newline\mathbf{R}}_{\ell
,m}\mathbf{\hat{Q}}_{m}(z_{2})\phi_{\ell}\phi_{m}\right]  =\frac{1}%
{1-\gamma_{\ell,m}\left(  z_{1},z_{2}\right)  }\frac{1}{N}\mathbb{E}%
\operatorname*{tr}\left[  \mathbf{\bar{Q}}_{\ell}(z_{1})\mathbf{\newline%
\mathbf{R}}_{\ell,m}\mathbf{\bar{Q}}_{m}(z_{2})\mathbf{\mathbf{\mathbf{R}}%
}_{m,\ell}\right]  +\mathcal{O}(N^{-1}).
\]
Inserting this into (\ref{eq:tempAQRQ}) and replacing $\mathbf{A}$ with
$\mathbf{\bar{Q}}_{m}(z_{2})\mathbf{\mathbf{A}}$ we get to
(\ref{eq:lemma2resolv_1}), and inserting it into (\ref{eq:tempAQRQRXX}) we
obtain (\ref{eq:lemma2resolv_2}).%
\end{IEEEproof}

\begin{lemma}
\label{lemma:auxiliar_3resolvents}Let $\mathbf{A}$ denote an $M\times M$
deterministic matrix with bounded spectral norm. Then, we can write%
\begin{multline}
\frac{1}{N}\mathbb{E}\operatorname*{tr}\left[  \mathbf{A\hat{Q}}_{\ell}%
(z_{1})\mathbf{\newline\mathbf{R}}_{\ell,m}\mathbf{\hat{Q}}_{m}^{2}(z_{2}%
)\phi_{\ell}\phi_{m}\right]  \label{eq:lemma3resolv}\\
=\frac{1}{\left(  1-\gamma_{\ell,m}\left(  z_{1},z_{2}\right)  \right)  ^{2}%
}\frac{\gamma_{\ell,m}^{(1,2)}\left(  z_{1},z_{2}\right)  }{1-\gamma
_{m,m}\left(  z_{2},z_{2}\right)  }\frac{1}{N}\operatorname*{tr}\left[
\mathbf{A\bar{Q}}_{\ell}(z_{1})\mathbf{\newline\mathbf{R}}_{\ell
,m}\mathbf{\bar{Q}}_{m}(z_{2})\right]  \\
-z_{2}^{-1}\left(  1-\frac{z_{2}}{\omega_{m}\left(  z_{2}\right)  }\frac
{1}{1-\gamma_{m,m}\left(  z_{2},z_{2}\right)  }\right)  \frac{1}%
{1-\gamma_{\ell,m}\left(  z_{1},z_{2}\right)  }\frac{1}{N}\operatorname*{tr}%
\left[  \mathbf{A\bar{Q}}_{\ell}(z_{1})\mathbf{\newline\mathbf{R}}_{\ell
,m}\mathbf{\bar{Q}}_{m}(z_{2})\right]  \\
+\frac{1}{1-\gamma_{\ell,m}\left(  z_{1},z_{2}\right)  }\frac{z_{2}}%
{\omega_{m}\left(  z_{2}\right)  }\frac{1}{1-\gamma_{m,m}\left(  z_{2}%
,z_{2}\right)  }\frac{1}{N}\operatorname*{tr}\left[  \mathbf{A\bar{Q}}_{\ell
}(z_{1})\mathbf{R}_{\ell,m}\mathbf{\bar{Q}}_{m}^{2}(z_{2})\right]
+\mathcal{O}(N^{-1})
\end{multline}
and%
\begin{multline}
\frac{1}{N}\mathbb{E}\operatorname*{tr}\left[  \mathbf{A\hat{Q}}_{\ell}%
(z_{1})\mathbf{\newline\mathbf{R}}_{\ell,m}\mathbf{\hat{Q}}_{m}^{2}%
(z_{2})\mathbf{R}_{m}^{1/2}\frac{\mathbf{XX}^{H}}{N}\mathbf{R}_{\ell}%
^{H/2}\phi_{\ell}\phi_{m}\right]  \label{eq:lemma3resolv2}\\
=\frac{z_{2}}{\omega_{m}\left(  z_{2}\right)  }\frac{1}{\left(  1-\gamma
_{\ell,m}\left(  z_{1},z_{2}\right)  \right)  ^{2}}\frac{\gamma_{\ell
,m}^{(1,2)}\left(  z_{1},z_{2}\right)  }{1-\gamma_{m,m}\left(  z_{2}%
,z_{2}\right)  }\frac{1}{N}\operatorname*{tr}\left[  \mathbf{A\bar{Q}}_{\ell
}(z_{1})\mathbf{\newline\mathbf{R}}_{\ell,m}\mathbf{\bar{Q}}_{m}%
(z_{2})\mathbf{R}_{m,\ell}\right]  \\
+\left(  \frac{z_{2}}{\omega_{m}\left(  z_{2}\right)  }\right)  ^{2}\frac
{1}{1-\gamma_{\ell,m}\left(  z_{1},z_{2}\right)  }\frac{1}{1-\gamma
_{m,m}\left(  z_{2},z_{2}\right)  }\frac{1}{N}\operatorname*{tr}\left[
\mathbf{A\bar{Q}}_{\ell}(z_{1})\mathbf{R}_{\ell,m}\mathbf{\bar{Q}}_{m}%
^{2}(z_{2})\mathbf{R}_{m,\ell}\right]  \\
-\frac{1}{\left(  1-\gamma_{\ell,m}\left(  z_{1},z_{2}\right)  \right)  ^{2}%
}\frac{\gamma_{\ell,m}^{(1,2)}\left(  z_{1},z_{2}\right)  }{1-\gamma
_{m,m}\left(  z_{2},z_{2}\right)  }\frac{1}{N}\operatorname*{tr}\left[
\mathbf{A\bar{Q}}_{\ell}(z_{1})\mathbf{\newline\mathbf{R}}_{\ell}\right]
+\mathcal{O}(N^{-1}).
\end{multline}
On the other hand, we have
\begin{align*}
\operatorname*{var}\frac{1}{N}\operatorname*{tr}\left[  \mathbf{\mathbf{A}%
\hat{Q}}_{\ell}(z_{1})\mathbf{\newline\mathbf{R}}_{\ell,m}\mathbf{\hat{Q}}%
_{m}^{2}(z_{2})\phi_{\ell}\phi_{m}\right]   &  =\mathcal{O}(N^{-2})\\
\operatorname*{var}\frac{1}{N}\mathbb{E}\operatorname*{tr}\left[
\mathbf{\hat{Q}}_{\ell}(z_{1})\mathbf{\newline\mathbf{R}}_{\ell,m}%
\mathbf{\hat{Q}}_{m}^{2}(z_{2})\mathbf{R}_{m}^{1/2}\frac{\mathbf{XX}^{H}}%
{N}\mathbf{R}_{\ell}^{H/2}\phi_{\ell}\phi_{m}\right]   &  =\mathcal{O}%
(N^{-2}).
\end{align*}

\end{lemma}

\begin{IEEEproof}%
The proof that the variance decays as $\mathcal{O}(N^{-2})$ follows the
conventional approach from the Nash-Poincar\'{e} inequality, and is therefore
omitted. To proof the first two identities, we proceed as in the proof of
Lemma \ref{lemma:auxiliar_2resolvents}. Using first the resolvent identity on
$\mathbf{\hat{Q}}_{m}(z_{2})$ together with the integration by parts formula
and Lemmas \ref{lemma:auxiliar_1resolvent} and
\ref{lemma:auxiliar_2resolvents}, we can write
\begin{align*}
&  \frac{1}{N}\mathbb{E}\operatorname*{tr}\left[  \mathbf{A\hat{Q}}_{\ell
}(z_{1})\mathbf{\newline\mathbf{R}}_{\ell,m}\mathbf{\hat{Q}}_{m}^{2}%
(z_{2})\phi_{\ell}\phi_{m}\right] \\
&  =z_{2}^{-1}\frac{1}{N}\mathbb{E}\operatorname*{tr}\left[  \mathbf{A\hat{Q}%
}_{\ell}(z_{1})\mathbf{\newline\mathbf{R}}_{\ell,m}\mathbf{\hat{Q}}_{m}%
^{2}(z_{2})\mathbf{R}_{m}\phi_{\ell}\phi_{m}\right] \\
&  -\frac{z_{1}z_{2}^{-1}}{\omega_{\ell}\left(  z_{1}\right)  }\frac{1}%
{N}\operatorname*{tr}\left[  \mathbf{A\bar{Q}}_{\ell}(z_{1})\mathbf{\newline%
\mathbf{R}}_{\ell,m}\right]  \frac{1}{N}\mathbb{E}\operatorname*{tr}\left[
\mathbf{\hat{Q}}_{\ell}(z_{1})\mathbf{\mathbf{R}}_{\ell,m}\mathbf{\hat{Q}}%
_{m}^{2}(z_{2})\mathbf{R}_{m,\ell}\phi_{\ell}\phi_{m}\right] \\
&  +z_{2}^{-1}\left(  1-\frac{\omega_{m}\left(  z_{2}\right)  }{z_{2}}\right)
\frac{1}{N}\mathbb{E}\operatorname*{tr}\left[  \mathbf{A\hat{Q}}_{\ell}%
(z_{1})\mathbf{\newline\mathbf{R}}_{\ell,m}\mathbf{\hat{Q}}_{m}^{2}%
(z_{2})\mathbf{R}_{m}^{1/2}\frac{\mathbf{XX}^{H}}{N}\mathbf{R}_{m}^{H/2}%
\phi_{\ell}\phi_{m}\right] \\
&  +z_{2}^{-2}\left(  1-\frac{z_{2}}{\omega_{m}\left(  z_{2}\right)  }\frac
{1}{1-\gamma_{m,m}\left(  z_{2},z_{2}\right)  }\right)  \frac{\omega
_{m}\left(  z_{2}\right)  }{z_{2}}\frac{1}{N}\operatorname*{tr}\left[
\mathbf{\mathbf{A}\bar{Q}}_{\ell}(z_{1})\mathbf{R}_{\ell,m}\right] \\
&  +z_{2}^{-1}\frac{1}{1-\gamma_{\ell,m}\left(  z_{1},z_{2}\right)  }\left(
\frac{\omega_{m}\left(  z_{2}\right)  }{z_{2}}-\frac{1}{1-\gamma_{m,m}\left(
z_{2},z_{2}\right)  }-1\right)  \frac{1}{N}\operatorname*{tr}\left[
\mathbf{\mathbf{A}\bar{Q}}_{\ell}(z_{1})\mathbf{R}_{\ell,m}\mathbf{\bar{Q}%
}_{m}(z_{2})\right] \\
&  +\mathcal{O}(N^{-1})
\end{align*}
where we have decorrelated the double terms using the fact that all variances
decay as $\mathcal{O}(N^{-2})$ and we have used the identity%
\[
\frac{1}{N}\operatorname*{tr}\left[  \mathbf{R}_{m}\mathbf{\bar{Q}}_{m}%
^{2}(z_{2})\right]  =\frac{1}{z_{2}}-\frac{\omega_{m}\left(  z_{2}\right)
}{z_{2}^{2}}\left(  1-\gamma_{m,m}\left(  z_{2},z_{2}\right)  \right)  .
\]
In a similar way, we can develop the term%
\begin{align}
&  \frac{1}{N}\mathbb{E}\operatorname*{tr}\left[  \mathbf{A\hat{Q}}_{\ell
}(z_{1})\mathbf{\newline\mathbf{R}}_{\ell,m}\mathbf{\hat{Q}}_{m}^{2}%
(z_{2})\mathbf{R}_{m}^{1/2}\frac{\mathbf{XX}^{H}}{N}\phi_{\ell}\phi_{m}\right]
\label{eq:tempAQRQ2RXX}\\
&  =\frac{z_{2}}{\omega_{m}\left(  z_{2}\right)  }\frac{1}{N}\mathbb{E}%
\operatorname*{tr}\left[  \mathbf{A\hat{Q}}_{\ell}(z_{1})\mathbf{\newline%
\mathbf{R}}_{\ell,m}\mathbf{\hat{Q}}_{m}^{2}(z_{2})\mathbf{R}_{m}^{1/2}%
\phi_{\ell}\phi_{m}\right] \nonumber\\
&  -\frac{z_{1}z_{2}}{\omega_{\ell}\left(  z_{1}\right)  \omega_{m}\left(
z_{2}\right)  }\frac{1}{N}\operatorname*{tr}\left[  \mathbf{A\bar{Q}}_{\ell
}(z_{1})\mathbf{\newline\mathbf{R}}_{\ell}^{1/2}\right]  \frac{1}{N}%
\mathbb{E}\operatorname*{tr}\left[  \mathbf{\hat{Q}}_{\ell}(z_{1}%
)\mathbf{\newline\mathbf{R}}_{\ell,m}\mathbf{\hat{Q}}_{m}^{2}(z_{2}%
)\mathbf{R}_{m,\ell}\phi_{\ell}\phi_{m}\right] \nonumber\\
&  +\frac{1}{1-\gamma_{\ell,m}\left(  z_{1},z_{2}\right)  }\frac{1}{\omega
_{m}\left(  z_{2}\right)  }\left(  1-\frac{z_{2}}{\omega_{m}\left(
z_{2}\right)  }\frac{1}{1-\gamma_{m,m}\left(  z_{2},z_{2}\right)  }\right)
\frac{1}{N}\operatorname*{tr}\left[  \mathbf{\mathbf{A}\bar{Q}}_{\ell}%
(z_{1})\mathbf{R}_{\ell,m}\mathbf{\bar{Q}}_{m}(z_{2})\mathbf{R}_{m}%
^{1/2}\right] \nonumber\\
&  -\frac{\gamma_{\ell,m}\left(  z_{1},z_{2}\right)  }{1-\gamma_{\ell
,m}\left(  z_{1},z_{2}\right)  }\frac{1}{z_{2}}\left(  1-\frac{z_{2}}%
{\omega_{m}\left(  z_{2}\right)  }\frac{1}{1-\gamma_{m,m}\left(  z_{2}%
,z_{2}\right)  }\right)  \frac{1}{N}\operatorname*{tr}\left[  \mathbf{A\bar
{Q}}_{\ell}(z_{1})\mathbf{R}_{\ell}^{1/2}\right]  +\mathcal{O}(N^{-1}%
)\nonumber
\end{align}
so that, inserting this back into the first equation and replacing
$\mathbf{A}$ with $\mathbf{R}_{m}^{H/2}\mathbf{\bar{Q}}_{m}(z_{2})\mathbf{A}$,
we obtain%
\begin{align*}
&  \frac{1}{N}\mathbb{E}\operatorname*{tr}\left[  \mathbf{A\hat{Q}}_{\ell
}(z_{1})\mathbf{\newline\mathbf{R}}_{\ell,m}\mathbf{\hat{Q}}_{m}^{2}%
(z_{2})\phi_{\ell}\phi_{m}\right] \\
&  =\frac{z_{1}z_{2}}{\omega_{\ell}\left(  z_{1}\right)  \omega_{m}\left(
z_{2}\right)  }\frac{1}{N}\mathbb{E}\operatorname*{tr}\left[  \mathbf{\hat{Q}%
}_{\ell}(z_{1})\mathbf{\newline\mathbf{R}}_{\ell,m}\mathbf{\hat{Q}}_{m}%
^{2}(z_{2})\mathbf{R}_{m.\ell}\phi_{\ell}\phi_{m}\right]  \frac{1}%
{N}\operatorname*{tr}\left[  \mathbf{A\bar{Q}}_{\ell}(z_{1})\mathbf{\newline%
\mathbf{R}}_{\ell,m}\mathbf{\bar{Q}}_{m}(z_{2})\right] \\
&  -z_{2}^{-1}\left(  1-\frac{z_{2}}{\omega_{m}\left(  z_{2}\right)  }\frac
{1}{1-\gamma_{m,m}\left(  z_{2},z_{2}\right)  }\right)  \frac{1}%
{N}\operatorname*{tr}\left[  \mathbf{A\bar{Q}}_{\ell}(z_{1})\mathbf{R}%
_{\ell,m}\mathbf{\bar{Q}}_{m}(z_{2})\right] \\
&  +\frac{1}{1-\gamma_{\ell,m}\left(  z_{1},z_{2}\right)  }\frac{z_{2}}%
{\omega_{m}\left(  z_{2}\right)  }\frac{1}{1-\gamma_{m,m}\left(  z_{2}%
,z_{2}\right)  }\frac{1}{N}\operatorname*{tr}\left[  \mathbf{A\bar{Q}}_{\ell
}(z_{1})\mathbf{R}_{\ell,m}\mathbf{\bar{Q}}_{m}^{2}(z_{2})\right]
+\mathcal{O}(N^{-1})
\end{align*}
Particularizing this expression for $\mathbf{A=R}_{m.\ell}$ we see that%
\begin{align*}
&  \frac{1}{N}\mathbb{E}\operatorname*{tr}\left[  \mathbf{R}_{m.\ell
}\mathbf{\hat{Q}}_{\ell}(z_{1})\mathbf{\newline\mathbf{R}}_{\ell
,m}\mathbf{\hat{Q}}_{m}^{2}(z_{2})\phi_{\ell}\phi_{m}\right] \\
&  =-\left(  1-\frac{z_{2}}{\omega_{m}\left(  z_{2}\right)  }\frac{1}%
{1-\gamma_{m,m}\left(  z_{2},z_{2}\right)  }\right)  \frac{\omega_{\ell
}\left(  z_{1}\right)  \omega_{m}\left(  z_{2}\right)  }{z_{1}z_{2}^{2}}%
\frac{\gamma_{\ell,m}\left(  z_{1},z_{2}\right)  }{1-\gamma_{\ell,m}\left(
z_{1},z_{2}\right)  }\\
&  +\frac{1}{\left(  1-\gamma_{\ell,m}\left(  z_{1},z_{2}\right)  \right)
^{2}}\frac{z_{2}}{\omega_{m}\left(  z_{2}\right)  }\frac{1}{1-\gamma
_{m,m}\left(  z_{2},z_{2}\right)  }\frac{1}{N}\operatorname*{tr}\left[
\mathbf{R}_{m.\ell}\mathbf{\bar{Q}}_{\ell}(z_{1})\mathbf{R}_{\ell
,m}\mathbf{\bar{Q}}_{m}^{2}(z_{2})\right]  +\mathcal{O}(N^{-1})
\end{align*}
so that inserting this back into the above expression, we obtain
(\ref{eq:lemma3resolv}). Finally, inserting the above expression into
(\ref{eq:tempAQRQ2RXX}) with $\mathbf{A}$ replaced by $\mathbf{R}_{\ell}%
^{H/2}\mathbf{A}$ and using (\ref{eq:lemma3resolv}) with $\mathbf{A}$ replaced
by $\mathbf{R}_{m,\ell}\mathbf{A}$ we obtain (\ref{eq:lemma3resolv2}).%
\end{IEEEproof}

\section{\label{sec:proofLemma2}Proof of Lemma \ref{lemma:polyident}}

The roots $\widetilde{\phi}_{k}(x),k=1,\ldots,\bar{M}$ are the solutions to
the polynomial equation $1=x\widetilde{\Phi}\left(  \phi\right)  .$ Therefore,
we have the identity
\begin{equation}
\prod\limits_{k=1}^{\bar{M}}\left(  \widetilde{\phi}_{k}(x)-\phi\right)
=\prod\limits_{k=1}^{\bar{M}}\left(  \gamma_{k}-\phi\right)  -x\frac{1}{N}%
\sum_{m=1}^{\bar{M}}\widetilde{K}_{m}\gamma_{m}\prod\limits_{\substack{k=1
\\k\neq m}}^{\bar{M}}\left(  \gamma_{k}-\phi\right)
\label{eq:polyindentphitilde}%
\end{equation}
and following the same procedure for the roots $\phi_{k},k=0,\ldots,\bar{M}$,
such that $\phi_{k}\neq0$ we obtain%
\begin{equation}
\prod\limits_{\substack{k=0 \\\phi_{k}\neq0}}^{\bar{M}}\left(  \phi_{k}%
-\phi\right)  =\prod\limits_{k=1}^{\bar{M}}\left(  \gamma_{k}-\phi\right)
-\frac{1}{N}\sum_{m=1}^{\bar{M}}K_{m}\gamma_{m}\prod\limits_{\substack{k=1
\\k\neq m}}^{\bar{M}}\left(  \gamma_{k}-\phi\right)  .
\label{eq:polyindentphi}%
\end{equation}
Now, forcing $\phi=\gamma_{\ell}$ for some $\ell$ on both
(\ref{eq:polyindentphitilde}) and (\ref{eq:polyindentphi}) and dividing the
result, we obtain
\begin{equation}
\frac{\prod\nolimits_{k=1}^{\bar{M}}\left(  \widetilde{\phi}_{k}%
(x)-\gamma_{\ell}\right)  }{\prod\nolimits_{\substack{k=0 \\\phi_{k}\neq
0}}^{\bar{M}}\left(  \phi_{k}-\gamma_{\ell}\right)  }=x\frac{\widetilde
{K}_{\ell}}{K_{\ell}}. \label{eq:KappaKappatilde}%
\end{equation}
Likewise, forcing $\phi=\widetilde{\phi}_{r}(x)$ for some $r$ in
(\ref{eq:polyindentphi}) and $\phi=\phi_{\ell}$ for some $\ell$ such that
$\phi_{\ell}\neq0$ in (\ref{eq:polyindentphitilde}) we readily see that
$\xi_{1}(x)=0$. \ On the other hand, forcing $\phi=\gamma_{\ell}$ for some
$\ell$ in (\ref{eq:polyindentphitilde}) directly shows that the argument of
the logarithm in the definition of $\xi_{2}(x)$ is equal to the identity,
leading to $\xi_{2}(x)=0$. In order to show that $\xi_{2}(x)=0$, we take
derivatives on both sides of (\ref{eq:polyindentphitilde}), namely%
\[
\sum_{m=1}^{\bar{M}}\prod\limits_{\substack{k=1 \\k\neq m}}^{\bar{M}}\left(
\widetilde{\phi}_{k}(x)-\phi\right)  =\sum_{m=1}^{\bar{M}}\prod
\limits_{\substack{k=1 \\k\neq m}}^{\bar{M}}\left(  \gamma_{k}-\phi\right)
-x\frac{1}{N}\sum_{m=1}^{\bar{M}}\widetilde{K}_{m}\gamma_{m}\sum
_{\substack{r=1 \\r\neq m}}^{\bar{M}}\prod\limits_{\substack{k=1 \\k\neq
m,r}}^{\bar{M}}\left(  \gamma_{k}-\phi\right)  .
\]
Forcing $\phi=\gamma_{\ell}$ for some $\ell$ in the above equation and using
the identity obtained by forcing $\phi=\gamma_{\ell}$ in
(\ref{eq:polyindentphitilde}) we obtain
\[
\frac{1}{N}\sum_{\substack{m=1 \\m\neq\ell}}^{\bar{M}}\frac{\widetilde{K}%
_{m}\gamma_{m}}{\gamma_{m}-\gamma_{\ell}}+\frac{1}{N}\sum_{\substack{r=1
\\r\neq\ell}}^{\bar{M}}\frac{\widetilde{K}_{\ell}\gamma_{\ell}}{\gamma
_{r}-\gamma_{\ell}}+\frac{1}{N}\sum_{m=1}^{\bar{M}}\frac{\widetilde{K}_{\ell
}\gamma_{\ell}}{\gamma_{\ell}-\phi_{m}(x)}=\frac{1}{x}.
\]
Multiplying this equation by $K_{\ell}/\widetilde{K}_{\ell}$ and summing over
$\ell$ we obtain $\xi_{3}(x)=0.$ Let us finally show (\ref{eq:identyphis}).
Assume first that $\phi_{0}\neq0$. By taking the derivatives of both sides
of\ (\ref{eq:polyindentphi}) and forcing $\phi=$ $\phi_{0}$ we directly obtain
the first identity. Regarding the identity for case $\phi_{0}=0$, it can be
readily obtained by forcing $\phi=0$ in (\ref{eq:polyindentphi}).

\section{\label{sec:proofLemma3}Proof of Lemma \ref{lemma:locnroots}}

The proof will be based on Rouch\'{e}'s theorem. Denote $f(\zeta
)=1-x\Psi_{\ell,m}\left(  \omega_{1},\zeta\right)  $ and observe that this
function is meromorphic on the complex plane. Using the Cauchy-Schwarz
inequality one can write%
\begin{align*}
\left\vert 1-f(\zeta)\right\vert ^{2}  &  =\left\vert x\sum_{r=0}^{\bar
{M}_{\ell}}\sum_{k=0}^{\bar{M}_{m}}\kappa_{rk}^{\ell m}\frac{\gamma_{r}%
^{(\ell)}\gamma_{k}^{(m)}}{\left(  \gamma_{r}^{(\ell)}-\omega_{1}\right)
\left(  \gamma_{k}^{(m)}-\zeta\right)  }\right\vert ^{2}\\
&  \leq\frac{1}{N}\sum_{r=0}^{\bar{M}_{\ell}}K_{r}^{(\ell)}\left\vert
\frac{\gamma_{r}^{(\ell)}}{\gamma_{r}^{(\ell)}-\omega_{1}}\right\vert
^{2}\frac{1}{N}\sum_{k=0}^{\bar{M}_{m}}K_{k}^{(m)}\left\vert \frac{\gamma
_{k}^{(m)}}{\gamma_{k}^{(m)}-\zeta}\right\vert ^{2}<\frac{1}{N}\sum
_{k=0}^{\bar{M}_{m}}K_{k}^{(m)}\left\vert \frac{\gamma_{k}^{(m)}}{\gamma
_{k}^{(m)}-\zeta}\right\vert ^{2}%
\end{align*}
where we have used the fact that $\omega_{1}\in\Omega_{\ell}$ and $x\leq1$.
Now, for all $\zeta\in\partial\Omega_{m}$ we clearly have $\left\vert
1-f(\zeta)\right\vert <1$. On the other hand, $f(\zeta)$ has no poles or zeros
along $\partial\Omega_{m}$. Indeed, the poles of $f(\zeta)$ are the
eigenvalues of $\mathbf{R}_{m}$, which are located inside $\partial\Omega_{m}
$. Furthermore, $f(\zeta)$ cannot have zeros on $\partial\Omega_{m}$ because
this would imply%
\[
1=\left\vert x\sum_{r=0}^{\bar{M}_{\ell}}\sum_{k=0}^{\bar{M}_{m}}\kappa
_{rk}^{\ell m}\frac{\gamma_{r}^{(\ell)}\gamma_{k}^{(m)}}{\left(  \gamma
_{r}^{(\ell)}-\omega_{1}\right)  \left(  \gamma_{k}^{(m)}-\zeta\right)
}\right\vert ^{2}<\frac{1}{N}\sum_{k=1}^{\bar{M}_{m}}K_{k}^{(m)}\left\vert
\frac{\gamma_{k}^{(m)}}{\gamma_{k}^{(m)}-\zeta}\right\vert ^{2}=1
\]
leading to contradiction. We can conclude that $f(\zeta)$ has the same number
of poles and zeros inside $\Omega_{m}$, i.e. it will have exactly $\bar{M}%
_{m}$ zeros that will be in direct correspondence with the positive
eigenvalues of $\mathbf{R}_{m}$ since $m\geq1$ by assumption.

\section{\label{sec:proofLemma4}Proof of Lemma\ref{lemma:ups1ups2}}

Using the fact that $\zeta_{j}^{(\ell,m)}\left(  \omega_{1},x\right)  $ are
the solutions to the equation in (\ref{eq:rootsPsilm}), we have the following
polynomial identity%
\begin{equation}
\prod\limits_{r=1}^{\bar{M}_{m}}\left(  \zeta_{r}^{(\ell,m)}\left(  \omega
_{1},x\right)  -\zeta\right)  =\prod\limits_{r=1}^{\bar{M}_{m}}\left(
\gamma_{r}^{(m)}-\zeta\right)  -x\sum_{k=1}^{\bar{M}_{m}}\sum_{r=1}^{\bar
{M}_{\ell}}\kappa_{rk}^{\ell m}\frac{\gamma_{r}^{(\ell)}}{\gamma_{r}^{(\ell
)}-\omega_{1}}\gamma_{k}^{(m)}\prod\limits_{\substack{i=1 \\i\neq k}}^{\bar
{M}_{m}}\left(  \gamma_{i}^{(m)}-\zeta\right)  . \label{eq:polyindentzetalm}%
\end{equation}
Now using the fact that $\widetilde{\phi}_{k}^{(m)}$ are the solutions to the
equation $\widetilde{\Phi}^{(m)}(\widetilde{\phi})=1$, we have the polynomial
identities%
\begin{equation}
\prod\limits_{k=1}^{\bar{M}_{m}}\left(  \widetilde{\phi}_{k}^{(m)}%
-\widetilde{\phi}\right)  =\prod\limits_{k=1}^{\bar{M}_{m}}\left(  \gamma
_{k}^{(m)}-\widetilde{\phi}\right)  -\frac{1}{N}\sum_{r=1}^{\bar{M}_{m}%
}\widetilde{K}_{r}^{(m)}\gamma_{r}^{(m)}\prod\limits_{\substack{k=1 \\k\neq
r}}^{\bar{M}_{m}}\left(  \gamma_{k}^{(m)}-\widetilde{\phi}\right)  .
\label{eq:polyindetpsitildeall}%
\end{equation}
Forcing $\zeta=\gamma_{j}^{(m)}$ in (\ref{eq:polyindentzetalm}),
$\widetilde{\phi}=\gamma_{j}^{(m)}$ in (\ref{eq:polyindetpsitildeall}) and
dividing the result, we obtain%
\begin{equation}
\prod\limits_{r=1}^{\bar{M}_{m}}\frac{\zeta_{r}^{(\ell,m)}\left(  \omega
_{1},x\right)  -\gamma_{j}^{(m)}}{\widetilde{\phi}_{r}^{(m)}-\gamma_{j}^{(m)}%
}=\frac{xN}{\widetilde{K}_{j}^{(m)}}\sum_{r=1}^{\bar{M}_{\ell}}\kappa
_{rj}^{\ell m}\frac{\gamma_{r}^{(\ell)}}{\gamma_{r}^{(\ell)}-\omega_{1}}
\label{eq:tempappend}%
\end{equation}
Likewise, taking $\zeta=\widetilde{\phi}_{j}^{(m)}$ in
(\ref{eq:polyindentzetalm}) and $\widetilde{\phi}=\zeta_{j}^{(\ell,m)}\left(
\omega_{1},x\right)  $ in (\ref{eq:polyindetpsitildeall}), we see that
\begin{gather*}
\prod\limits_{r=1}^{\bar{M}_{m}}\frac{\zeta_{r}^{(\ell,m)}\left(  \omega
_{1},x\right)  -\widetilde{\phi}_{j}^{(m)}}{\gamma_{r}^{(m)}-\widetilde{\phi
}_{j}^{(m)}}=1-x\Psi_{\ell,m}\left(  \omega_{1},\widetilde{\phi}_{j}%
^{(m)}\right) \\
\prod\limits_{k=1}^{\bar{M}_{m}}\frac{\widetilde{\phi}_{k}^{(m)}-\zeta
_{j}^{(\ell,m)}\left(  \omega_{1},x\right)  }{\gamma_{k}^{(m)}-\zeta
_{j}^{(\ell,m)}\left(  \omega_{1},x\right)  }=1-\widetilde{\Phi}^{(m)}\left(
\zeta_{j}^{(\ell,m)}\left(  \omega_{1},x\right)  \right)  .
\end{gather*}
Multiplying for all $j=1,\ldots,\bar{M}_{m},$dividing the two resulting
equations and using (\ref{eq:tempappend}) we obtain%
\[
\prod\limits_{j=1}^{\bar{M}_{m}}\frac{xN}{\widetilde{K}_{j}^{(m)}}\sum
_{r=1}^{\bar{M}_{\ell}}\kappa_{rj}^{\ell m}\frac{\gamma_{r}^{(\ell)}}%
{\omega_{1}-\gamma_{r}^{(\ell)}}=\prod\limits_{j=1}^{\bar{M}_{m}}\frac
{1-x\Psi_{\ell,m}\left(  \omega_{1},\widetilde{\phi}_{j}^{(m)}\right)
}{1-\widetilde{\Phi}^{(m)}\left(  \zeta_{j}^{(\ell,m)}\left(  \omega
_{1},x\right)  \right)  }%
\]
and therefore $\xi_{1}(\omega_{1},x)=0$. On the other hand, forcing
$\zeta=\phi_{0}^{(m)}$ in (\ref{eq:polyindentzetalm}) we directly obtain
$\xi_{2}(\omega_{1},x)=0$.

\section{\label{sec:prooflemma5}Proof of Lemma \ref{lemma:possols2}}

To see this, consider the function
\[
h^{\ell m}(\nu)=1-\sum_{r=0}^{\bar{M}_{\ell}}\widetilde{\kappa}_{r}^{\ell
m}\frac{\gamma_{r}^{(\ell)}}{\gamma_{r}^{(\ell)}-\nu}%
\]
and observe that, using the Cauchy-Schwarz inequality, we may write
\[
\left\vert 1-h^{\ell m}(\nu)\right\vert ^{2}\leq\sum_{k=1}^{\bar{M}_{m}}%
\frac{K_{k}^{(m)}}{N}\left\vert \frac{\gamma_{k}^{(m)}}{\gamma_{k}^{(m)}%
-\phi_{0}^{(m)}}\right\vert ^{2}\sum_{r=0}^{\bar{M}_{\ell}}\sum_{k=1}^{\bar
{M}_{\ell}}\kappa_{rk}^{\ell m}\left\vert \frac{\gamma_{r}^{(\ell)}}%
{\gamma_{r}^{(\ell)}-\nu}\right\vert ^{2}<\sum_{r=0}^{\bar{M}_{\ell}}%
\frac{K_{r}^{(\ell)}}{N}\left\vert \frac{\gamma_{r}^{(\ell)}}{\gamma
_{r}^{(\ell)}-\nu}\right\vert ^{2}%
\]
where the second inequality follows from $\kappa_{rk}^{\ell m}\leq\frac
{K_{r}^{(\ell)}}{N}$ and the fact that $\phi_{0}^{(m)}$ is never enclosed by
the contour $\mathcal{C}_{\omega_{m}}^{+}$. It can readily be seen that
$h^{\ell m}(\nu)$ cannot have zeros or poles on the contour $\mathcal{C}%
_{\omega_{\ell}}^{+}$. Hence, when $\nu$ is located in $\mathcal{C}%
_{\omega_{\ell}}^{+}$, we can write $\left\vert 1-h^{\ell m}(\nu)\right\vert
<1$ and by the Rouch\'{e}'s theorem we conclude that $h^{\ell m}(\nu)$ has the
same number of poles and zeros inside $\mathcal{C}_{\omega_{\ell}}^{+}$, as we
wanted to show.

\section{\label{sec:prooflemma6}Proof of Lemma \ref{lemma_sum_mus}}

By using the fact that $\nu_{j}^{(\ell,m)}$ are the solutions to
(\ref{eq:equation_mus}), we can identify the two polynomials
\begin{equation}
\prod\limits_{r=1}^{\bar{M}_{\ell}}\left(  \nu_{r}^{(\ell,m)}-\nu\right)
=\prod\limits_{r=1}^{\bar{M}_{\ell}}\left(  \gamma_{r}^{(\ell)}-\nu\right)
-\sum_{r=1}^{\bar{M}_{\ell}}\widetilde{\kappa}_{r}^{\ell m}\gamma_{r}^{(\ell
)}\prod\limits_{\substack{k=1 \\k\neq r}}^{\bar{M}_{\ell}}\left(  \gamma
_{k}^{(\ell)}-\nu\right)  . \label{eq:polynus}%
\end{equation}
By identifying the coefficients of the terms in $\nu^{\bar{M}_{m}-1}$ we
obtain (\ref{eq:sum_gamma_nus}). On \ the other hand, forcing $\nu=\gamma
_{j}^{(\ell)}$ in (\ref{eq:polynus}) allows us to write
\[
\prod\limits_{r=1}^{\bar{M}_{m}}\left(  \nu_{r}^{(\ell,m)}-\gamma_{j}^{(\ell
)}\right)  =-\widetilde{\kappa}_{j}^{\ell m}\gamma_{j}^{(\ell)}\prod
\limits_{\substack{k=1 \\k\neq j}}^{\bar{M}_{\ell}}\left(  \gamma_{k}^{(\ell
)}-\gamma_{j}^{(\ell)}\right)  .
\]
Taking derivatives on both sides of (\ref{eq:polynus}) and forcing $\nu
=\gamma_{j}^{(\ell)}$,
\[
\frac{\sum_{i=1}^{\bar{M}_{\ell}}\prod\nolimits_{\substack{r=1 \\r\neq
i}}^{\bar{M}_{\ell}}\left(  \nu_{r}^{(\ell,m)}-\gamma_{j}^{(\ell)}\right)
}{\prod\nolimits_{\substack{k=1 \\k\neq j}}^{\bar{M}_{\ell}}\left(  \gamma
_{k}^{(\ell)}-\gamma_{j}^{(\ell)}\right)  }=1-\sum_{\substack{r=1 \\r\neq j
}}^{\bar{M}_{\ell}}\widetilde{\kappa}_{r}^{\ell m}\frac{\gamma_{r}^{(\ell)}%
}{\gamma_{r}^{(\ell)}-\gamma_{j}^{(\ell)}}-\sum_{\substack{i=1 \\i\neq
j}}^{\bar{M}_{\ell}}\widetilde{\kappa}_{j}^{\ell m}\frac{\gamma_{j}^{(\ell)}%
}{\gamma_{i}^{(\ell)}-\gamma_{j}^{(\ell)}}%
\]
and using the above identity, we obtain (\ref{eq:sum_inv_nu_minus_gamma}).

\section{\label{sec:prooflemma7}Proof of Lemma \ref{lemma:transform_final}.}

By forcing $\nu=\widetilde{\phi}_{j}^{(\ell)}$ and $\nu=\gamma_{j}^{(\ell)}$
on both sides of the polynomial identity in (\ref{eq:polynus}) we directly
obtain the first and second equations. On the other hand, forcing
$\widetilde{\phi}=\nu_{j}^{(\ell,m)}$ and $\widetilde{\phi}=\gamma_{j}%
^{(\ell)}$\ in the polynomial identity in (\ref{eq:polyindetpsitildeall}) we
obtain the third and fourth equations. Finally, forcing $\nu=\phi_{0}^{(\ell
)}$ in (\ref{eq:polynus}) we obtain the last equation.

\end{document}